\documentclass[%
reprint,
 amsmath,amssymb,
 aps,
 superscriptaddress,
prc
]{revtex4-1}

\usepackage{graphicx}
\usepackage{dcolumn}
\usepackage{bm}
\usepackage{hyperref}

\usepackage[capitalise]{cleveref}
\usepackage{comment}

\begin{document}


\title{Nonequilibrium kinetic freeze-out properties in relativistic heavy ion collisions from energies employed at the RHIC beam energy scan to those available at the LHC}

\author{Jia Chen}\affiliation{Institute of Frontier and Interdisciplinary Science, Shandong University, Qingdao, Shandong, 266237, China}\affiliation{Key Laboratory of Particle Physics and Particle Irradiation of Ministry of Education, Shandong University, Qingdao, Shandong, 266237, China}
\author{Jian Deng}\affiliation{Institute of Frontier and Interdisciplinary Science, Shandong University, Qingdao, Shandong, 266237, China}\affiliation{Key Laboratory of Particle Physics and Particle Irradiation of Ministry of Education, Shandong University, Qingdao, Shandong, 266237, China}
\author{Zebo Tang}\affiliation{State Key Laboratory of Particle Detection and Electronics, University of Science and Technology of China, Hefei, Anhui, 230026, China}
\author{Zhangbu Xu}\affiliation{Brookhaven National Laboratory, Upton, New York, 11973, USA}
\author{Li Yi}~\email{li.yi@sdu.edu.cn}\affiliation{Institute of Frontier and Interdisciplinary Science, Shandong University, Qingdao, Shandong, 266237, China}\affiliation{Key Laboratory of Particle Physics and Particle Irradiation of Ministry of Education, Shandong University, Qingdao, Shandong, 266237, China}

\date{\today}

\begin{abstract}
In this paper, we investigate the kinetic freeze-out properties in relativistic heavy ion collisions at different collision energies. We present a study of standard Boltzmann-Gibbs Blast-Wave (BGBW) fits and Tsallis Blast-Wave (TBW) fits performed on the transverse momentum spectra of identified hadrons produced in Au + Au collisions at collision energies of $\sqrt{s_{\rm{NN}}}=$ 7.7 - 200~GeV at the Relativistic Heavy Ion Collider (RHIC), and in Pb + Pb collisions at collision energies of $\sqrt{s_{\rm{NN}}}=$ 2.76 and 5.02~TeV at the Large Hadron Collider (LHC). The behavior of strange and multi-strange particles is also investigated. We found that the TBW model describes data better than the BGBW one overall, and the contrast is more prominent as the collision energy increases as the degree of non-equilibrium of the produced system is found to increase. From TBW fits, the kinetic freeze-out temperature at the same centrality shows a weak dependence of collision energy between 7.7 and 39~GeV, while it decreases as collision energy continues to increase up to 5.02~TeV. The radial flow is found to be consistent with zero in peripheral collisions at RHIC energies but sizable at LHC energies and central collisions at all RHIC energies.
We also observed that the strange hadrons, with higher temperature and similar radial flow, approach equilibrium more quickly from peripheral to central collisions than light hadrons.
The dependence of temperature and flow velocity on non-equilibrium parameter ($q-1$) is characterized by two second-order polynomials. Both $a$ and $d\xi$ from the polynomials fit, related to the influence of the system bulk viscosity, increase toward lower RHIC energies.
\end{abstract}


\maketitle


\section{\label{sec:level1}Introduction}

Relativistic heavy ion collisions can create extreme hot and dense matter and reach a phase transition to Quark-Gluon Plasma (QGP). As QGP expands rapidly, its temperature drops and the system starts to enter the hadronic phase. Eventually the system reaches kinetic freeze-out when all particle interactions cease.
The particle spectra are thus frozen, which carry information about system dynamics at that freeze-out and even earlier.
Therefore, the study of differential transverse momentum ($p_{T}$) distributions of hadron particles is a useful tool to look into the evolution of the system, especially to extract system properties at its freeze-out phase space.
The Boltzmann-Gibbs Blast-Wave (BGBW) model~\cite{Schnedermann:1993ws,Schnedermann:1994gc} has been widely used to describe the produced system at kinetic freeze-out with its systemwise parameters characterizing the system radial flow velocity and temperature.

The BGBW model assumes that the produced system has reached local thermal equilibrium so that a Boltzmann distribution with a radial flow profile can be used to describe the particle spectra~\cite{Schnedermann:1993ws}.
However, the equilibrium distribution can only describe the very limited low $p_{T}$ spectra and is sensitive to the choice of a specific $p_T$ range. Tsallis statistics was introduced later in the literature to describe the particle production for an extended $p_{T}$ range in high energy collisions~\cite{De:2007zza,Wilk:2008ue,Alberico:1999nh,Osada:2008sw,Biro:2003vz,Bhattacharyya:2015hya}.
One advantage of those Tsallis statistical analyses compared to the Boltzmann-Gibbs statistical one is that a new parameter is introduced in the model to describe the degree of non-equilibrium in the system, which is especially important for $p$+$p$ collisions~\cite{Jiang:2013gxa} and peripheral A+A collisions~\cite{Tang:2008ud}.
It has been shown~\cite{STAR:2004bgh,Wilk:1999dr,Wong:2015mba,Urmossy:2011xk} that hard
processes dominate the particle production for $p_{T}$ $\textgreater$ 1.5~GeV/$c$ and the relativistic hard scattering for $p$+$p$ collisions leads to a transverse momentum distribution that resembles the Tsallis distribution.
As pointed out in Ref.~\cite{Wong:2015mba}, the transverse momentum spectra of jets or the following hadron spectra from hard scattering satisfy a power law distribution, and the power index is closely related to the new degree of non-equilibrium parameter in Tsallis statistics.
Ref.~\cite{Wong:2015mba,ALICE:2012aqc,Urmossy:2014gpa,Urmossy:2015kva,Rybczynski:2014ura} have depicted a synthesizing evolution from primary $p+p$ and peripheral A+A to central A+A collisions with the non-extensive statistical mechanical Tsallis distribution.
A possible microcanonical generalization of the Tsallis distribution has been proposed~\cite{Urmossy:2011xk}  which gives a good fit to data on fragmentation functions measured in $e^{+}e^{-}$ collisions for 0.01 $\leqslant$ $x$ $\leqslant$ 1.

In relativistic heavy-ion collisions, the system evolution is usually characterized by two stages of freeze-outs: the chemical freeze-out when further interactions (if any at all) do not alter the particle composition and the kinetic freeze-out when the momentum distribution of particles ceases to change. One could assume that the chemical and kinetic freeze-outs happen simultaneously at the hadronic and QGP phase boundary. At that moment, the system is at chemical and kinetic equilibrium with sudden freeze-out~\cite{Broniowski:2001we}.
The short-lived resonances would decay and alter the kinetic spectra of the stable particles observed by experiments~\cite{Mazeliauskas:2019ifr}. However, direct measurements
of resonance suppression in the experiments at RHIC~\cite{STAR:2004bgh,AggarwalPRC.84.034909,AbelevPRL.97.132301} and LHC~\cite{AcharyaPRC.99.024905} are not compatible with such a scenario. At the other extreme, one could assume that the chemical and kinetic freeze-outs occur at very different times.
The resonances created at chemical freeze-out would decay quickly but the system continues to evolve with elastic collisions among hadrons (e.g., $\pi^+\pi^-\leftrightarrow\rho$) and is at local thermal equilibrium until kinetic freeze-out (BGBW)~\cite{Retiere:2003kf}. In this implementation, the stable hadrons are at kinetic equilibrium (with flow) and its kinematic distribution is indistinguishable from the resonance decay because they are at kinetic equilibrium (local detail balance)~\cite{Abelev:2008ab}. The non-equilibrium TBW is in-between these two extremes. TBW attempts to take non-equilibrium fluctuation and possible resonance decay (or the kinetic detail balance) in a consistent macroscopic approach. It is also possible to treat such a two-stage freeze-out in a microscopic model taking into account the resonance yields at the kinetic freeze-out and not from the resonance yields at the chemical freeze-out~\cite{Motornenko:2019jha}.

The collision energy dependence of radial flow velocity and kinetic freeze-out temperature in high energy heavy ion collisions has been an interesting subject in the community and been extensively studied for all available collision energies.
In the energy range of the Heavy Ion Synchrotron to Super Proton Synchrotron, multiple studies agreed on an increasing trend for those two variables with an increase of the collision energy~\cite{Adamczyk:2017iwn,Andronic:2014zha,Zhang:2016tbf}.
From RHIC to the LHC energy range, however, the interpretation of the experimental results is model dependent to date.
For radial flow velocity, most studies found an increasing trend of flow velocity with increasing collision energy~\cite{Andronic:2014zha,Abelev:2012wca,Lao:2016gxv,Lao:2017skd,Zhang:2014jug,Zhang:2016tbf,Adamczyk:2017iwn} but the quantitative value and whether there is sizable flow in $p$+$p$ and peripheral A+A collisions are model dependent. For kinetic freeze-out temperature, some studies claimed an increasing trend of kinetic freeze-out temperature with increasing collision energy~\cite{Lao:2016gxv,Lao:2017skd} while others stated a decreasing trend~\cite{Das:2014qca,Zhang:2016tbf,Adamczyk:2017iwn,Luo:2015doi,Chatterjee:2015fua}, and some concluded little dependence on collision energy~\cite{Andronic:2014zha,Abelev:2012wca}.

In this paper, to extract kinetic freeze-out temperature and radial flow velocity, we use the blast-wave model with Tsallis statistics~\cite{Tang:2008ud,Shao:2009mu,Tang:2011xq} and compare to the Boltzmann-Gibbs statistics one~\cite{Schnedermann:1993ws,Abelev:2008ab,Abelev:2009bw} to simultaneously fit all the transverse momentum spectra of hadrons produced at mid-(pseudo)rapidity in $\sqrt{s_{\rm{NN}}}=$ 7.7, 11.5, 14.5, 19.6, 27, 39, 62.4, and 200~GeV Au + Au collisions at RHIC~\cite{Adamczyk:2017iwn,Adam:2019koz,Adam:2019dkq,Adare:2012uk,Abelev:2008ab,Abelev:2007ra,Aggarwal:2010ig,Abelev:2008aa,Abelev:2006jr,Adams:2003xp,Adare:2013esx,Adams:2006ke,Abelev:2007rw} and in $\sqrt{s_{\rm{NN}}}=$ 2.76~TeV~\cite{Abelev:2013vea,Abelev:2013xaa,ABELEV:2013zaa} and 5.02~TeV~\cite{Acharya:2019yoi} Pb + Pb collisions at the LHC. Such a systematic study on collision energy and centrality dependence of radial flow velocity, kinetic freeze-out temperature, and the degree of non-equilibrium from RHIC to the LHC energy range may shed light on the underlying physics in these collisions.
Strange and multi-strange particles, with smaller hadronic interaction cross-sections, are believed to decouple from the system earlier than hadrons with only light valence quarks~\cite{Shao:2009mu,Adams:2005dq,vanHecke:1998yu,Adams:2003fy,Petrovici:2009pd}. The kinetic freeze-out behaviors of strange and multi-strange particles are therefore also investigated separately.

This paper is organized as follows. In \Cref{sec:formalism}, we describe the analysis method used in this paper. Results and discussions are given in \Cref{sec:results}. The conclusion is summarized in the last section.

\section{\label{sec:formalism}Analysis method}
\subsection{Blast-Wave model}
BGBW is a phenomenological model for hadron spectra based on flowing local thermal sources with global variables of temperature $T$ and transverse flow profile $\beta$ \cite{Schnedermann:1993ws,Schnedermann:1994gc}. $T$ is the temperature of the local thermal sources which particles radiate from. While the longitudinal expansion is assumed to be boost invariant, the transverse radial flow velocity of the thermal source
is parametrized as $\beta(r)=\beta_{S}(\frac{r}{R})^n$ at radius $0\leqslant r\leqslant R$ with surface velocity $\beta_{S}$ and exponent $n$. The average radial flow velocity then can be written as $\langle\beta\rangle=\beta_{S}\cdot2/(2+n)$ \cite{Ristea:2013ara}. For an emitting source with Boltzmann-Gibbs distribution, the produced particle spectrum is therefore written in the form of
\begin{equation}
\begin{split}
   \frac{d^{2}N}{2\pi p_{T}dp_{T}dy}|_{y=0}  =&A\int^{R}_{0}rdrm_{T}I_{0}(\frac{p_{T}\sinh(\rho)}{T})\\
   &\cdot K_{1}(\frac{m_{T}\cosh(\rho)}{T}),
\end{split}
\end{equation}
where $A$ is a normalization factor.
$m_{T}=\sqrt{p^{2}_{T}+m^{2}}$ is the transverse mass of a particle.
$I_{0}$ and $K_{1}$ are the modified Bessel functions.
$\rho=\tanh^{-1}\beta$.
$T$ is the kinetic freeze-out temperature.
In order to compare with TBW results, we take $n=1$ for the BGBW model in this paper. With common freeze-out temperature $T$ and average radial flow velocity $\langle \beta\rangle$, the shape of the spectrum for each particle species is determined by its mass in BGBW.

\subsection{Tsallis Blast-Wave model}

TBW~\cite{Tang:2008ud,Shao:2009mu,Tang:2011xq,Ristea:2013ara}  is modified from the standard BGBW model when a Tsallis statistics replaces the conventional Boltzmann-Gibbs statistics for the particle emission distribution. The invariant differential particle yield in TBW is then written in the form of

\begin{small}
\begin{equation}\label{eq:TBW}
\begin{split}
   \frac{d^{2}N}{2\pi m_{T}dm_{T}dy}|_{y=0} & =A\int^{+y_{b}}_{-y_{b}}e^{\sqrt{y^{2}_{b}-y^{2}_{s}}}m_{T}\cosh(y_{s})dy_{s} \\
  & \times\int^{R}_{0}rdr\int^{\pi}_{-\pi}[1+\frac{q-1}{T}E_{T}]^{-1/(q-1)}d\phi,
\end{split}
\end{equation}
\end{small}

where
\begin{equation}
\begin{split}
   E_{T} & =m_{T}\cosh(y_{s})\cosh(\rho)-p_{T}\sinh(\rho)\cos(\phi).
\end{split}
\end{equation}
$y_{s}$ is the source rapidity.
$y_{b}$ is the beam rapidity.
$\phi$ is the particle emission angle in the rest frame of the thermal source.
$q$ is the parameter characterizing the degree of non-equilibrium of the produced system, which is the new parameter introduced in TBW compared to the BGBW model.
Although the applicability of such a model to high energy nuclear collisions is still under investigation, possible physics implications are available in the literature. The initial energy density in heavy ion collision has multiple hot spots caused by Color-Glass Condensate formation in a nucleon or individual nucleon-nucleon collisions.
Those hot spots are dissipated into the system, producing more particles, generating collective flow, and resulting in a temperature fluctuation at the initial state~\cite{Wilk:2008ue,Wilk:2009nn}. The initial state fluctuation is not guaranteed to be completely washed out by the medium, in QGP or hadron gas phases.
The survived fluctuation will leave imprints in spectra at low and intermediate $p_{T}$ range.
Such features in spectra will lead to $q$ values larger than 1 in the TBW model~\cite{Ristea:2013ara}.
When $q=1$, Eq.~\eqref{eq:TBW} recovers its familiar Boltzmann-Gibbs form.

In this paper, we also use a Tsallis blast-wave model with four fit parameters with different $q$ for mesons and baryons separately, referred to as TBW4. TBW4 was first proposed in reference \cite{Tang:2008ud} for a better description of meson and baryon spectra at $p$+$p$ collisions while the TBW fits with one single $q$ for all particles gave a very poor $\chi^{2}/nDoF$.
TBW without further description in this paper refers to the default one with three fit parameters that is the one using the same $q$ for both mesons and baryons.

\section{\label{sec:results}Results and discussions}
\subsection{Transverse momentum spectra}

\begin{figure*}[htbp]
\begin{center}
\includegraphics[width=1.0\textwidth]{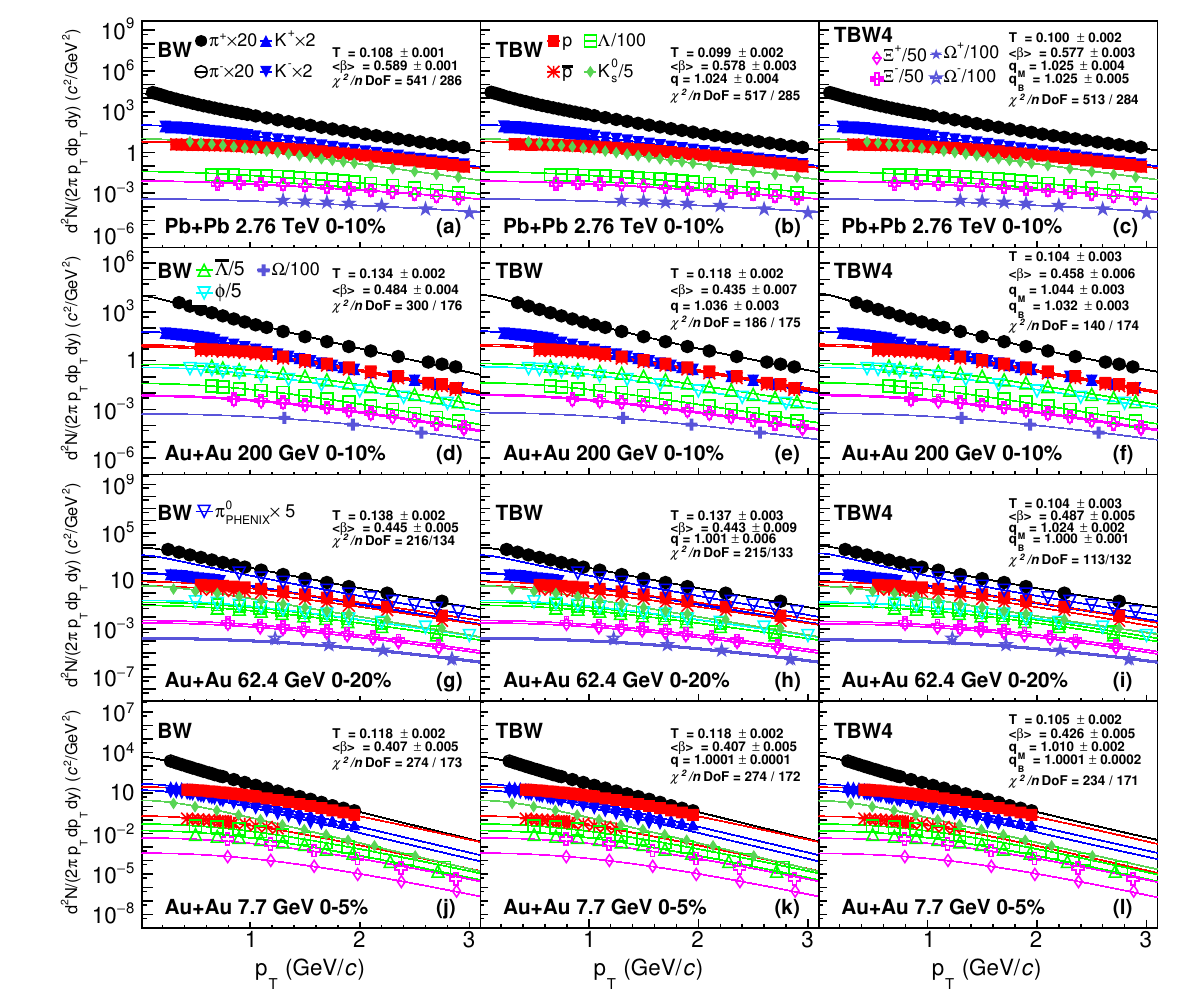}
\caption{Blast-wave model fits to hadron spectra in most central Pb + Pb and Au + Au collisions at $\sqrt{s_{\rm{NN}}}=$ 2.76~TeV, 200~GeV, 62.4~GeV, and 7.7~GeV from top to bottom panels.  The different symbols represent experimental data of different particle species. Uncertainties on experimental data represent quadratic sums of statistical and systematic uncertainties. The solid curves represent fit results for BGBW (left column), TBW (middle column), and TBW4 (right column).}\label{fig:yieldcentralall}
\end{center}
\end{figure*}

\begin{figure*}[htbp]
\begin{center}
\includegraphics[width=1.0\textwidth]{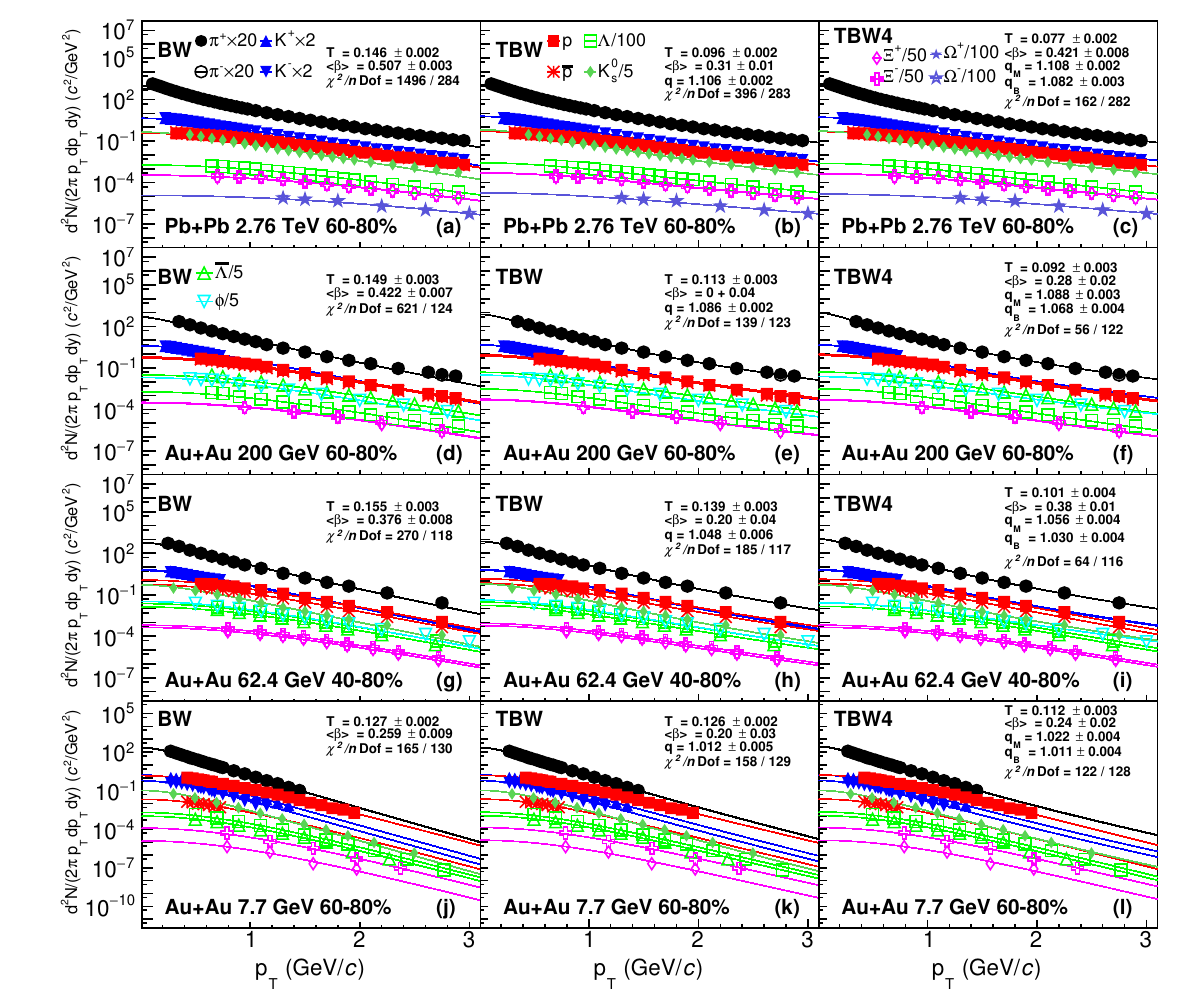}
\caption{Same as \cref{fig:yieldcentralall}, but for most peripheral collisions.}\label{fig:yieldperipheralall}
\end{center}
\end{figure*}

\begin{figure*}[htbp]
\begin{center}
\includegraphics[width=1.0\textwidth]{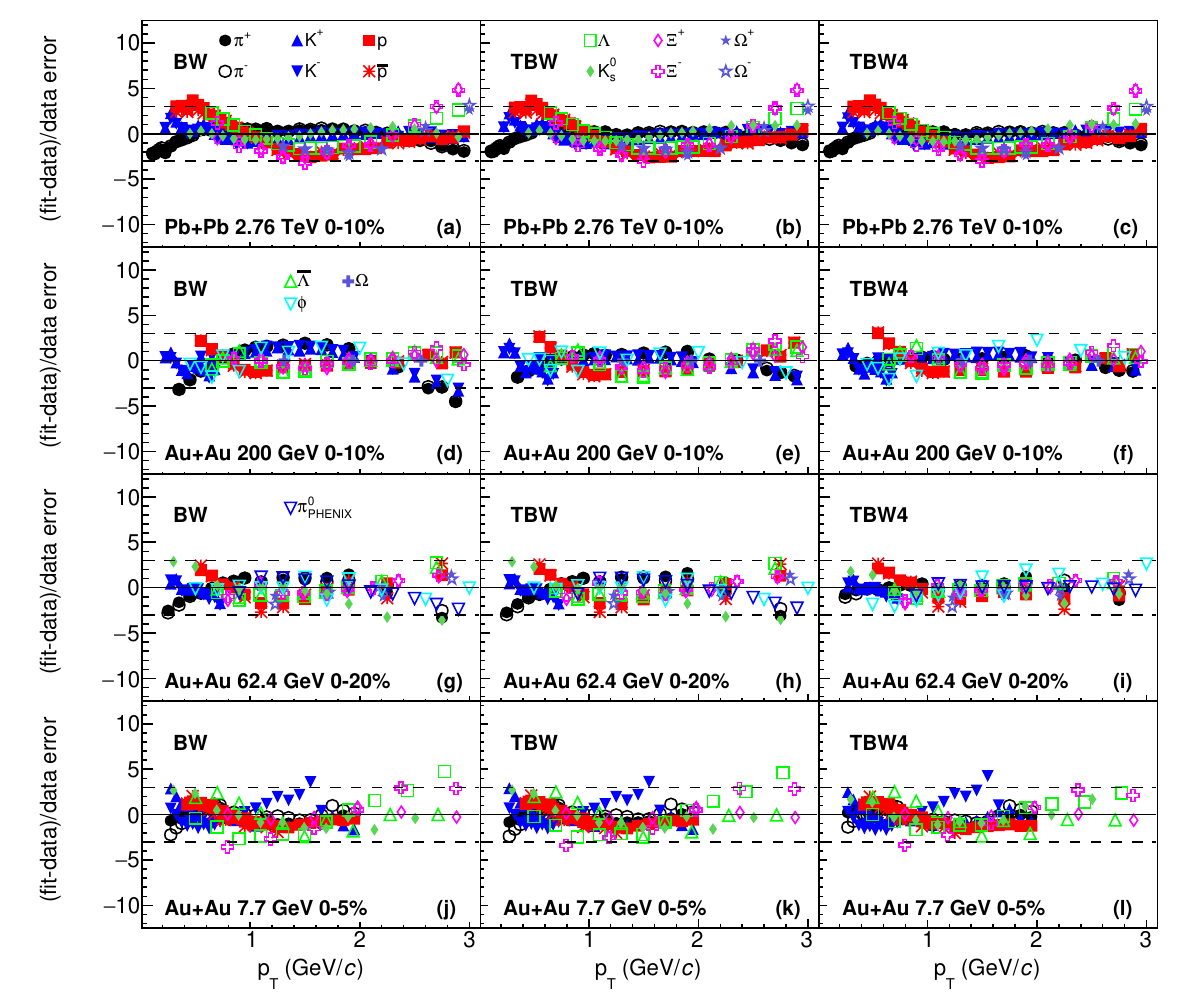}
\caption{The deviations of BGBW (left column), TBW (middle column), and TBW4 (right column) model fits to hadron spectra divided by data uncertainties in most central Pb + Pb and Au + Au collisions at $\sqrt{s_{\rm{NN}}}=$ 2.76~TeV, 200~GeV, 62.4~GeV, and 7.7~GeV from top to bottom panels. The different symbols are used to distinguish particle species. The dashed lines represent where the difference between model and experiment data is three times the error of data.}\label{fig:yieldcentralallSTDDEV}
\end{center}
\end{figure*}

\begin{figure*}[htbp]
\begin{center}
\includegraphics[width=1.0\textwidth]{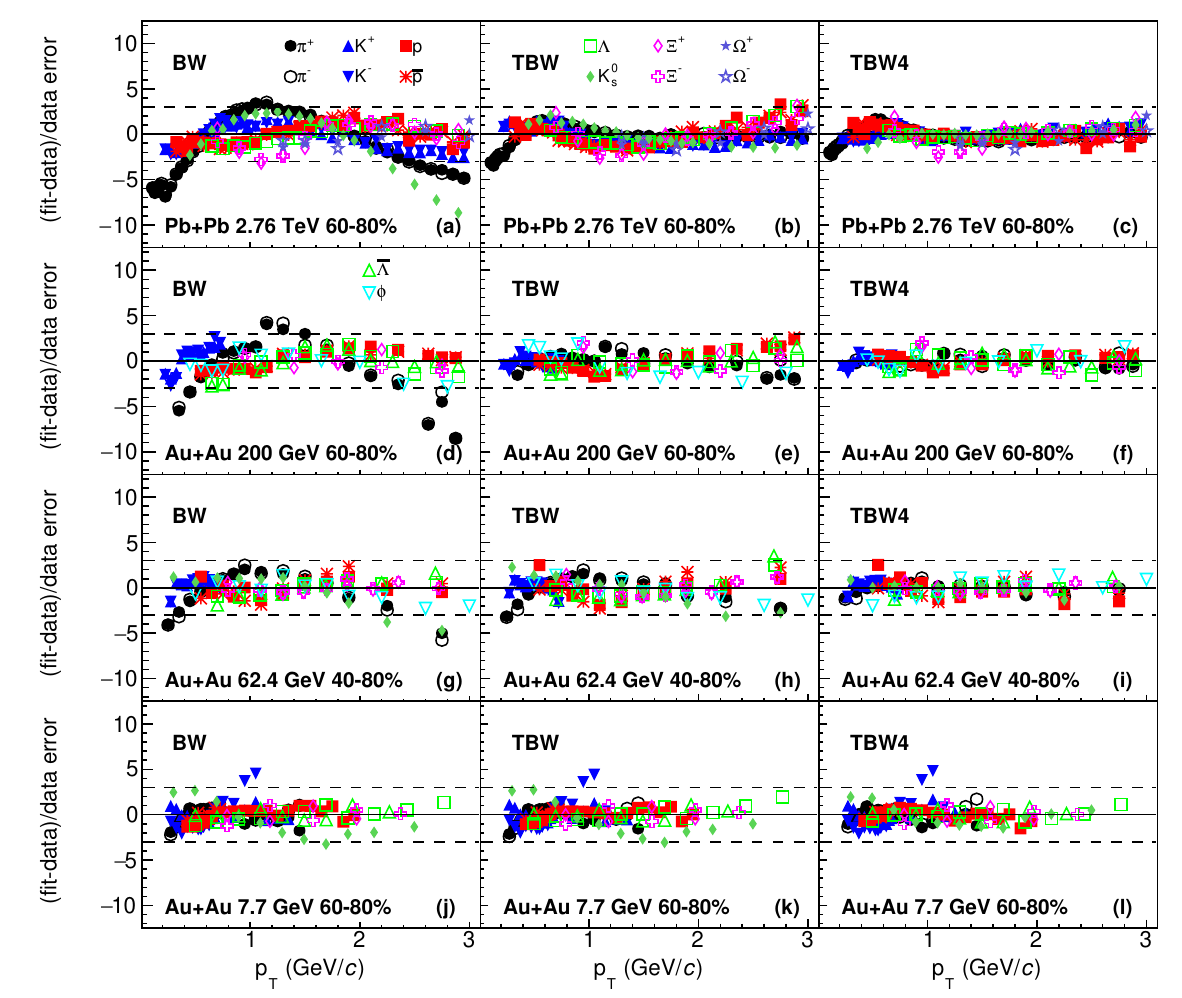}
\caption{Same as \cref{fig:yieldcentralallSTDDEV}, but for most peripheral collisions. }\label{fig:yieldperipheralallSTDDEV}
\end{center}
\end{figure*}

This section compares three blast-wave model fits of the transverse momentum spectra. \Cref{tabpaper} lists the particle spectra data used in this paper.
Those particle spectra form two species groups for the fit procedure: one with all available  hadrons, and another with charged pion, kaon, proton, and antiproton only. The first group aims to identify the common freeze-out properties for all particles. The latter is chosen to be consistent with previous publications \cite{Adamczyk:2017iwn} for an apple-to-apple comparison.
The reported experimental systematic and statistical uncertainties are combined as a quadratic sum for the fitting procedures.
For all fit procedures, the average flow velocity $\langle\beta\rangle$ is limited to the range of $0\leqslant\langle\beta\rangle\leqslant2/3$~\cite{Abelev:2008ab} for a better convergence in fitting and to avoid nonphysical results (negative value or faster than the speed of light)~\cite{Tang:2008ud}.
Furthermore, spectra fit range is limited to $p_{T}\leqslant$ 3~GeV/$c$ in order to have a comparable $p_T$ range for all energies and to focus on the bulk properties.
For the sake of concision, this section only shows the fits to spectra for all particles in most central and most peripheral centrality classes at four collision energies as examples in {\Crefname{figure}{Fig.}{Figs.}\Cref{fig:yieldcentralall,fig:yieldperipheralall,fig:yieldcentralallSTDDEV,fig:yieldperipheralallSTDDEV}}. \Cref{fig:yieldcentralall} (\Cref{fig:yieldperipheralall}) shows blast-wave fits to identified particle transverse momentum spectra in most central (most peripheral) collisions, with corresponding deviations of those fits to experimental data divided by data uncertainties shown in {\Crefname{figure}{Fig.}{Figs.}\Cref{fig:yieldcentralallSTDDEV}} ({\Crefname{figure}{Fig.}{Figs.}\Cref{fig:yieldperipheralallSTDDEV}}).
The fit results of kinetic freeze-out parameters for both species groups at various centrality classes and collision energies are discussed in the next section.
Those extracted fit parameters and $\chi^{2}/nDoF$ of TBW models are also summarized in  \Cref{tabTBW3pa1,tabTBW3pa2,tabTBW3pa3,tabTBW3pa4,tabTBW3pa5,tabTBW4pa1,tabTBW4pa2,tabTBW4pa3,tabBW1,tabBW2,tabBW3}.

\begin{table*}[htbp]
\caption{\label{tabpaper}Spectra data references}
\begin{ruledtabular}
\begin{tabular}{ccccc}

   $\rm system$  & $\rm\sqrt{s_{NN}}\; (\rm{GeV})$  & $\rm particle$  & $\rm collaboration$ &  $\rm {\it reference}$ \\
  \hline
  $\rm Au+Au$ &  $7.7, 11.5, 19.6, 27$     & $\pi^{\pm}, \, K^{\pm}, \, p, \, \bar{p} $  & ${\rm STAR}$ & ~\cite{Adamczyk:2017iwn}  \\
         &  & $K^{0}_{s} , \, \Lambda  , \, \bar{\Lambda} , \, \Xi^{+} , \, \Xi^{-}$  & ${\rm STAR}$ & ~\cite{Adam:2019koz} \\ \hline

  $\rm Au+Au$ &  $14.5$     & $\pi^{\pm}, \, K^{\pm}, \, p, \, \bar{p} $  & ${\rm STAR}$ & ~\cite{Adam:2019dkq}  \\ \hline

  $\rm Au+Au$ &  $39$     & $\pi^{\pm}, \, K^{\pm}, \, p, \, \bar{p} $  & ${\rm STAR}$ & ~\cite{Adamczyk:2017iwn}  \\
         &  & $K^{0}_{s} , \, \Lambda  , \, \bar{\Lambda} , \, \Xi^{+} , \, \Xi^{-}$  & ${\rm STAR}$ & ~\cite{Adam:2019koz} \\
           &  & $\pi^{0} $  & ${\rm PHENIX}$ &       ~\cite{Adare:2012uk}  \\  \hline

  $\rm Au+Au$ &  $62.4$     & $\pi^{\pm}, \, K^{\pm}, \, p, \, \bar{p} $  & ${\rm STAR}$ &   ~\cite{Abelev:2008ab}  \\

         &     & $\pi^{\pm},  \, p, \, \bar{p} $  & ${\rm STAR}$ &     ~\cite{Abelev:2007ra}  \\

         &  & $K^{0}_{s} , \, \Lambda  , \, \bar{\Lambda} , \, \Xi^{+} , \, \Xi^{-} , \, \Omega^{+} , \, \Omega^{-}   $  & ${\rm STAR}$ &   ~\cite{Aggarwal:2010ig}  \\

         &     & $\phi $  & ${\rm STAR}$ &       ~\cite{Abelev:2008aa}  \\

         &     & $\pi^{0} $  & ${\rm PHENIX}$ &       ~\cite{Adare:2012uk}  \\  \hline

  $\rm Au+Au$ &  $200$    & $\pi^{\pm}, \,  p, \, \bar{p} $  & ${\rm STAR}$ &
  ~\cite{Abelev:2006jr}  \\

        &   & $ K^{\pm} $  & ${\rm STAR}$ &
  ~\cite{Adams:2003xp}  \\

     &   & $ K^{\pm} $  & ${\rm PHENIX}$ &         ~\cite{Adare:2013esx}  \\

         &  & $ \Lambda  , \, \bar{\Lambda} , \, \Xi^{+} , \, \Xi^{-}, \, \Omega$  & ${\rm STAR}$ &
  ~\cite{Adams:2006ke}  \\

        &     & $\phi $  & ${\rm STAR}$ &
  ~\cite{Abelev:2007rw}  \\  \hline

  $\rm Pb+Pb$ &  $2760$     & $\pi^{\pm}, \, K^{\pm}, \, p, \, \bar{p} $  & ${\rm ALICE}$ &   ~\cite{Abelev:2013vea}  \\

         &  & $K^{0}_{s} , \, \Lambda  $  & ${\rm ALICE}$ & ~\cite{Abelev:2013xaa}  \\

         &  & $\Xi^{+} , \, \Xi^{-} , \, \Omega^{+} , \, \Omega^{-}$  & ${\rm ALICE}$ &
  ~\cite{ABELEV:2013zaa} \\\hline

  $\rm Pb+Pb$ &  $5020$     & $\pi^{\pm}, \, K^{\pm}, \, p, \, \bar{p} $  & ${\rm ALICE}$ &   ~\cite{Acharya:2019yoi}  \\

\end{tabular}
\end{ruledtabular}
\end{table*}

BGBW and TBW models are compared in the left and middle panels of {\Crefname{figure}{Fig.}{Figs.}\Cref{fig:yieldcentralall,fig:yieldperipheralall,fig:yieldcentralallSTDDEV,fig:yieldperipheralallSTDDEV}}. From top to bottom panels of each figure, the spectra ({\Crefname{figure}{Fig.}{Figs.}\Cref{fig:yieldcentralall,fig:yieldperipheralall}}) or difference between model and experiment data divided by the error of data ({\Crefname{figure}{Fig.}{Figs.}\Cref{fig:yieldcentralallSTDDEV,fig:yieldperipheralallSTDDEV}}) in Pb + Pb or Au+Au collisions at $\sqrt{s_{\rm{NN}}}=$ 2.76~TeV, 200~GeV, 62.4~GeV and 7.7~GeV are presented.
At LHC and top RHIC energies, the deviation of the BGBW model from experimental data is larger in peripheral than in central collisions.
As beam energy decreases, the deviation of BGBW model fit from data decreases. We note that there are fewer experimental data at $p_{T}>2$~GeV/$c$ for energies below 39 GeV.
Overall, TBW has better fits and has smaller $\chi^{2}/nDoF$ than BGBW.
TBW agrees with most data points within three-$\sigma$ standard deviation from experimental data.
TBW yields a smaller $q$ toward lower beam energy, which indicates that the system is closer to equilibrium state toward lower energy. For all LHC and RHIC energies, TBW fits in peripheral collisions have larger $q$ values than those in central collisions at the same collision energy.
In short, the TBW model performs much better than BGBW and the non-equilibrium seems to be necessary for peripheral collisions at high energies. As the BGBW model assumes thermal equilibrium and the TBW model uses non-equilibrium statistics, the above observations suggest that the collision system deviates more from thermal equilibrium at higher energy, especially in peripheral collisions.

The comparison of TBW (with a single $q$) and TBW4 (with separate $q$ for meson and baryon) is shown in the middle and right panels of \cref{fig:yieldcentralall,fig:yieldperipheralall,fig:yieldcentralallSTDDEV,fig:yieldperipheralallSTDDEV}.
TBW4 has even smaller $\chi^{2}/nDoF$ than TBW for all LHC and RHIC energies, while the improvement is larger in peripheral than central collisions.
For TBW4, the non-equilibrium parameter $q$ of baryons is found to be smaller than that of mesons, as baryons have steeper spectra than mesons.
Baryons used in the fitting include mostly strange ($\Lambda$) and multi-strange ($\Xi$ and $\Omega$) particle species and strangeness has smaller $q$ value and higher freeze-out temperature. More details on different freeze-out will be discussed later in \Cref{sec:par}.

\subsection{\label{sec:par}Kinetic freeze-out parameters}

\begin{figure*}[htbp]
\begin{center}
\includegraphics[width=1.0\textwidth]{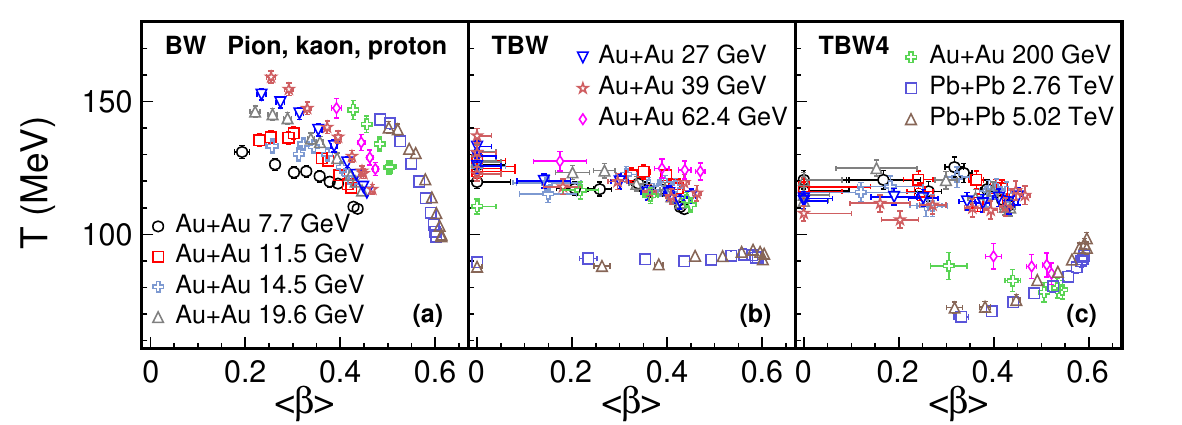}
\caption{Variation of $T$ with $\langle\beta\rangle$ for different energies and centralities from BGBW (left panel), TBW (middle panel), and TBW4 (right panel) fits to $p_{T}$ spectra of only positive and negative pions, kaons, and protons.
Symbols with the same style represent different centrality classes at the same colliding energy. For a given energy, from left to right, the centrality moves from peripheral to central collision.
} \label{fig:parameterTbetacomparison}
\end{center}
\end{figure*}

\begin{figure*}[htbp]
\begin{center}
\includegraphics[width=1.0\textwidth]{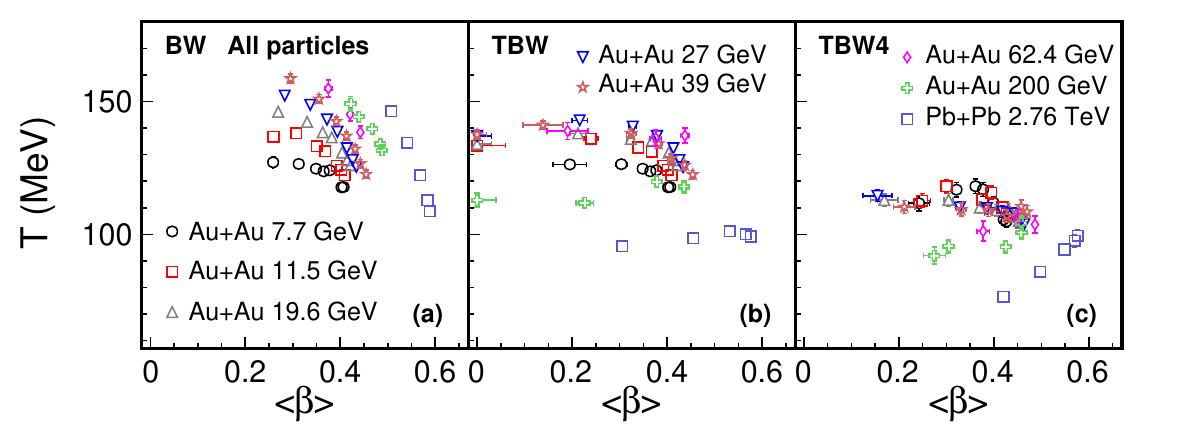}
\caption{Same as \cref{fig:parameterTbetacomparison}, but for all hadrons including strange and multi-strange particles.
} \label{fig:parameterTbetaallcomparison}
\end{center}
\end{figure*}

\begin{figure*}[htbp]
\begin{center}
\includegraphics[width=0.68\textwidth]{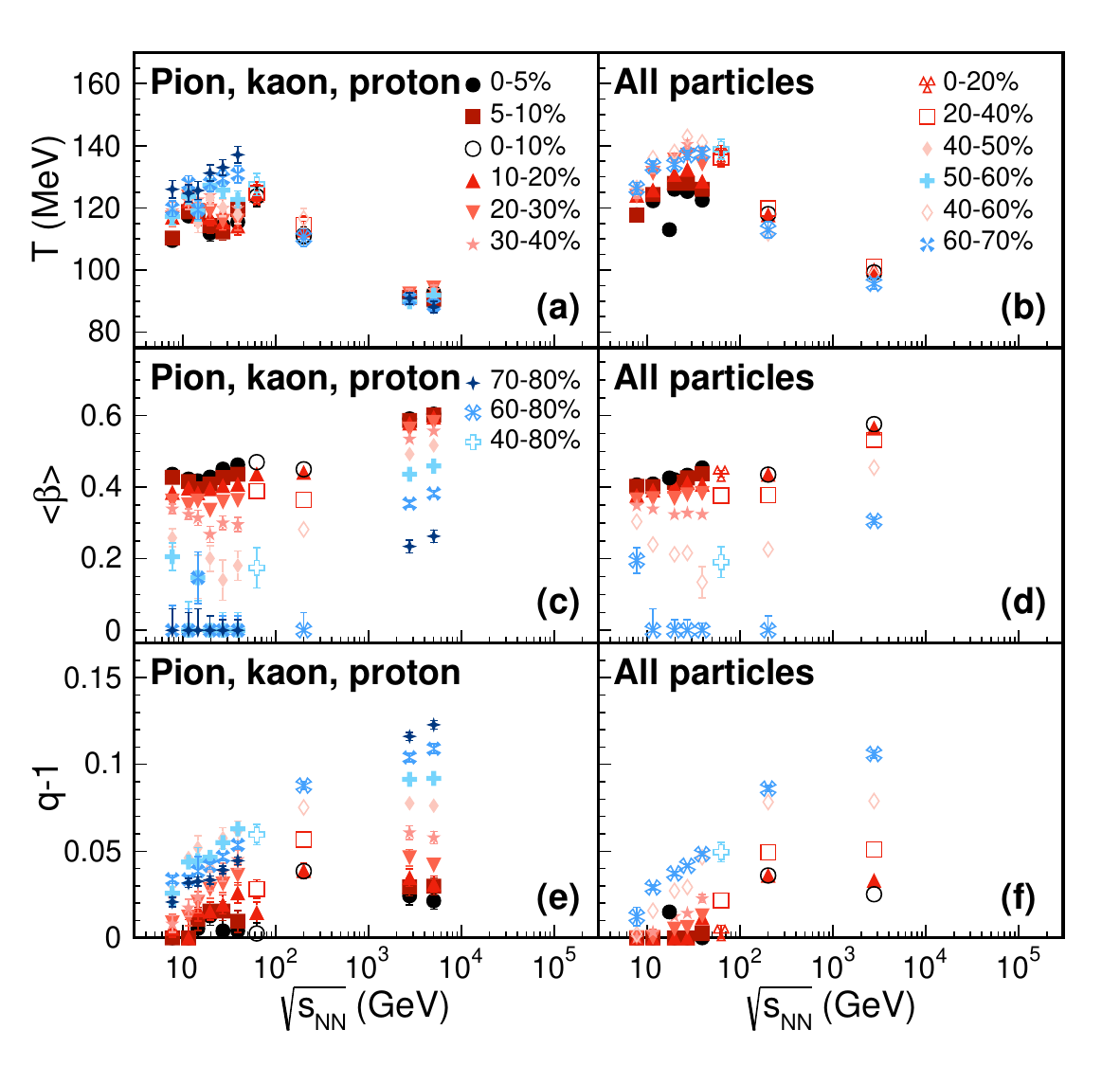}
\caption{Collision energy dependence of the extracted kinetic freeze-out parameters for heavy ion collisions of different centralities in TBW fit of $p_{T}$ spectra for only positive and negative pions, kaons, and protons (left column), and all particles (right column). The kinetic freeze-out temperature $T$, average transverse radial flow velocity $\langle\beta\rangle$, and  non-equilibrium parameter $q-1$ are shown in the top, middle and bottom panels, respectively. The results for all particles in most central Pb+Pb collisions at $\sqrt{s_{\rm{NN}}}=$ 17.3~GeV are from Ref.~\cite{Shao:2009mu}.
} \label{fig:parametercomparison}
\end{center}
\end{figure*}

\begin{figure*}[htbp]
\begin{center}
\includegraphics[width=0.68\textwidth]{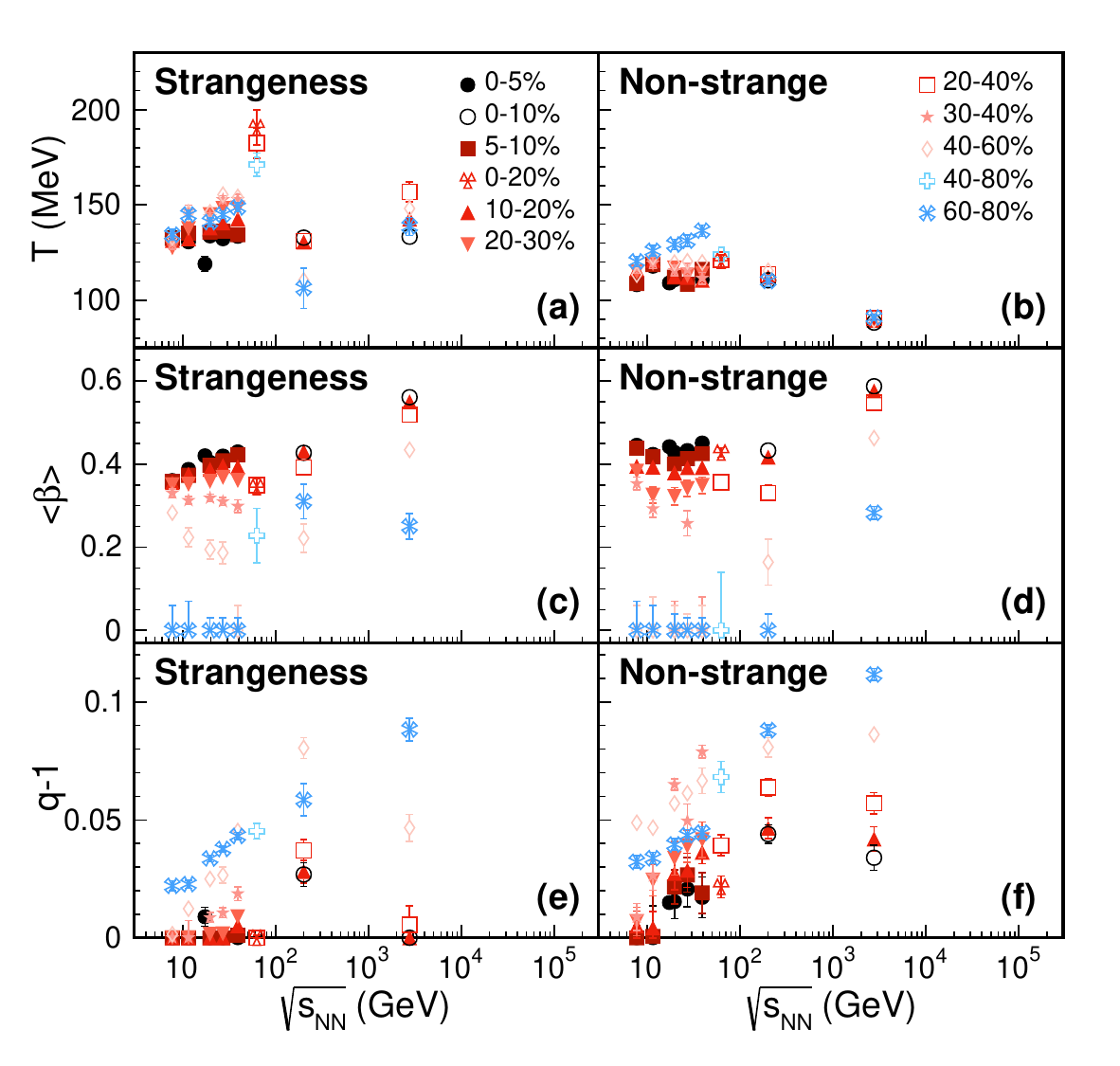}
\caption{Same as \cref{fig:parametercomparison}, but for strangeness only (left column), and non-strange particles (right column). The results in most central Pb+Pb collisions at $\sqrt{s_{\rm{NN}}}=$ 17.3~GeV are from Ref.~\cite{Shao:2009mu}.
} \label{fig:parameterstrangecomparison}
\end{center}
\end{figure*}

 The extracted results for temperature $T$ and average radial flow velocity $\langle \beta\rangle $ from BGBW, TBW, and TBW4 are compared in {\Crefname{figure}{Fig.}{Figs.}\Cref{fig:parameterTbetacomparison,fig:parameterTbetaallcomparison}}. The beam energy, centrality, and particle species dependences of $T$, $\langle\beta\rangle$, and $q$ from the TBW model are investigated in {\Crefname{figure}{Fig.}{Figs.}\Cref{fig:parametercomparison,fig:parameterstrangecomparison}}.

The dependence of $T$ on $\langle\beta\rangle$ of BGBW, TBW, and TBW4 is shown in {\Crefname{figure}{Fig.}{Figs.}\Cref{fig:parameterTbetacomparison}} for charged pions, kaons, and protons, and in {\Crefname{figure}{Fig.}{Figs.}\Cref{fig:parameterTbetaallcomparison}} for all available hadrons including strange and multi-strange particles.
Symbols with same color represent A+A collisions at same beam energy for different centrality classes.
In general, fit parameters in {\Crefname{figure}{Fig.}{Figs.}\Cref{fig:parameterTbetaallcomparison}} for all particles have smaller fit uncertainties than those in {\Crefname{figure}{Fig.}{Figs.}\Cref{fig:parameterTbetacomparison}} for charged pions, kaons, and protons only, as more particles are used to study their common freeze-out properties. Other than that, these two species groups give similar results for $T$ dependence on $\langle\beta\rangle$.
BGBW results shown on the left panel display an anti-correlation between $T$ and $\langle\beta\rangle$. At the same collision energy, $T$ decreases and $\langle\beta\rangle$ increases as the system moves from peripheral to central collisions.  As collision energy increases, the anti-correlation curve moves toward high $\langle\beta\rangle$. Such anti-correlation behavior was also reported in Ref.~\cite{Adamczyk:2017iwn}.
TBW results in the middle panel, however, are different from BGBW ones. $T$ from TBW has much weaker dependence on centrality than the BGBW one. For example, at the LHC energies, the increase of $T$ from most central to most peripheral collisions is around 40\% in BGBW. In contrast, the variation of $T$ is only about 5\% in TBW. Similar behavior of weak centrality dependence for temperature was also reported in Ref.~\cite{Mazeliauskas:2019ifr}, when further resonance decay after sudden equilibrium freeze-out is considered in BGBW. The resonance decay is one of the microscopic non-equilibrium sources in the macroscopic TBW approach. The common observation from these two studies supports the expectation that the strong centrality dependence of freeze-out temperature in the BGBW model is rooted in its incapability to describe the non-equilibrium system.
The parameter $\langle\beta\rangle$ in most central collisions from TBW is between 0.4 and 0.5 for RHIC energies and around 0.6 at the LHC, similar with those in BGBW. In peripheral collisions $\langle\beta\rangle$ is lower in TBW than that in BGBW, while it even reaches zero value for the most peripheral collisions at RHIC energies. It seems that from the TBW model's viewpoint hadron scatterings (or QGP droplets if any) are not sufficient to produce a large collective radial flow or to maintain a thermal equilibrium  in peripheral collisions at RHIC. The BGBW model, while lacking a knob for non-equilibrium degree, has to boost its radial flow parameter in a struggle to fit the high yields of the spectra at intermediate $p_T$ in the peripheral collisions.
In {\Crefname{figure}{Fig.}{Figs.}\Cref{fig:parameterTbetacomparison,fig:parameterTbetaallcomparison}}, $T$ and $ \langle\beta\rangle$ from TBW4 on the right panels are similar to those from TBW with one single $q$ in the middle panels and different from those from BGBW on the left panels. There is a weaker centrality dependence for $T$ and lower $\langle\beta\rangle$ values at peripheral collisions for both TBW models than BGBW.
We observed that at the LHC energies, $T$ and $\langle\beta\rangle$ from TBW4 tend to have a positive correlation rather than anti-correlation as in BGBW fits or a lack of correlation as in default TBW.
The  fit parameter values are slightly different for the two TBW models as discussed below.
In {\Crefname{figure}{Fig.}{Figs.}\Cref{fig:parameterTbetacomparison}}, for charged pions, kaons, and protons, $T$ from TBW4 is lower than the one in default TBW at $\sqrt{s_{\rm{NN}}}$ above 62.4~GeV. For lower beam energies,  within the uncertainties, $T$ values from these two models appear to be consistent. For the cases including all hadrons and shown in {\Crefname{figure}{Fig.}{Figs.}\Cref{fig:parameterTbetaallcomparison}}, the results are in general with smaller uncertainties of all the fit parameters, and $T$ from the TBW4 are lower than the default TBW for all the energies.
The main observation from the comparison of BGBW and two TBW models is that $T$ in TBW models has weaker centrality dependence and $\langle\beta\rangle$ at peripheral collisions is lower than those in BGBW.

\Cref{fig:parametercomparison} shows the energy and centrality dependence of kinetic freeze-out parameters $T$, $\langle\beta\rangle$, and $(q-1)$ from the TBW model for charged pions, kaons, and protons in the left column and for all particles in the right column.
The kinetic freeze-out temperature $T$ in panel (a) and (b) of {\Crefname{figure}{Fig.}{Figs.}\Cref{fig:parametercomparison}} shows weak collision energy dependence at $\sqrt{s_{\rm{NN}}}$ of 7.7 - 39~GeV, while it drops from $\sqrt{s_{\rm{NN}}} =$  62.4~GeV to 5.02~TeV in panel (a) and 2.76~TeV in panel (b).
At 7.7 - 39~GeV, at any given collision energy, $T$ only decreases marginally from peripheral to central collisions. At 62.4~GeV to 5.02~TeV, the centrality dependence of $T$ is even smaller.
In contrast, as discussed in the previous section, in BGBW, $T$  decreases notably from peripheral to central collisions. In most of the peripheral collisions, BGBW deviates significantly from data with larger $\chi^2/nDoF$. Meanwhile, the non-equilibrium parameter $q$ in TBW for peripheral collisions also increases with increasing collision energy as shown in the bottom two panels of {\Crefname{figure}{Fig.}{Figs.}\Cref{fig:parametercomparison}}. The strong centrality dependence of $T$ in BGBW may be synthetic to the model's incapacity to incorporate the large non-equilibrium effect of the system in the peripheral collisions.
The average transverse radial flow velocity $\langle\beta\rangle$ shown in panel (c) and (d) of {\Crefname{figure}{Fig.}{Figs.}\Cref{fig:parametercomparison}}, for most central collisions, is between 0.4 and 0.5 at RHIC energies, and around 0.6 at the LHC energies. $\langle\beta\rangle$ decreases from central to peripheral collisions.  In most peripheral collisions, $\langle\beta\rangle$ drops to zero at RHIC energies and is less than 0.3 at the LHC energies. For the most peripheral collisions at RHIC, the system in general fails to generate a rapid radial expansion.
The non-equilibrium parameter ($q-1$) in panel (e) and (f) of {\Crefname{figure}{Fig.}{Figs.}\Cref{fig:parametercomparison}} is small in central Au+Au collisions, suggesting that the produced particles are approaching thermal equilibrium. In peripheral collisions, ($q-1$) increases from less than 0.04 at 7.7~GeV to more than 0.1 at 5.02~TeV, indicating an increasing deviation from Boltzmann statistics as collision energy increases.
The centrality dependence of the ($q-1$) parameter suggests an evolution from an almost thermalized system in the central collisions towards a highly off-equilibrium system in the peripheral collisions.
Such large ($q-1$) is also found in the study of $p$+$p$ collision~\cite{Jiang:2013gxa}. This may be because the energy density fluctuations at initial state due to Color-Glass Condensate formation or individual hard scattering (minijets) inside a nucleus-nucleus collision increase as collision energy increases. Such fluctuations are not completely washed out by subsequent QGP evolution or hadronic interactions and leave footprints in final state particle spectra at the $p_{T}$ range in our paper~\cite{Tang:2008ud}.

In general, the group of charged pion, kaon, and proton and the group of all particles as shown in {\Crefname{figure}{Fig.}{Figs.}\Cref{fig:parametercomparison}} produce similar kinetic freeze-out parameters in TBW fits. Small difference can be identified with slightly higher $T$ and lower $q$ for the group with all particles than that with only the $\pi/K/p$. Such difference may come from the influence of particle species as the group of all particles contains more strange particles.
The direct comparison of non-strange and strangeness in {\Crefname{figure}{Fig.}{Figs.}\Cref{fig:parameterstrangecomparison}} confirms that the strange hadrons have higher temperature ($T$) and a smaller non-equilibrium degree ($q$) than those of non-strange hadrons, while their radial flow values ($\langle\beta\rangle$) are similar. A higher temperature indicates an earlier decoupling of strange hadrons from the system. The smaller $q$ in the strangeness group than the non-strangeness group and a similar $\langle\beta\rangle$ between those two groups suggest that the system is closer to an equilibrium state when the strangeness hadrons decouple from the system and  further hadronic interactions do not increase the system's radial flow velocity.
A possible conclusion is that the hadronic phase does not increase radial flow of light hadrons significantly at RHIC and LHC energies, and instead drives the system toward non-equilibrium: the system in central collisions has approached thermal equilibrium at the partonic phase, and the later hadronic scattering drives the system off equilibrium and does not increase the radial flow of copiously produced light hadrons~\cite{Shao:2009mu}.
Another interesting observation is that in {\Crefname{figure}{Fig.}{Figs.}\Cref{fig:parameterstrangecomparison}} (b) for non-strange particles the kinetic freeze-out temperature of the central collisions decreases from RHIC to LHC energies in the TBW model, while in {\Crefname{figure}{Fig.}{Figs.}\Cref{fig:parameterstrangecomparison}} (a)  strangeness does not show this behavior. A possible explanation is that the system at the LHC has higher flow velocity and larger volume than that at RHIC and maybe needs more time for all particles to kinetic freeze-out (``cool'') in the expansion during the hadronic phase.

\begin{figure*}[htbp]
\begin{center}
\includegraphics[width=0.8\textwidth]{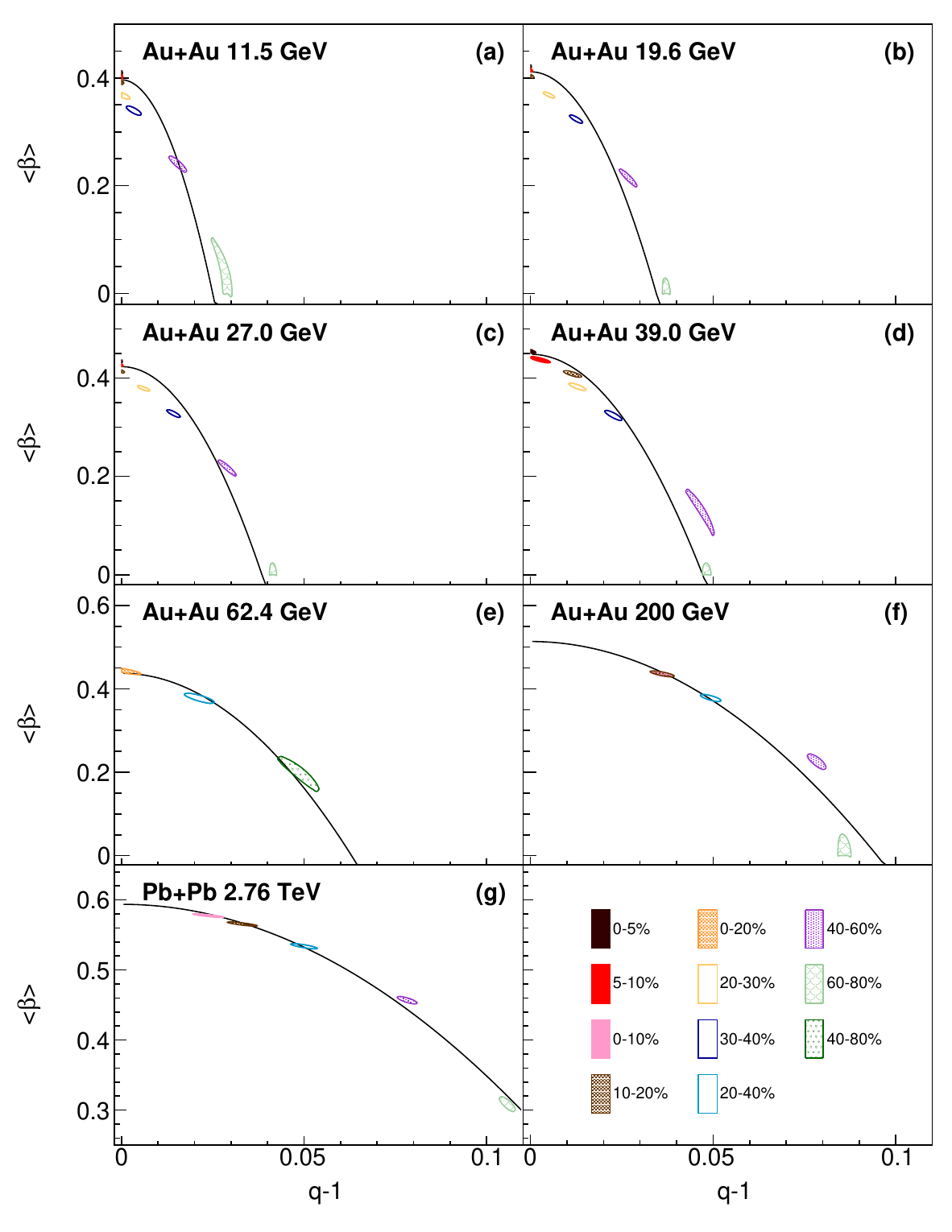}
\caption{Extracted average radial flow velocity $\langle\beta\rangle$ as a function of non-equilibrium degree $(q-1)$ obtained in TBW fits of $p_{T}$ spectra of all particles. Each block is a one-$\sigma$ contour from the error matrix of the TBW fit for a given centrality of Au + Au (Pb + Pb) collisions. The curves represent quadratics fits in the form of $\langle\beta\rangle =\langle\beta\rangle_0-a(q-1)^{2}$.
}\label{fig:pabetaqall}
\end{center}
\end{figure*}

\begin{figure*}[htbp]
\begin{center}
\includegraphics[width=0.9\textwidth]{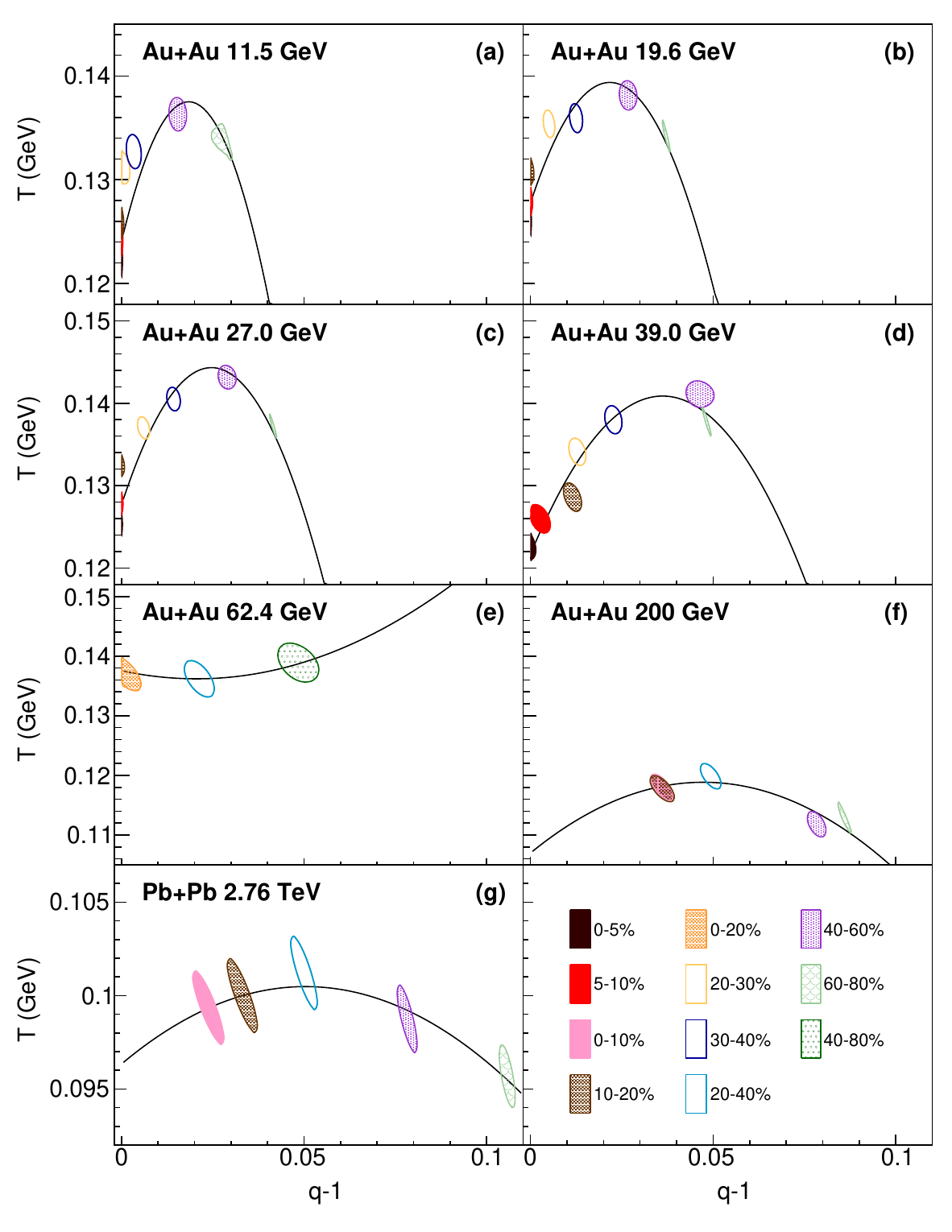}
\caption{Extracted kinetic freeze-out temperature $T$ as a function of non-equilibrium degree $(q-1)$ obtained in TBW fits of $p_{T}$ spectra of all particles. Each block is a one-$\sigma$ contour from the error matrix of the TBW fit for a given centrality of Au + Au (Pb + Pb) collisions.  The curves represent quadratics fits in the form of $T=T_0+b(q-1)-d\xi(q-1)^2$.
}\label{fig:paTqall}
\end{center}
\end{figure*}

\begin{figure*}[h]
\begin{center}
\includegraphics[width=0.42\textwidth]{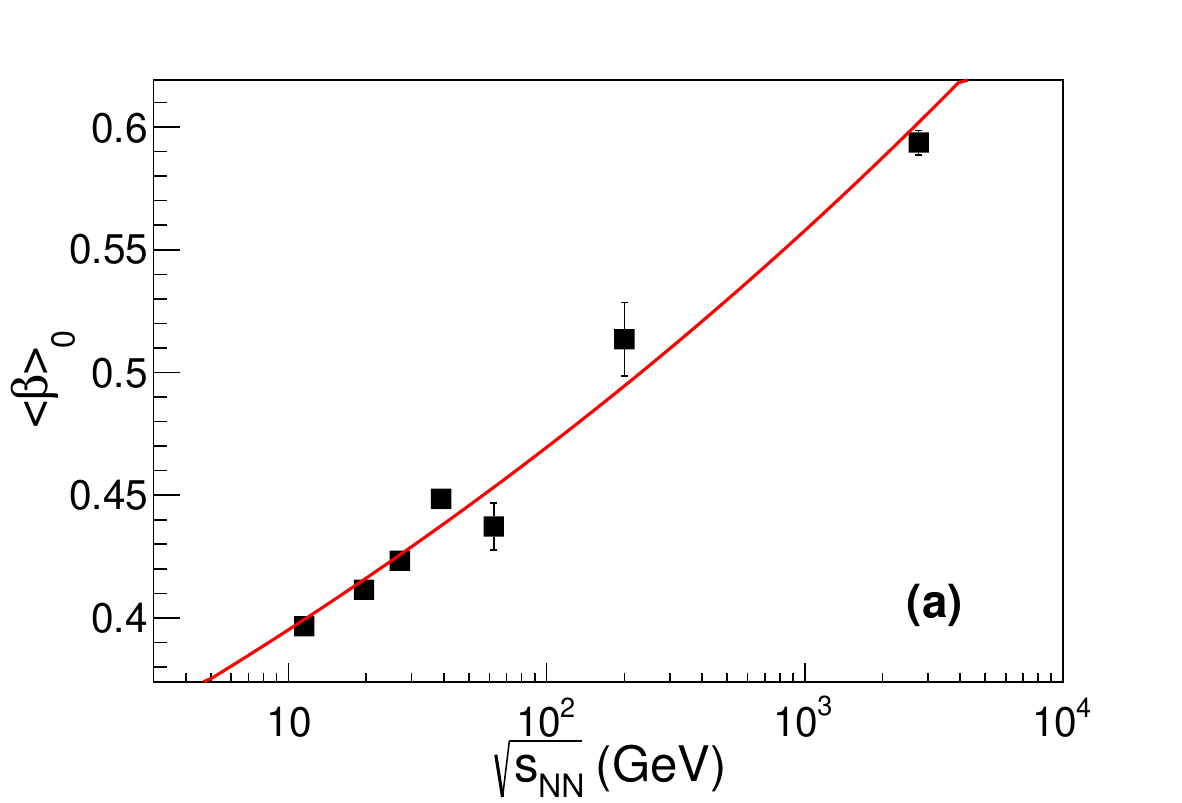}
\includegraphics[width=0.42\textwidth]{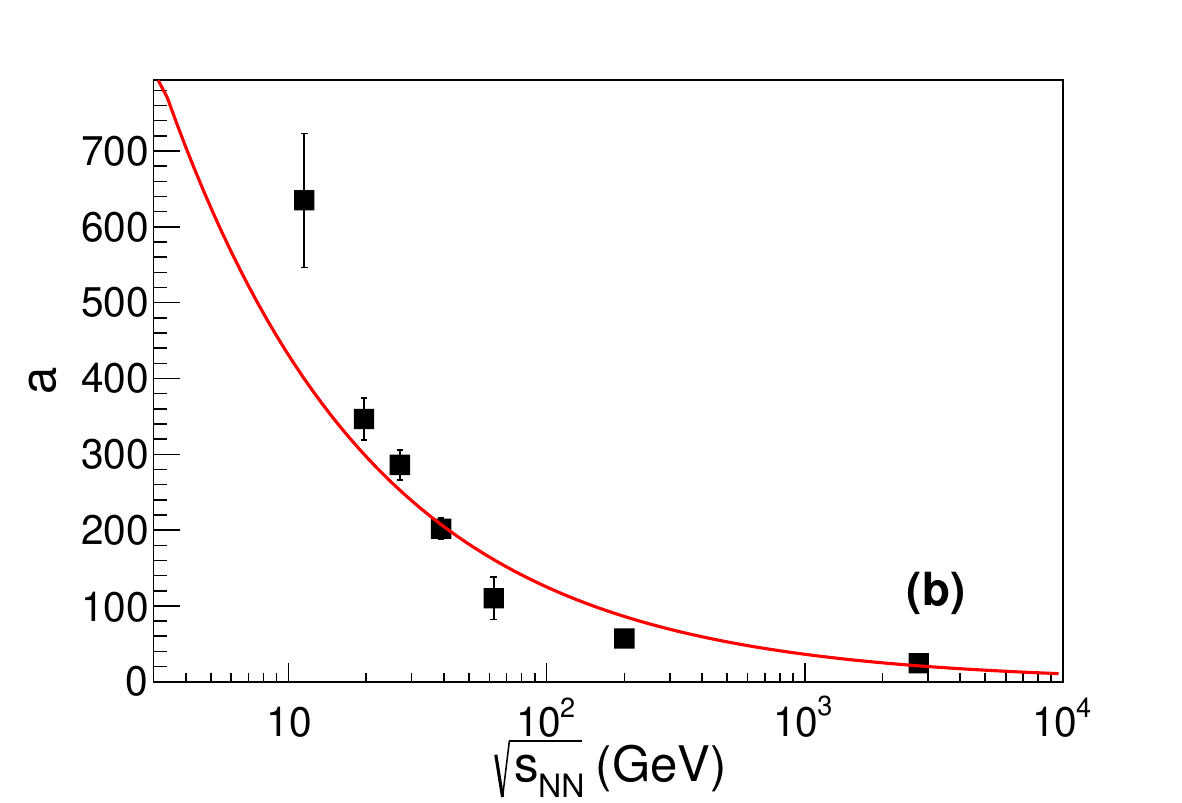}
\caption{Collision energy dependence of parameters $\langle\beta\rangle_0$ and $a$ in $\langle\beta\rangle =\langle\beta\rangle_0-a(q-1)^{2}$. Curves are to guide the eye.}\label{fig:pabetaqall_enedep}
\end{center}
\end{figure*}

\begin{figure*}[h]
\begin{center}
\includegraphics[width=0.42\textwidth]{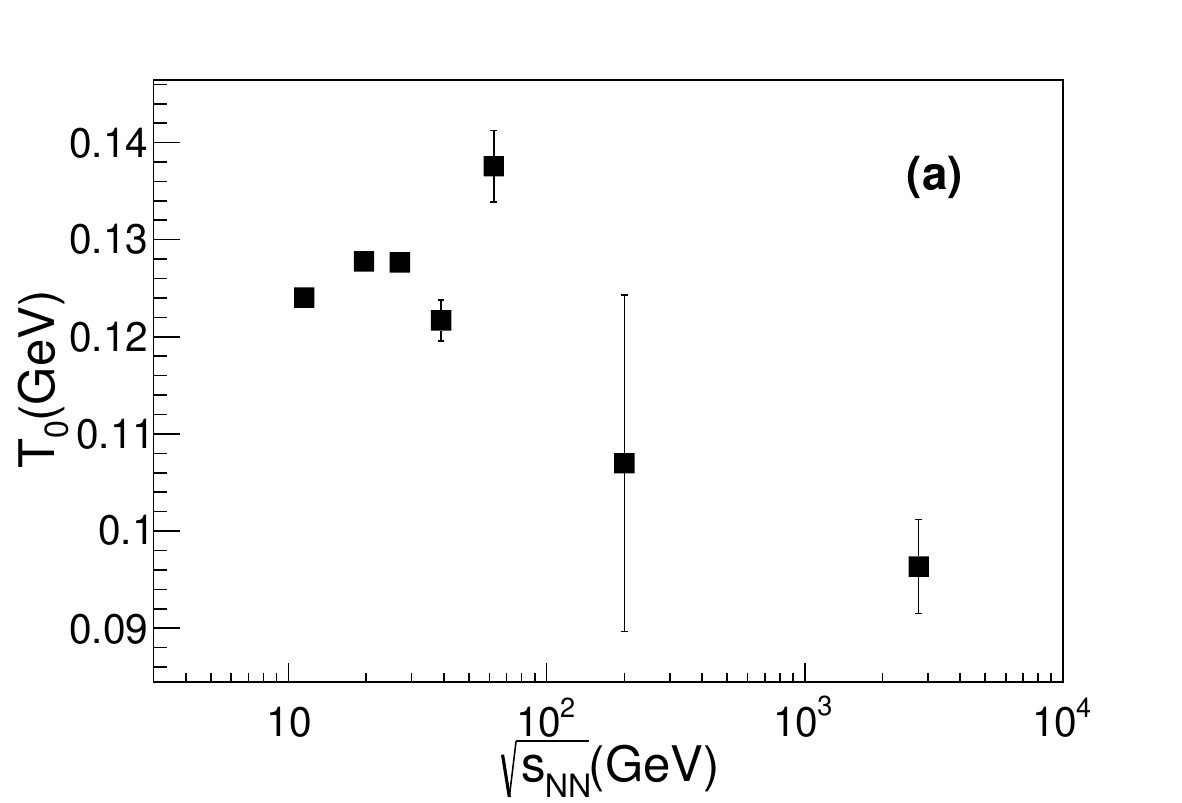}
\includegraphics[width=0.42\textwidth]{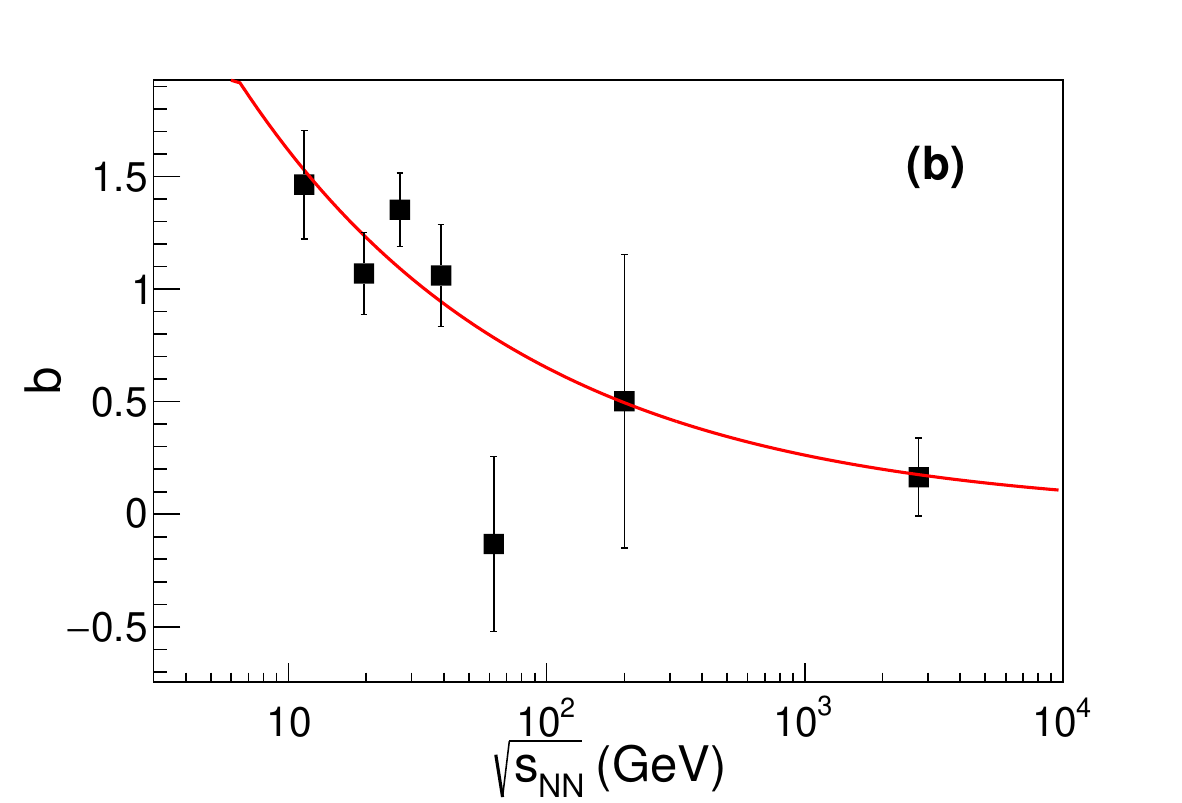}
\includegraphics[width=0.42\textwidth]{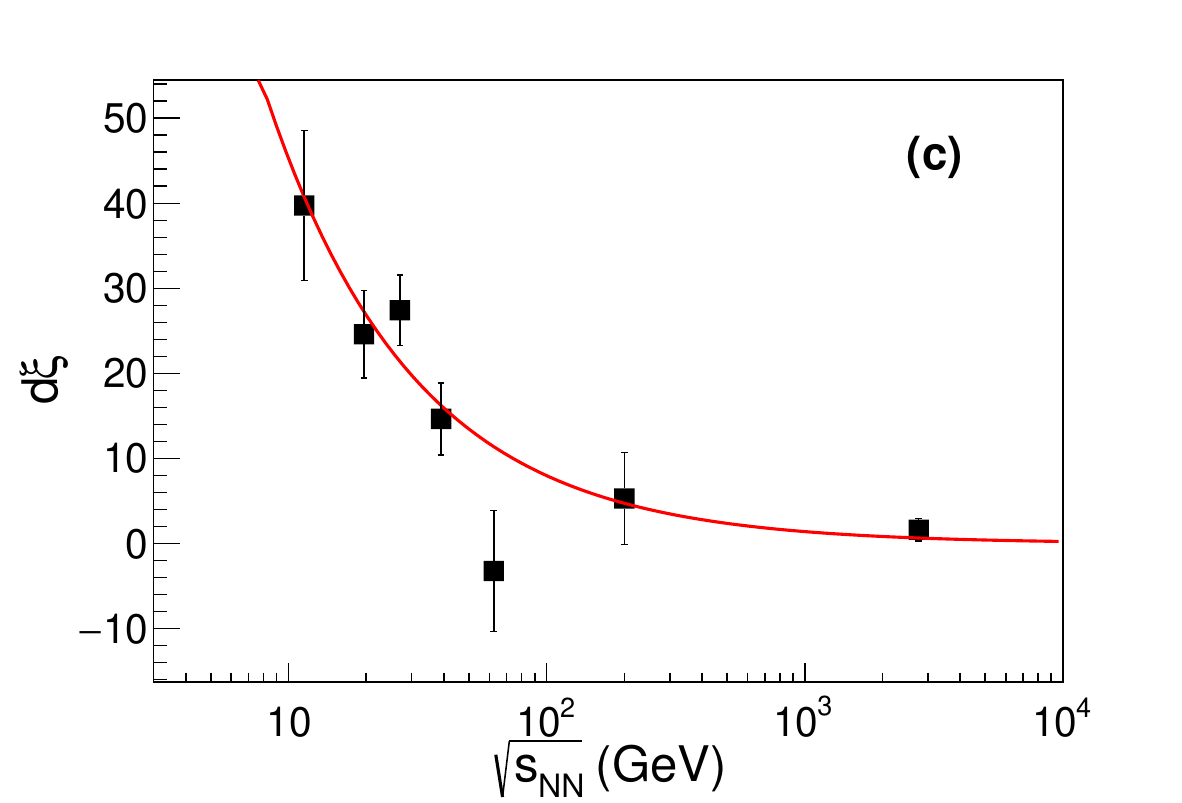}
\caption{Collision energy dependence of parameters $T_0$, $b$ and $d\xi$  in $T=T_0+b(q-1)-d\xi(q-1)^2$. Curves are to guide the eye.}\label{fig:paTqall_enedep}
\end{center}
\end{figure*}

It has been argued within the framework of non-equilibrium statistics that the dependence of temperature and flow velocity on the non-equilibrium factor ($q-1$) is related to the shear and bulk $\xi$ viscosity in linear or quadratic proportion~\cite{Wilk:1999dr,Wilk:2008ue}. This hypothesis is examined by quadratic fits of  $\langle\beta\rangle = \langle\beta\rangle_0-a(q-1)^{2}$ and $T=T_0+b(q-1)-d\xi(q-1)^2$ (where $\xi$ is the bulk viscosity) to the inclusive hadron group as shown in {\Crefname{figure}{Fig.}{Figs.}\Cref{fig:pabetaqall,fig:paTqall}}. Data at 7.7 GeV are close to equilibrium and do not provide a significant variation of the parameters, and are not included in this examination. From 11.5 to 2.76~TeV collision energy, there displays a clear evolution of $\langle\beta\rangle$ vs $(q-1)$ and $T$ vs $(q-1)$ relationships on collision energy.  A summary of parameters $\langle\beta\rangle_0$, $a$, $T_{0}$, $b$, and $d\xi$ dependence on collision energy is depicted in {\Crefname{figure}{Fig.}{Figs.}\Cref{fig:pabetaqall_enedep,fig:paTqall_enedep}}. As energy increases, $\langle\beta\rangle_0$ increases and the coefficient of the squared term $a$ decreases. A similar feature is observed for the $T$ vs $(q-1)$. With only three available centrality classes, the fitting procedure at 62.4~GeV is found to be not constrained. The relationship of $T$ vs $(q-1)$ was previously inspected in Ref.~\cite{Tang:2008ud} for 200~GeV where only a squared term (with a constant) is used. Our paper shows that both linear and quadratic terms are needed to describe the $T$ vs $(q-1)$ relationship for lower collision energies. The linear term parameter $b$ and quadratic term related to viscosity parameter $d\xi$ show a trend of decrease with collision energy. It is interesting to note that it has been argued that the bulk viscosity increases dramatically toward the phase transition~\cite{Kharzeev:2007wb,Karsch:2007jc}, coinciding with the feature we observed of $d\xi$ shown in {\Crefname{figure}{Fig.}{Figs.}\Cref{fig:paTqall_enedep}}.

\begin{table*}[h]
\caption{\label{tabpa}The fitting parameters of $\langle\beta\rangle =\langle\beta\rangle_0-a(q-1)^{2}$ in \cref{fig:pabetaqall} and $T=T_0+b(q-1)-d\xi(q-1)^2$ in \cref{fig:paTqall}.}
\begin{ruledtabular}
\begin{tabular}{ccccccc}

   $\rm system$  & $\rm\sqrt{s_{NN}}\; (\rm {GeV})$  & $\langle\beta\rangle_0 \;$ & $a$ &  $T_0 \;(\rm{GeV})$  & $b$ &  $ d\xi$\\
  \hline
  $\rm Au+Au$ &  $11.5$     & $0.397\pm0.002$  & $635\pm88$ & $0.1240\pm0.0009$ & $1.5\pm0.2$ &  $40\pm9$\\

  $\rm Au+Au$ &  $19.6$     & $0.411\pm0.002$  & $347\pm27 $ & $ 0.1278\pm0.0008 $ & $  1.1\pm0.2$ & $ 25\pm5$\\

  $\rm Au+Au$ &  $27$     & $0.423\pm0.002$  & $286\pm20$ & $ 0.1277\pm0.0008$ & $ 1.4\pm0.2$ &  $ 27\pm4$\\

  $\rm Au+Au$ &  $39$     & $0.448\pm0.002$  & $202\pm14 $ & $0.122\pm0.002 $ & $1.1\pm0.2$ & $ 15\pm4$ \\

  $\rm Au+Au$  & $62.4$    & $0.44\pm0.01$  &   $110\pm28 $ & $0.138\pm0.004 $ & $ -(0.1\pm0.4)$ & $-(3\pm7)$ \\

  $\rm Au+Au$ &  $200$       & $0.51\pm0.02$  & $ 57\pm6$ & $ 0.11\pm0.02 $ & $0.5\pm0.7$ & $ 5\pm5$ \\

  $\rm Pb + Pb$ &  $2760$     & $0.594\pm0.005$  & $25\pm1$ & $ 0.096\pm0.005$ & $ 0.2\pm0.2$ & $ 2\pm1$ \\

\end{tabular}
\end{ruledtabular}
\end{table*}

In a short summary, BGBW and TBW have been used to explore the beam energy dependence of  kinetic freeze-out properties of the system created in relativistic heavy ion collisions. The BGBW model is designed to describe the system in local thermal equilibrium. However, as collision energy increases, the produced system in peripheral collisions deviates far from equilibrium state, and can not be described well with a BGBW fit. An additional parameter $q$ is introduced in the TBW model to characterize the degree of non-equilibrium. The divergence between BGBW and TBW escalates with an increasing $q$ value as collision energy increases, especially in peripheral collisions.
For 7.7 - 39~GeV collision energies, the increase of temperature in the TBW model from central to peripheral collisions is much less than that in BGBW. For 62.4~GeV to 5.02~TeV, the temperature in the TBW model stays almost constant from central to peripheral collisions.
Meanwhile, the radial flow value in central collisions is around 0.4-0.5$c$ at RHIC energies, and becomes larger at the LHC energies.

\section{Conclusion}

In this paper, we have used the blast-wave model with Boltzmann-Gibbs statistics and with Tsallis statistics to fit the transverse momentum spectra of hadrons produced at mid-(pseudo)rapidity in Au + Au collisions at $\sqrt{s_{\rm{NN}}}=$ 7.7, 11.5, 14.5, 19.6, 27, 39, 62.4, and 200~GeV at RHIC and in Pb+Pb collisions at $\sqrt{s_{\rm{NN}}}=$ 2.76~TeV and 5.02~TeV at LHC to extract kinetic freeze-out temperature and transverse flow velocity and to study their collision centrality and energy dependence.
The hadrons containing strangeness at those collision energies were also examined to study their impact on the freeze-out properties.
For centrality dependence, the results show that the average transverse radial flow velocity decreases and the degree of non-equilibrium $q$ increases from central to peripheral collisions in the TBW model.
The kinetic freeze-out temperature shows weak dependence on centrality in the TBW model, while in the BGBW model there is a clear increase from central to peripheral in A + A collisions.
This finding suggests that a change in non-equilibrium degree of the system in TBW is reflected as a change in freeze-out temperature in the language of transitional BGBW. One should take caution when interpreting temperature behavior in BGBW for beam energy scan results.
For energy dependence in TBW fits, the average transverse radial flow velocity and the degree of non-equilibrium $q$ both increase with the increase of the collision energy, which suggests a stronger expansion with larger deviation from thermal equilibrium at higher energy.
The kinetic freeze-out temperature at the same centrality shows a weak collision energy  dependence for 7.7 to 39~GeV, while it decreases from 62.4~GeV to 5.02~TeV with an increase of non-equilibrium degree. A dependence of temperature and radial flow on non-equilibrium is observed and may be related to the bulk viscosity.
Finally, we find that strange hadrons have a higher kinetic freeze-out temperature than that for light hadrons. The strange hadrons approach equilibrium more quickly from peripheral to central A + A collisions than non-strange hadrons.

\begin{acknowledgments}
We appreciate valuable discussions with Zhenyu Chen, Xiaofeng Luo,  Nihar Ranjan Sahoo, Qinghua Xu, Chi Yang and Qian Yang. This work was partly supported by the National Natural Science Foundation of China under Grants No. 11890710, No.11890713, and No.11720101001. This work was also supported in part by the Office of Science of the U.S. Department of Energy.
\end{acknowledgments}



\clearpage

\onecolumngrid
\appendix

\section{}

\begin{table*}[htbp]
\caption{\label{tabTBW3pa1}Extracted kinetic freeze-out parameters and $\chi^{2}/nDoF$ from TBW fits to identified particle transverse spectra in heavy ion collisions of different centralities at $\sqrt{s_{\rm{NN}}}=$ 7.7 and 11.5~GeV. Results for charged pions, kaons, and protons have labels `$(\pi, K, p)$' behind their collision energy. All available hadrons including strange and multi-strange particles are labeled as `(all)'.  We also fit the spectra separately for strangeness with label `(strange)' and non-strange particles with label `(non-strange)'.}
\begin{ruledtabular}
\begin{tabular}{ccccccc}

   $\rm system$  & $\rm\sqrt{s_{NN}}\; (\rm {GeV})$  & centrality  & $\langle\beta\rangle \;$ &  $T \;(\rm{MeV})$  & $ q$ & $\chi^{2}/nDoF$ \\
  \hline
  $\rm Au+Au$ &  $7.7 \;(\pi, K, p)$     & $0-5\%$  & $0.436\pm 0.005$ & $110\pm 2$ & $1.000^{+ 0.002}_{-0}$ & $113 / 133$ \\
          &  & $5-10\%$  & $0.428\pm 0.006$ & $110\pm 2$ & $1.000^{+ 0.003}_{-0}$ & $106 / 134$ \\
          &  & $10-20\%$  & $0.39\pm 0.01$ & $117\pm 3$ & $1.006\pm 0.005$ & $86 / 138$ \\
          &  & $20-30\%$  & $0.36\pm 0.01$ & $117\pm 3$ & $1.009\pm 0.005$ & $129 / 136$ \\
         &   & $30-40\%$  & $0.34\pm 0.01$ & $119\pm 3$ & $1.008\pm 0.005$ & $124 / 135$ \\
         &   & $40-50\%$  & $0.26\pm 0.03$ & $117\pm 3$ & $1.024\pm 0.006$ & $110 / 125$ \\
         &   & $50-60\%$  & $0.21\pm 0.04$ & $117\pm 3$ & $1.026\pm 0.007$ & $132 / 122$ \\
         &   & $60-70\%$  & $0^{+0.07}_{-0}$ & $120\pm 3$ & $1.034\pm 0.003$ & $95 / 117$ \\
         &   & $70-80\%$  & $0^{+0.06}_{-0}$ & $126\pm 3$ & $1.021\pm 0.003$ & $88 / 97$ \\
         &   & $0-80\%$  &   $  0.36\pm 0.02 $ & $111 \pm 3 $ & $1.019 \pm 0.008 $ & $48 / 90$ \\ \hline

  $\rm Au+Au$ &  $7.7 \;(\rm non-strange)$     & $0-5\%$  & $0.445 \pm 0.006 $ & $108 \pm 2$ & $ 1.000^{+ 0.003}_{-0}$ & $ 34 / 89$ \\

        &  & $5-10\%$  & $0.438 \pm 0.006$ & $ 109 \pm 2$ & $ 1.000^{+ 0.005}_{-0}$ & $ 45 / 88$ \\

        &  & $10-20\%$  & $0.39 \pm 0.01$ & $ 116 \pm 3$ & $ 1.005^{+ 0.006}_{-0}$ & $ 37 / 92$ \\

       &  & $20-30\%$  & $0.38 \pm 0.01$ & $ 115 \pm 3$ & $ 1.007\pm 0.006$ & $ 31 / 90$ \\

        &   & $30-40\%$  & $0.35 \pm 0.02$ & $ 117 \pm 3$ & $ 1.009 \pm 0.006$ & $ 36 / 90$ \\

          &   & $40-60\%$  & $0^{+0.06}_{-0} $ & $113 \pm 3 $ & $1.049 \pm 0.002$ & $ 52 / 80$ \\

          &   & $60-80\%$  & $ 0^{+0.07}_{-0} $ & $ 120 \pm 3$ & $ 1.032 \pm 0.003 $ & $31 / 62$ \\ \hline

  $\rm Au+Au$ &  $7.7 \;(\rm strange)$     & $0-5\%$  & $0.361 \pm 0.01$ & $ 133 \pm 4$ & $ 1.0001 ^{+ 0.0002}_{-0.0001}$ & $ 178 / 80$ \\

         &  & $5-10\%$  & $ 0.358 \pm 0.01$ & $ 132 \pm 4 $ & $1.0001^{+ 0.0002}_{-0.0001} $ & $156 / 82$ \\

         &  & $10-20\%$  & $ 0.356 \pm 0.009 $ & $131 \pm 3 $ & $1.0001^{+ 0.0002}_{-0.0001} $ & $160 / 82$ \\

             &  & $20-30\%$  & $0.351 \pm 0.009$ & $ 128 \pm 3 $ & $1.0001^{+ 0.0004}_{-0.0001} $ & $160 / 82$ \\

           &   & $30-40\%$  & $ 0.33 \pm 0.01 $ & $130\pm 3$ & $ 1.000 ^{+ 0.001}_{-0} $ & $139 / 81$ \\

            &   & $40-60\%$  & $0.28 \pm 0.02 $ & $133 \pm 3$ & $ 1.002^{+ 0.003}_{-0.002} $ & $174 / 75$ \\

            &   & $60-80\%$  & $0^{+0.06}_{-0} $ & $134 \pm 3$ & $ 1.022 \pm 0.002 $ & $119 / 64$ \\ \hline

  $\rm Au+Au$ &  $7.7 \;(\rm all)$     & $0-5\%$  & $0.407 \pm 0.005 $ & $118 \pm 2 $ & $1.0001 \pm 0.0001$ & $ 274 / 172$ \\
                    &  & $5-10\%$  & $ 0.402 \pm 0.005 $ & $118 \pm 2 $ & $1.0001 ^{+ 0.0002}_{-0.0001} $ & $251 / 173$ \\
                    &  & $10-20\%$  & $ 0.378 \pm 0.005 $ & $124 \pm 2 $ & $1.0001^{+ 0.0002}_{-0.0001} $ & $218 / 177$ \\
                    &  & $20-30\%$  & $ 0.365 \pm 0.005 $ & $124 \pm 2 $ & $1.0001^{+ 0.0003}_{-0.0001} $ & $216 / 175$ \\
                    &   & $30-40\%$  & $ 0.349 \pm 0.006 $ & $125 \pm 2 $ & $1.0001 ^{+ 0.0007}_{-0.0001} $ & $189 / 174$ \\
                     &   & $40-60\%$  & $0.31 \pm 0.01 $ & $127  \pm 2$ & $ 1.002 \pm 0.002$ & $ 229 / 158$ \\
                      &   & $60-80\%$  & $0.20 \pm 0.03 $ & $126 \pm 2 $ & $1.012  \pm0.005$ & $ 158 / 129$ \\ \hline

  $\rm Au+Au$  & $11.5 \;(\pi, K, p)$    & $0-5\%$  &   $0.423 \pm 0.005 $ & $117  \pm 2 $ & $ 1.000 ^{+ 0.002}_{-0}$ & $104 / 142$ \\
         &   & $5-10\%$  &   $0.416 \pm 0.006  $ & $119\pm 2$ & $ 1.000 ^{+ 0.002}_{-0}$ & $79 / 145$ \\
         &   & $10-20\%$  &   $0.399 \pm 0.006 $ & $ 122 \pm 2 $ & $ 1.000^{+ 0.010}_{-0}$ & $92 / 145$ \\
         &   & $20-30\%$  &   $0.35 \pm 0.01  $ & $ 124\pm 3 $ & $ 1.012\pm  0.005$ & $90 / 145$ \\
         &   & $30-40\%$  &   $0.33 \pm 0.02  $ & $ 122 \pm 3  $ & $ 1.017 \pm 0.005$ & $109 / 144$ \\
         &   & $40-50\%$  & $0^{+0.08}_{-0}$ & $  126 \pm 3  $ & $ 1.046 \pm 0.002$ & $114 / 140$ \\
         &   & $50-60\%$  &  $0^{+0.08}_{-0} $ & $  124 \pm 3  $ & $  1.044 \pm 0.002$ & $96 / 138$ \\
         &   & $60-70\%$  &   $0^{+0.06}_{-0} $ & $  128 \pm 3  $ & $ 1.034 \pm 0.002$ & $108 / 124$ \\
         &   & $70-80\%$  & $0^{+0.05}_{-0} $ & $  125 \pm 3 $ & $  1.032 \pm 0.003$ & $117 / 120$ \\
         &   & $0-80\%$  &   $ 0.37 \pm 0.01$ & $ 114  \pm 3 $ & $1.019 \pm 0.006 $ & $33 / 118$ \\ 









\end{tabular}
\end{ruledtabular}
\end{table*}

\begin{table*}[htbp]
\caption{\label{tabTBW3pa2}
Same as \cref{tabTBW3pa1}, but for $\sqrt{s_{\rm{NN}}}=$ 11.5~GeV (continued), 14.5~GeV, 17.3~GeV, and 19.6~GeV. The results at $\sqrt{s_{\rm{NN}}}=$ 17.3~GeV are from Ref.~\cite{Shao:2009mu}.
}
\begin{ruledtabular}
\begin{tabular}{ccccccc}

   $\rm system$  & $\rm\sqrt{s_{NN}}\; (\rm {GeV})$  & centrality  & $\langle\beta\rangle \; $ &  $T \;(\rm{MeV})$  & $ q$ & $\chi^{2}/nDoF$ \\
  \hline

  $\rm Au+Au$ &  $11.5 \;(\rm non-strange)$       & $0-5\%$  & $ 0.423 \pm 0.005$ & $ 118 \pm 2 $ & $1.00 ^{+ 0.01}_{-0}$ & $ 63 / 96$ \\

        &   & $5-10\%$  &   $0.42 \pm 0.01$ & $ 119 \pm 3 $ & $1.00 ^{+ 0.03}_{-0} $ & $49 / 97$ \\

        &   & $10-20\%$  &   $0.39 \pm 0.01 $ & $121 \pm 3 $ & $1.004 ^{+ 0.007}_{-0.004}$ & $ 68 / 97$ \\

        &   & $20-30\%$  &   $0.33 \pm 0.02 $ & $119 \pm 3 $ & $1.025 \pm 0.007 $ & $56 / 97$ \\

        &   & $30-40\%$  &   $0.29 \pm 0.02$ & $ 120 \pm 3 $ & $1.026 \pm 0.006$ & $ 56 / 97$ \\

        &   & $40-60\%$  & $0^{+0.08}_{-0}$ & $ 123 \pm 3 $ & $1.047 \pm 0.002$ & $ 61 / 92$ \\

         &   & $60-80\%$  & $0^{+0.06}_{-0}$ & $ 126 \pm 3$ & $ 1.034 \pm 0.002 $ & $65 / 82$ \\

        &   & $0-80\%$  &   $0.37\pm 0.02$ & $ 113\pm 3 $ & $1.021 \pm0.008$ & $ 21 / 82$
\\ \hline

  $\rm Au+Au$ &  $11.5 \;(\rm strange)$       & $0-5\%$  & $    0.387 \pm 0.007 $ & $131 \pm 3 $ & $1.0001 ^{+ 0.0002}_{-0.0001}$ & $   148 / 84$ \\
     &   & $5-10\%$  & $ 0.373 \pm 0.008$ & $ 135 \pm 3 $ & $1.0001 ^{+ 0.0004}_{-0.0001} $ & $154 / 86$ \\

     &  & $10-20\%$  & $0.376 \pm 0.007 $ & $132 \pm 2 $ & $1.0001 ^{+ 0.0005}_{-0.0001}$ & $ 134 / 86$ \\

     &  & $20-30\%$  & $0.352 \pm 0.007$ & $ 138 \pm 3 $ & $1.000 ^{+ 0.001}_{-0} $ & $107 / 86$ \\

      &   & $30-40\%$  & $0.313 \pm 0.009 $ & $147 \pm 3 $ & $1.000 ^{+ 0.007}_{-0}$ & $ 92 / 85$ \\

      &   & $40-60\%$  & $0.22 \pm 0.02$ & $ 146 \pm 3$ & $ 1.012 \pm 0.003$ & $ 126 / 84$ \\

      &   & $60-80\%$  & $0^{+0.07}_{-0} $ & $145 \pm 3 $ & $1.023 \pm0.002 $ & $132 / 73$ \\ \hline

  $\rm Au+Au$ &  $11.5 \;(\rm all)$     & $0-5\%$  & $0.409 \pm 0.004$ & $ 122 \pm 1$ & $ 1.0001 ^{+ 0.0002}_{-0.0001}$ & $ 228 / 183$ \\
          &  & $5-10\%$  & $ 0.402 \pm 0.004$ & $ 124 \pm 2$ & $ 1.0001^{+ 0.0003}_{-0.0001} $ & $227 / 186$ \\
            &  & $10-20\%$  & $ 0.392 \pm 0.004$ & $ 126 \pm 1 $ & $1.0001^{+ 0.0006}_{-0.0001} $ & $214 / 186$ \\
           &  & $20-30\%$  & $ 0.368 \pm 0.007$ & $ 131 \pm 2 $ & $1.000^{+ 0.005}_{-0} $ & $185 / 186$ \\
            &   & $30-40\%$  & $ 0.340 \pm 0.009$ & $ 133 \pm 2 $ & $1.003 \pm0.002 $ & $205 / 185$ \\
           &   & $40-60\%$  & $0.24 \pm 0.01$ & $ 136 \pm 2$ & $ 1.015 \pm 0.002$ & $ 228 / 179$ \\
            &   & $60-80\%$  & $0^{+0.06}_{-0} $ & $134 \pm 2 $ & $1.029 \pm 0.001 $ & $240 / 158$ \\ \hline

    $\rm Au+Au$  &$14.5 \;(\pi, K, p)$    & $0-5\%$  &   $0.42 \pm 0.01 $ & $118 \pm 3 $ & $1.006 ^{+ 0.007}_{-0.006} $ & $ 56 / 149$ \\
         &   & $5-10\%$  &   $0.40 \pm 0.01$ & $ 119 \pm 3 $ & $1.011 \pm 0.007 $ & $ 58 / 149$ \\
         &   & $10-20\%$  &   $0.38 \pm 0.02$ & $ 117 \pm 3 $ & $1.016 \pm 0.006 $ & $ 53 / 149$ \\
         &   & $20-30\%$  &   $0.36 \pm 0.02$ & $ 118 \pm 3 $ & $1.021 \pm 0.006 $ & $ 33 / 149$ \\
         &   & $30-40\%$  &   $0.31 \pm 0.02 $ & $123 \pm 3 $ & $1.024 \pm 0.006 $ & $ 57 / 149$ \\
         &   & $40-50\%$  & $0.15 \pm 0.06 $ & $115 \pm 3 $ & $1.052 \pm 0.007$ & $ 83 / 143$ \\
         &   & $50-60\%$  &  $0.15 \pm 0.06$ & $ 119 \pm 3$ & $ 1.045 \pm 0.008 $ & $ 110 / 139$ \\
         &   & $60-70\%$  &   $0.15 \pm 0.07$ & $ 119 \pm 3$ & $ 1.039 \pm 0.008$ & $ 79 / 131$ \\
         &   & $70-80\%$  & $0^{+0.06}_{-0} $ & $126 \pm 3 $ & $1.033 \pm 0.003 $ & $93 / 127$ \\\hline

  $\rm Pb+Pb$  &$ 17.3$ \;(\rm non-strange){\rm~\cite{Shao:2009mu}}     & $0-5\%$  & $0.442\pm0.005  $ & $  109\pm 1  $ & $ 1.015 \pm  0.001$ & $102/86$ \\ \hline

  $\rm Pb+Pb$  &$ 17.3$ \;(\rm strange){\rm~\cite{Shao:2009mu}}    & $0-5\%$  & $0.420\pm0.007  $ & $  119\pm 4  $ & $ 1.009\pm0.004$ & $137/70$ \\ \hline

  $\rm Pb+Pb$  &$ 17.3$ \;(\rm all){\rm~\cite{Shao:2009mu}}    & $0-5\%$  & $0.426 \pm  0.004  $ & $  113 \pm 1  $ & $ 1.015 \pm  0.001$ & $267/159$ \\ \hline

  $\rm Au+Au$  &$19.6\; (\pi, K, p)$    & $0-5\%$  &   $0.428  \pm0.009  $ & $ 112 \pm 3 $ & $  1.013\pm  0.005$ & $52 / 146$ \\
         &   & $5-10\%$  &   $0.41  \pm0.01 $ & $  114 \pm 3 $ & $  1.016 \pm 0.005$ & $155 / 142$ \\
         &   & $10-20\%$  &   $0.40 \pm 0.01 $ & $  117 \pm 3 $ & $  1.015 \pm 0.005$ & $73 / 142$ \\
         &   & $20-30\%$  &   $0.34 \pm 0.02 $ & $  119 \pm 3$ & $   1.028\pm  0.005$ & $71 / 142$ \\
         &   & $30-40\%$  &   $0.27 \pm 0.02 $ & $  124 \pm 3 $ & $  1.033  \pm0.006$ & $84 / 143$ \\
         &   & $40-50\%$  & $0.20  \pm0.04$ & $  123 \pm 3 $ & $  1.041 \pm 0.006$ & $88 / 141$ \\
         &   & $50-60\%$  &  $0^{+0.05}_{-0} $ & $  127 \pm 2 $ & $  1.047 \pm 0.002$ & $128 / 141$ \\
         &   & $60-70\%$  &   $0^{+0.04}_{-0} $ & $  128 \pm 3 $ & $  1.042  \pm0.002$ & $192 / 135$ \\
         &   & $70-80\%$  & $0^{+0.04}_{-0} $ & $  131 \pm 3 $ & $  1.033 \pm 0.002$ & $234 / 130$ \\
           &  & $0-80\%$  &   $0.35 \pm 0.01 $ & $111 \pm 3 $ & $1.036 \pm 0.006$ & $ 33 / 127$ \\\hline

  $\rm Au+Au$ &  $19.6 \;(\rm non-strange)$     & $0-5\%$  & $0.43 \pm  0.01 $ & $111 \pm  3 $ & $1.015 \pm  0.007 $ & $40 / 96$ \\

          &  & $5-10\%$  & $0.40 \pm  0.01 $ & $112 \pm  3$ & $ 1.022 \pm  0.007$ & $ 53 / 92$ \\

           &  & $10-20\%$  & $0.38 \pm  0.02$ & $ 112 \pm  3$ & $ 1.027 \pm  0.007$ & $ 36 / 92$ \\

           &  & $20-30\%$  & $0.32 \pm  0.02 $ & $117 \pm  3$ & $ 1.034 \pm  0.007 $ & $53 / 92$ \\

           &   & $30-40\%$  & $0^{+0.07}_{-0} $ & $116 \pm  3$ & $ 1.065 \pm  0.002 $ & $68 / 93$ \\

            &   & $40-60\%$  & $0^{+0.06}_{-0}$ & $ 120 \pm 3$ & $ 1.057 \pm 0.002$ & $ 48 / 93$ \\

              &   & $60-80\%$  & $0^{+0.04}_{-0} $ & $129 \pm 3$ & $ 1.040 \pm 0.002 $ & $125 / 88$ \\

               &  & $0-80\%$  &   $ 0.34 \pm 0.02$ & $ 110 \pm 3 $ & $1.038 \pm 0.007 $ & $24 / 86$ \\\hline

   $\rm Au+Au$ &  $19.6 \;(\rm strange)$     & $0-5\%$  & $0.404 \pm 0.004 $ & $134 \pm 2 $ & $1.0001 \pm 0.0001 $ & $201 / 88$ \\

    &  & $5-10\%$  & $0.397 \pm 0.004$ & $ 136 \pm 2 $ & $1.0001 ^{+0.0003}_{-0.0001}$ & $ 181 / 88$ \\

    &  & $10-20\%$  & $0.388 \pm 0.004 $ & $138 \pm 2$ & $ 1.0001^{+0.0004}_{-0.0001}$ & $ 188 / 88$ \\

     &  & $20-30\%$  & $0.359 \pm0.007$ & $ 145 \pm 2$ & $ 1.002 \pm 0.002 $ & $182 / 88$ \\

     &   & $30-40\%$  & $0.319 \pm 0.009 $ & $144 \pm 2$ & $ 1.009 \pm 0.002 $ & $224 / 88$ \\

      &   & $40-60\%$  & $ 0.19 \pm 0.02$ & $ 146 \pm 2$ & $ 1.025 \pm 0.003$ & $ 271 / 86$ \\

     &   & $60-80\%$  & $ 0^{+0.03}_{-0} $ & $141 \pm 2$ & $ 1.034 \pm 0.001 $ & $265 / 79$\\

\end{tabular}
\end{ruledtabular}
\end{table*}

\begin{table*}[htbp]
\caption{\label{tabTBW3pa3}
Same as \cref{tabTBW3pa1}, but for $\sqrt{s_{\rm{NN}}}=$ 19.6~GeV (continued), 27~GeV and 39~GeV.
}
\begin{ruledtabular}
\begin{tabular}{ccccccc}

   $\rm system$  & $\rm \sqrt{s_{NN}} \;(\rm{GeV})$  & centrality  & $\langle\beta\rangle \; $ &  $T \;(\rm{MeV})$  & $q$ & $\chi^{2}/nDoF$ \\
  \hline

  $\rm Au+Au$ &  $19.6 \;(\rm all)$     & $0-5\%$  & $0.421 \pm 0.003 $ & $126 \pm 1$ & $ 1.0001 ^{+0.0002}_{-0.0001} $ & $293 / 187$ \\
         &  & $5-10\%$  & $ 0.414 \pm 0.003$ & $ 128 \pm 1 $ & $1.0001 ^{+0.0005}_{-0.0001} $ & $282 / 183$ \\
         &  & $10-20\%$  & $ 0.404 \pm 0.003 $ & $131 \pm 1$ & $ 1.000^{+0.001}_{-0}   $ & $278 / 183$ \\
          &  & $20-30\%$  & $0.369 \pm 0.006$ & $ 135 \pm 1 $ & $1.005 \pm 0.002 $ & $312 / 183$ \\
         &   & $30-40\%$  & $ 0.324 \pm 0.008 $ & $136 \pm 1$ & $ 1.013 \pm 0.002 $ & $343 / 184$ \\
          &   & $40-60\%$  & $0.22 \pm 0.02$ & $138  \pm 1 $ & $1.027 \pm 0.002 $ & $374 / 182$ \\
          &   & $60-80\%$  & $ 0^{+0.03}_{-0} $ & $134 \pm 2 $ & $1.037 \pm 0.001 $ & $411 / 170$\\ \hline

  $\rm Au+Au$  &$27\; (\pi, K, p)$      & $0-5\%$  &   $0.451 \pm 0.009 $ & $  114 \pm 3 $ & $   1.004^{+0.006}_{-0.004}$ & $86 / 139$ \\
         &   & $5-10\%$  &   $0.43\pm  0.01 $ & $  112 \pm 3 $ & $  1.016 \pm 0.006$ & $66 / 140$ \\
         &   & $10-20\%$  &   $0.40 \pm 0.01 $ & $  116 \pm 3 $ & $  1.019 \pm 0.005$ & $61 / 140$ \\
         &   & $20-30\%$  &   $0.36 \pm 0.01 $ & $  116 \pm 3  $ & $ 1.031 \pm 0.005$ & $54 / 140$ \\
         &   & $30-40\%$  &   $0.30 \pm 0.02 $ & $  120 \pm 3  $ & $ 1.038\pm  0.005$ & $57 / 140$ \\
         &   & $40-50\%$  & $0.14 \pm 0.06 $ & $  120 \pm 3  $ & $ 1.058 \pm 0.006$ & $48 / 140$ \\
         &   & $50-60\%$  &  $0^{+0.05}_{-0} $ & $  126 \pm 3 $ & $  1.055  \pm0.002$ & $102 / 140$ \\
         &   & $60-70\%$  &   $0^{+0.04}_{-0} $ & $  130 \pm 3 $ & $  1.047 \pm 0.002$ & $162 / 140$ \\
         &   & $70-80\%$  & $0^{+0.03}_{-0}  $ & $ 133 \pm 3  $ & $ 1.039 \pm 0.002$ & $267 / 138$ \\
         &   & $0-80\%$  &   $0.38\pm 0.01 $ & $115\pm 3 $ & $1.027 \pm0.005 $ & $49 / 137$ \\  \hline

   $\rm Au+Au$ &  $27 \;(\rm non-strange)$     & $0-5\%$  & $0.43 \pm 0.01 $ & $109 \pm 3 $ & $1.021 \pm 0.008 $ & $38 / 90$ \\

          &  & $5-10\%$  & $ 0.41 \pm 0.01$ & $ 108\pm 3$ & $ 1.027 \pm 0.007 $ & $39 / 90$ \\

      &  & $10-20\%$  & $0.39 \pm 0.02$ & $ 112\pm 3 $ & $1.029 \pm 0.007 $ & $32 / 90$ \\

     &  & $20-30\%$  & $0.34 \pm 0.02$ & $ 113 \pm 3$ & $ 1.039 \pm 0.007$ & $ 29 / 90$ \\

          &   & $30-40\%$  & $0.26 \pm 0.03 $ & $116 \pm 3 $ & $1.050 \pm 0.007$ & $ 31 / 90$ \\

           &   & $40-60\%$  & $0^{+0.04}_{-0}$ & $ 120 \pm 3$ & $ 1.061 \pm 0.003 $ & $38 / 90$ \\

           &   & $60-80\%$  & $0^{+0.03}_{-0}$ & $ 131 \pm 3 $ & $1.044 \pm 0.003$ & $ 125 / 88$ \\

           &   & $0-80\%$  &   $0.36 \pm 0.02$ & $ 110 \pm 3 $ & $1.039 \pm 0.007$ & $ 22 / 88$ \\   \hline

  $\rm Au+Au$ &  $27 \;(\rm strange)$     & $0-5\%$  & $0.419 \pm 0.004$ & $ 132 \pm 2$ & $ 1.0001 ^{+0.0002}_{-0.0001}$ & $ 263 / 87$ \\

      &  & $5-10\%$  & $ 0.408 \pm 0.004$ & $ 136 \pm 2$ & $ 1.0001^{+0.0002}_{-0.0001}$ & $ 199 / 88$ \\

     &  & $10-20\%$  & $0.396 \pm 0.004 $ & $141 \pm 2$ & $ 1.0001^{+0.0004}_{-0.0001}$ & $ 189 / 88$ \\

     &  & $20-30\%$  & $0.368 \pm 0.006$ & $ 148 \pm 2$ & $ 1.002 \pm 0.002$ & $ 177 / 88$ \\

     &   & $30-40\%$  & $0.31 \pm 0.01$ & $ 153 \pm 2$ & $ 1.011 \pm 0.002$ & $ 154 / 88$ \\

      &   & $40-60\%$  & $  0.19 \pm 0.03$ & $ 155 \pm 2$ & $ 1.027\pm0.003 $ & $201 / 88$ \\

      &   & $60-80\%$  & $  0^{+0.03}_{-0}$ & $ 145 \pm 2$ & $ 1.038 \pm 0.001 $ & $253 / 88$ \\ \hline

  $\rm Au+Au$ &  $27 \;(\rm all)$     & $0-5\%$  & $0.434 \pm 0.003$ & $ 125 \pm 1 $ & $1.0001 ^{+0.0002}_{-0.0001} $ & $350 / 180$ \\
                &  & $5-10\%$  & $ 0.426 \pm 0.003$ & $ 128 \pm 1 $ & $1.0001 ^{+0.0003}_{-0.0001} $ & $302 / 181$ \\
                 &  & $10-20\%$  & $ 0.414 \pm 0.003 $ & $132 \pm 1 $ & $1.0001 ^{+0.0009}_{-0.0001} $ & $290 / 181$ \\
                 &  & $20-30\%$  & $ 0.379 \pm 0.005 $ & $137 \pm 1 $ & $1.006 \pm 0.002 $ & $316 / 181$ \\
                &   & $30-40\%$  & $ 0.328 \pm 0.008 $ & $141 \pm 1 $ & $1.014 \pm 0.002 $ & $302 / 181$ \\
              &   & $40-60\%$  & $0.22 \pm 0.02 $ & $143  \pm 1$ & $ 1.029 \pm 0.003 $ & $348 / 181$ \\
              &   & $60-80\%$  & $0^{+0.03}_{-0} $ & $137 \pm 1 $ & $1.042\pm 0.001 $ & $422 / 179$ \\ \hline

  $\rm Au+Au$  &$39\; (\pi, K, p)$      & $0-5\%$  &   $0.463\pm  0.009 $ & $  116 \pm 3 $ & $  1.004 ^{+0.007}_{-0.004}$ & $57 / 140$ \\
         &   & $5-10\%$  &   $0.44 \pm 0.01  $ & $ 120\pm 3  $ & $ 1.010 \pm 0.006$ & $61 / 140$ \\
         &   & $10-20\%$  &   $0.41 \pm 0.01  $ & $ 114 \pm 3 $ & $  1.026 \pm 0.006$ & $48 / 140$ \\
         &   & $20-30\%$  &   $0.36 \pm 0.01  $ & $ 116 \pm 3 $ & $  1.036 \pm 0.006$ & $63 / 140$ \\
         &   & $30-40\%$  &   $0.30 \pm 0.02  $ & $ 120  \pm 3 $ & $  1.045 \pm 0.006$ & $53 / 140$ \\
        &    & $40-50\%$  & $0.18 \pm 0.04  $ & $ 118 \pm 3 $ & $  1.062 \pm 0.006$ & $55 / 140$ \\

         &   & $50-60\%$  &  $0^{+0.05}_{-0}  $ & $ 123 \pm 3 $ & $  1.063 \pm 0.002$ & $94 / 140$ \\
         &   & $60-70\%$  &   $0^{+0.04}_{-0} $ & $  131 \pm 3  $ & $ 1.054 \pm 0.002$ & $226 / 140$ \\
          &  & $70-80\%$  & $0^{+0.03}_{-0} $ & $  137 \pm 3  $ & $ 1.045\pm  0.002$ & $341 / 140 $ \\
          &   & $0-80\%$  &   $ 0.38\pm 0.01$ & $ 117 \pm 3 $ & $1.030 \pm 0.006 $ & $46 / 140$ \\

\end{tabular}
\end{ruledtabular}
\end{table*}

\begin{table*}[htbp]
\caption{\label{tabTBW3pa4}
Same as \cref{tabTBW3pa1}, but for $\sqrt{s_{\rm{NN}}}=$ 39~GeV (continued), 62.4~GeV and 200~GeV.
}
\begin{ruledtabular}
\begin{tabular}{ccccccc}

   $\rm system$  & $\rm \sqrt{s_{NN}} \;(\rm{GeV})$  & centrality  & $\langle\beta\rangle \;$ &  $T \;(\rm{MeV})$  & $q$ & $\chi^{2}/nDoF$ \\
  \hline

   $\rm Au+Au$ &  $39 \;(\rm non-strange)$     & $0-5\%$\footnotemark[1]  & $0.45 \pm 0.01$ & $ 111 \pm 4$ & $ 1.017 \pm 0.009 $ & $39 / 90$ \\

        &   & $5-10\%$\footnotemark[1]  &   $0.43 \pm 0.01 $ & $116 \pm 4 $ & $1.019 \pm 0.009 $ & $45 / 90$ \\

        &  & $10-20\%$  & $0.39 \pm 0.01 $ & $111 \pm 3 $ & $1.036 \pm 0.005 $ & $40 / 100$ \\

         &  & $20-30\%$\footnotemark[1]  & $0.35 \pm 0.02 $ & $114 \pm 4$ & $ 1.042 \pm0.008 $ & $50 / 90$ \\

         &   & $30-40\%$\footnotemark[1]  & $0^{+0.08}_{-0} $ & $112 \pm 3$ & $ 1.079 \pm 0.003 $ & $49 / 90$ \\

         &   & $40-60\%$  & $0^{+0.06}_{-0} $ & $120 \pm 3 $ & $1.067 \pm 0.005 $ & $43 / 100$ \\

         &   & $60-80\%$\footnotemark[1]  & $0^{+0.03}_{-0} $ & $137 \pm 3$ & $ 1.045 \pm 0.003 $ & $158 / 90$ \\

          &   & $0-80\%$\footnotemark[1]  &   $0.37 \pm 0.02$ & $ 114 \pm 3 $ & $1.039 \pm 0.008$ & $ 35 / 90$ \\  \hline

   $\rm Au+Au$ &  $39 \;(\rm strange)$     & $0-5\%$  & $0.430 \pm 0.007$ & $ 134 \pm 3 $ & $1.000 ^{+0.003}_{-0} $ & $65 / 88$ \\

        &   & $5-10\%$  &   $0.42 \pm 0.01 $ & $134 \pm 3$ & $ 1.001 ^{+0.003}_{-0.001}$ & $ 73 / 88$ \\

      &  & $10-20\%$  & $ 0.39 \pm 0.01$ & $ 143 \pm 3$ & $ 1.005 \pm 0.003$ & $ 85 / 88$ \\

      &  & $20-30\%$  & $0.36 \pm 0.01 $ & $148 \pm 3 $ & $1.009 \pm 0.003$ & $ 110 / 88$ \\

       &   & $30-40\%$  & $0.30 \pm 0.02 $ & $153 \pm 3$ & $ 1.019 \pm 0.003$ & $ 93 / 88$ \\

      &   & $40-60\%$  & $ 0^{+0.06}_{-0} $ & $154 \pm 2 $ & $1.045 \pm 0.001$ & $ 166 / 88$ \\

        &   & $60-80\%$  & $0^{+0.03}_{-0} $ & $149 \pm 2$ & $ 1.043 \pm 0.002$ & $ 221 / 88$ \\ \hline

  $\rm Au+Au$ &  $39 \;(\rm all)$     & $0-5\%$\footnotemark[1]  & $0.454 \pm0.004 $ & $123 \pm 2 $ & $1.000 ^{+0.002}_{-0}$ & $ 131 / 181$ \\

        &   & $5-10\% $\footnotemark[1]  &   $ 0.438 \pm 0.007 $ & $126 \pm 2 $ & $1.003 \pm 0.003 $ & $133 / 181$ \\

          &  & $10-20\%$  & $0.408 \pm 0.007 $ & $129 \pm 2 $ & $1.012 \pm 0.003$ & $ 190 / 191$ \\

           &  & $20-30\% $\footnotemark[1]  & $ 0.382 \pm 0.008 $ & $134 \pm 2 $ & $1.013 \pm 0.002 $ & $234 / 181$ \\

           &   & $30-40\% $\footnotemark[1]  & $  0.32 \pm 0.01 $ & $138 \pm 2 $ & $1.023 \pm 0.002 $ & $214 / 181$ \\

            &   & $40-60\%$  &   $ 0.14 \pm 0.04 $ & $141 \pm 2 $ & $1.046 \pm 0.004 $ & $309 / 191$ \\

             &   & $60-80\% $\footnotemark[1]  & $ 0^{+0.02}_{-0} $ & $138 \pm 2 $ & $1.048 \pm 0.001 $ & $464 / 181$ \\ \hline

  $\rm Au+Au$  &  $62.4 \;(\pi, K, p)$    & $0-10\%$  &   $0.47 \pm 0.01$ & $ 124 \pm 3$ & $ 1.002 ^{+0.006}_{-0.002} $ & $104 / 65$ \\

          &  & $10-20\%$  &   $0.44 \pm 0.01$ & $ 124 \pm 3 $ & $1.015 \pm 0.006 $ & $96 / 65$ \\

            &  & $20-40\%$  &   $ 0.39 \pm 0.02 $ & $125 \pm 4 $ & $1.028 \pm 0.005$ & $ 85 / 65$ \\

           &  & $40-80\%$  &   $0.18 \pm 0.05 $ & $128 \pm 4$ & $ 1.059 \pm0.006 $ & $91 / 65$ \\ \hline

   $\rm Au+Au$  &  $62.4\; (\rm non-strange)$    & $0-20\%$  &   $0.43 \pm 0.01$ & $ 121 \pm 3$ & $ 1.022 \pm 0.005 $ & $93 / 57$ \\

        &  & $20-40\%$  &   $0.36 \pm 0.02 $ & $121 \pm 3 $ & $1.039 \pm 0.004$ & $ 69 / 57$ \\

        &  & $40-80\%$\footnotemark[1]  &   $0^{+0.14}_{-0} $ & $124 \pm 4 $ & $1.068 \pm 0.007$ & $ 73 / 47$ \\  \hline

  $\rm Au+Au$  &  $62.4 \;(\rm strange)$    & $0-20\%$  &   $0.34 \pm 0.02 $ & $190 \pm 10 $ & $1.000 ^{+0.001}_{-0}$ & $ 50 / 73$ \\

         &  & $20-40\%$  &   $0.35 \pm0.02$ & $ 182 \pm 8 $ & $1.000 ^{+0.003}_{-0}$ & $ 43 / 73$ \\

        &  & $40-80\%$  &   $0.23 \pm 0.07$ & $ 169 \pm 7 $ & $1.03 \pm 0.01$ & $ 70 / 67$ \\  \hline

  $\rm Au+Au$  &  $62.4 \;(\rm all)$    & $0-20\%$  &   $0.443 \pm 0.009 $ & $137 \pm 3 $ & $1.001 ^{+0.006}_{-0.001} $ & $215 / 133$ \\

      &  & $20-40\%$  &   $ 0.38 \pm 0.01 $ & $136 \pm 3 $ & $1.021 \pm 0.004 $ & $181 / 133$ \\

       &  & $40-80\%$\footnotemark[1]  &   $0.20 \pm 0.04 $ & $139 \pm 3$ & $ 1.048 \pm 0.006 $ & $185 / 117$ \\  \hline

  $\rm Au+Au$  &  $200\; (\pi, K, p)$     & $0-10\%$\footnotemark[2]  &   $0.45 \pm 0.01 $ & $111 \pm 3$ & $ 1.039 \pm 0.004 $ & $105 / 79$ \\
             &  & $10-20\%$  &   $ 0.44\pm 0.01 $ & $113 \pm 3 $ & $1.039 \pm 0.005   $ & $98 / 79$ \\
              &   & $20-40\%$  &   $ 0.36 \pm 0.02 $ & $115 \pm 3  $ & $ 1.057 \pm 0.004 $ & $  112 / 81$ \\
             &   & $40-60\%$  &   $ 0.22 \pm 0.04 $ & $117 \pm 3 $ & $1.075 \pm 0.004 $ & $87 / 81$ \\
              &   & $60-80\%$  &   $0^{+0.04}_{-0} $ & $111 \pm 3 $ & $1.088 \pm 0.002 $ & $79 / 81$\\   \hline

  $\rm Au+Au$  &  $200\; (\rm non-strange)$     & $0-10\%$\footnotemark[2]  &   $0.43 \pm 0.01$ & $ 110 \pm 3 $ & $1.044 \pm 0.004 $ & $73 / 61$ \\

      &  & $10-20\%$  &   $0.42 \pm 0.01$ & $ 112 \pm 3$ & $ 1.047 \pm 0.005 $ & $67 / 59$ \\

       &   & $20-40\%$  &   $0.33 \pm 0.02$ & $ 114 \pm 3$ & $ 1.064 \pm 0.004 $ & $75 / 61$ \\

        &   & $40-60\%$  &   $0.16 \pm 0.05$ & $ 115 \pm 3$ & $ 1.081 \pm 0.004$ & $ 68 / 61$ \\

        &   & $60-80\%$  &   $0^{+0.04}_{-0}$ & $ 110 \pm 3$ & $ 1.088 \pm 0.002$ & $ 74 / 61$ \\  \hline

  $\rm Au+Au$  &  $200\; (\rm strange)$     & $0-10\%$\footnotemark[3]  &   $0.43 \pm 0.01 $  &   $133 \pm 4$  &   $ 1.027 \pm 0.005$  &   $ 92 / 111$ \\

        &  & $10-20\%$\footnotemark[4]  &   $ 0.43 \pm 0.01 $  &   $131 \pm 4 $  &   $1.028 \pm 0.005 $  &   $112 / 111$ \\

       &   & $20-40\%$  &   $ 0.39 \pm 0.01 $  &   $131 \pm 3 $  &   $1.037 \pm 0.005 $  &   $ 171 / 113$ \\

      &   & $40-60\%$  &   $ 0.22 \pm 0.03 $  &   $110 \pm 3 $  &   $1.081 \pm 0.004 $  &   $ 165 / 113$ \\

        &   & $60-80\%$\footnotemark[5]  &   $0.31 \pm 0.04$ & $ 110 \pm 10 $ & $1.059 \pm 0.007$ & $ 54 / 59$ \\   \hline

\end{tabular}
\end{ruledtabular}
\footnotetext[1]{Lack of measurements of $\pi^{0}$ at this centrality class \cite{Adare:2012uk}.}
\footnotetext[2]{The measurements of $\pi^{\pm}$, $p$ and $\bar{p}$ for centrality 0-12\% \cite{Abelev:2006jr} are used as 0-10\%.}
\footnotetext[3]{The measurements of $\Lambda$, $\bar{\Lambda}$, $\Xi^{+}$, $\Xi^{-}$ and $\Omega$ for centrality 0-5\% \cite{Adams:2006ke} are used as 0-10\%.}
\footnotetext[4]{Lack of measurements of $\Omega$ at this centrality class \cite{Adams:2006ke}.}
\footnotetext[5]{Lack of measurements of $\Omega$ \cite{Adams:2006ke} and intermediate $p_{T}$ $K^{\pm}$ \cite{Adare:2013esx} at this centrality class.}
\end{table*}

\begin{table*}[htbp]
\caption{\label{tabTBW3pa5}
Same as \cref{tabTBW3pa1}, but for $\sqrt{s_{\rm{NN}}}=$ 200~GeV (continued), 2760~GeV and 5020~GeV. Previous studies of 200~GeV from Refs.~\cite{Tang:2008ud,Shao:2009mu} are also listed. The difference between our results at 200~GeV from previous ones is due to the fact that more data points at 2 $\leqslant p_{T}\leqslant$ 3~GeV/$c$ are available now and are thus used in fitting, mainly $K^{\pm}$ data from PHENIX~\cite{Adare:2013esx} which provide better constraint on $q$, while data for $K^{\pm}$ used in Ref.~\cite{Tang:2008ud} are only at $p_{T}\leqslant$ 0.8~GeV/$c$ and data for $K^{\pm}$ used in Ref.\cite{Shao:2009mu} are only at $p_{T}\leqslant$ 2~GeV/$c$.
}
\begin{ruledtabular}
\begin{tabular}{ccccccc}

   $\rm system$  & $\rm \sqrt{s_{NN}} \;(\rm{GeV})$  & centrality  & $\langle\beta\rangle \; $ &  $T \;(\rm{MeV})$  & $q$ & $\chi^{2}/nDoF$ \\
  \hline

  $\rm Au+Au$  &  $200 \;(\rm all)$     & $0-10\%$\footnotemark[1]  &   $0.435 \pm 0.007$ & $ 118 \pm 2 $ & $1.036 \pm 0.003 $ & $186 / 175$ \\
         &  & $10-20\%$\footnotemark[2]  &   $0.436 \pm 0.007$ & $ 118 \pm 2 $ & $1.036 \pm 0.003 $ & $198 / 173$ \\
         &   & $20-40\%$  &   $0.378 \pm 0.009$ & $ 120 \pm 2 $ & $1.049 \pm 0.003 $ & $278 / 177$ \\
         &   & $40-60\%$  &   $0.23 \pm 0.02$ & $ 112 \pm 2$ & $ 1.078 \pm 0.003 $ & $237 / 177$ \\
         &   & $60-80\%$\footnotemark[3]  &   $0^{+0.04}_{-0}$ & $ 113 \pm 3 $ & $1.086 \pm 0.002 $ & $139 / 123$ \\   \hline

  $\rm Au+Au$  &  $200$ \;(\rm all){\rm~\cite{Tang:2008ud}}     & $0-10\%$  &   $0.470 \pm0.009  $ & $ 122 \pm 2 $ & $ 1.018 \pm 0.005$ & $130/125$ \\
          &  & $10-20\%$  &   $0.475 \pm 0.008  $ & $  122 \pm 2 $ & $  1.015\pm 0.005$ & $119/127$ \\
         &   & $20-40\%$  &   $0.441 \pm 0.009 $ & $  124 \pm 2  $ & $ 1.024\pm 0.004$ & $159/127$ \\
         &   & $40-60\%$  &   $0.282 \pm 0.017 $ & $  119 \pm 2 $ & $  1.066\pm 0.003$ & $165/135$ \\
         &   & $60-80\%$  &   $0^{+0.05}_{-0}   $ & $   114 \pm 3 $ & $  1.086 \pm 0.002$ & $138/123$ \\   \hline

  $\rm Au+Au$  &  $200$ \;(\rm all){\rm~\cite{Shao:2009mu}}     & $0-10\%$  &   $0.472 \pm0.009  $ & $ 122 \pm 3 $ & $ 1.017 \pm 0.006$ & $140/155$ \\  \hline

  $\rm Pb+Pb$  &$2760 \;(\pi, K, p)$    & $0-5\%$  &   $  0.591 \pm 0.003  $ & $  91 \pm 2  $ & $ 1.024 \pm 0.005  $ & $  247 / 213$ \\
               &   & $5-10\%$  &   $    0.587 \pm 0.003  $ & $ 91 \pm 2 $ & $  1.029 \pm 0.005  $ & $  247 / 213$ \\
                &   & $10-20\%$  &   $      0.580 \pm 0.003 $ & $  92 \pm 2 $ & $  1.035 \pm 0.005  $ & $  230 / 213$ \\
                &   & $20-30\%$  &   $        0.563 \pm 0.004 $ & $  92 \pm 2 $ & $  1.046 \pm 0.005 $ & $ 207 / 213$ \\
                &   & $30-40\%$  &   $      0.535 \pm 0.005$ & $ 92 \pm 2 $ & $ 1.061 \pm 0.004 $ & $ 201 / 213$ \\
                &   & $40-50\%$  & $        0.493 \pm 0.005 $ & $ 90 \pm 2 $ & $  1.078 \pm 0.003 $ & $ 185 / 213$ \\
                &   & $50-60\%$  &  $         0.437 \pm 0.007 $ & $ 90 \pm 2 $ & $ 1.091 \pm 0.003 $ & $ 196 / 213$ \\
                &   & $60-70\%$  &   $          0.35 \pm 0.01 $ & $ 91 \pm 2 $ & $ 1.104 \pm 0.002 $ & $ 244 / 213$ \\
                &   & $70-80\%$  & $            0.23 \pm 0.02 $ & $ 91 \pm 2$ & $ 1.116 \pm 0.002  $ & $ 299 / 213$ \\
                 &   & $80-90\%$  &   $         0^{+0.01}_{-0}   $ & $ 90 \pm 2$ & $ 1.122 \pm 0.001 $ & $ 344 / 213$ \\  \hline

  $\rm Pb+Pb$  &$2760 \;(\rm non-strange)$    & $0-10\%$  &   $0.587 \pm 0.003 $ & $88 \pm 2 $ & $1.034 \pm0.005 $ & $209 / 143$ \\

        &   & $10-20\%$  &   $0.576 \pm 0.004 $ & $89 \pm 2 $ & $1.042 \pm 0.005 $ & $201 / 143$ \\

        &   & $20-40\%$  &   $0.548 \pm0.005 $ & $91 \pm 2 $ & $1.057 \pm0.005 $ & $189 / 143$ \\

        &   & $40-60\%$  &   $0.463 \pm 0.007 $ & $90 \pm 2 $ & $1.086\pm 0.003 $ & $168 / 143$ \\

        &   & $60-80\%$  &   $0.28 \pm 0.01 $ & $91 \pm 2 $ & $1.112\pm 0.003 $ & $198 / 143$ \\

         &   & $80-90\%$  &   $0^{+0.01}_{-0} $ & $90 \pm 2$ & $ 1.120 \pm 0.001 $ & $225 / 143$\\  \hline

  $\rm Pb+Pb$  &$2760 \;(\rm strange)$    & $0-10\%$  &   $0.561 \pm0.003$ & $ 133 \pm 3 $ & $1.000 ^{+0.001}_{-0} $ & $125 / 139$ \\

         &   & $10-20\%$  &   $0.550 \pm 0.003 $ & $143 \pm 3 $ & $1.000 ^{+0.002}_{-0} $ & $74 / 139$ \\

       &   & $20-40\%$  &   $0.519 \pm 0.008$ & $ 157 \pm 5 $ & $1.006 ^{+0.008}_{-0.006}$ & $ 61 / 139$ \\

       &   & $40-60\%$  &   $0.43 \pm 0.01 $ & $148\pm 5$ & $ 1.047 \pm 0.006$ & $ 53 / 139$ \\

      &   & $60-80\%$  &   $0.25 \pm 0.03 $ & $139 \pm 5 $ & $1.088 \pm0.005 $ & $72 / 137$ \\\hline

   $\rm Pb+Pb$  &$2760 \;(\rm all)$    & $0-10\%$  &   $0.578 \pm 0.003 $ & $99 \pm 2$ & $ 1.024 \pm 0.004$ & $ 517 / 285$ \\

            &   & $10-20\%$  &   $ 0.566 \pm 0.003 $ & $100 \pm 2 $ & $1.033 \pm 0.004 $ & $467 / 285$ \\

           &   & $20-40\%$  &   $0.534 \pm 0.004 $ & $101 \pm 2 $ & $1.050 \pm 0.004$ & $ 470 / 285$ \\

        &   & $40-60\%$  &   $ 0.457 \pm 0.005 $ & $99 \pm 2$ & $ 1.078 \pm 0.003$ & $ 373 / 285$ \\

            &   & $60-80\%$  &   $0.31 \pm 0.01 $ & $96\pm 2 $ & $1.106 \pm 0.002$ & $ 396 / 283$ \\ \hline

    $\rm Pb+Pb$  &$5020\; (\pi, K, p)$    & $0-5\%$  &   $ 0.605 \pm 0.002$ & $ 93 \pm 2$ & $ 1.021 \pm 0.005$ & $ 316 / 90$ \\
        &   & $5-10\%$  &   $ 0.602 \pm 0.003$ & $ 91 \pm 2 $ & $1.030 \pm 0.005 $ & $303 / 90$ \\
       &   & $10-20\%$  &   $0.596 \pm 0.003 $ & $93 \pm 2$ & $ 1.031 \pm 0.005 $ & $317 / 90$ \\
        &   & $20-30\%$  &   $ 0.581 \pm 0.003 $ & $94 \pm 2$ & $ 1.042 \pm 0.004 $ & $268 / 90$ \\
       &   & $30-40\%$  &   $  0.557 \pm 0.003 $ & $93 \pm 2 $ & $1.058 \pm 0.004 $ & $217 / 90$ \\
       &   & $40-50\%$  & $ 0.517 \pm 0.004 $ & $92 \pm 2 $ & $1.076 \pm 0.003 $ & $189 / 90$ \\
       &   & $50-60\%$  &  $ 0.459 \pm 0.005 $ & $92 \pm 2 $ & $1.092 \pm 0.003 $ & $192 / 90$ \\
       &   & $60-70\%$  &   $    0.383 \pm 0.009$ & $ 89 \pm 2$ & $ 1.109 \pm 0.003 $ & $ 177 / 90$ \\
       &   & $70-80\%$  & $ 0.26 \pm 0.02$ & $ 88 \pm 2 $ & $1.123 \pm 0.003$ & $ 189 / 90$ \\
       &   & $80-90\%$  &   $ 0^{+0.01}_{-0} $ & $88 \pm 2$ & $ 1.131 \pm 0.003 $ & $174 / 90$ \\

\end{tabular}
\end{ruledtabular}
\footnotetext[1]{The measurements of $\pi^{\pm}$, $p$ and $\bar{p}$ for centrality 0-12\% \cite{Abelev:2006jr} are used as 0-10\%. $\Lambda$, $\bar{\Lambda}$, $\Xi^{+}$, $\Xi^{-}$ and $\Omega$ for centrality 0-5\% \cite{Adams:2006ke} are used as 0-10\%.}
\footnotetext[2]{Lack of measurements of $\Omega$ at this centrality class \cite{Adams:2006ke}.}
\footnotetext[3]{Lack of measurements of $\Omega$ \cite{Adams:2006ke} and intermediate $p_{T}$ $K^{\pm}$ \cite{Adare:2013esx} at this centrality class.}
\end{table*}

\begin{table*}[htbp]
\caption{\label{tabTBW4pa1}
Extracted kinetic freeze-out parameters and $\chi^{2}/nDoF$ from TBW4 fit to identified particle transverse spectra in heavy ion collisions of different centralities at $\sqrt{s_{\rm{NN}}}=$ 7.7, 11.5, 14.5, and 19.6~GeV.  Results for charged pions, kaons, and protons have labels `$(\pi, K, p)$' behind their collision energy. All available hadrons including strange and multi-strange particles are labeled as `(all)'.  
}
\begin{ruledtabular}
\begin{tabular}{cccccccc}

  $\rm system$  & $\rm \sqrt{s_{NN}} \;(\rm{GeV})$  & centrality  & $\langle\beta\rangle \;$ &  $T \;(\rm{MeV})$  & $q_{M}$ &  $q_{B}$ &  $\chi^{2}/nDoF$ \\
  \hline
  $\rm Au+Au$  &$7.7 \;(\pi, K, p)$    & $0-5\%$  & $0.431\pm  0.008 $ & $ 111\pm  2 $ & $ 1.000 ^{+0.002}_{-0} $ & $ 1.003 \pm 0.003$ & $112 / 132$ \\
        &   & $5-10\%$  & $0.418 \pm 0.009 $ & $ 113  \pm 3 $ & $ 1.000 ^{+0.003}_{-0} $ & $ 1.005 \pm 0.003$ & $103 / 133$ \\
        &   & $10-20\%$  & $0.39 \pm 0.01 $ & $ 117 \pm 4 $ & $ 1.006 \pm 0.005 $ & $  1.006 \pm 0.006$ & $86 / 137$ \\
         &  & $20-30\%$  & $0.34 \pm 0.02 $ & $ 123 \pm 4 $ & $ 1.008 \pm 0.005 $ & $ 1.015 \pm 0.006$ & $120 / 135$ \\
        &   & $30-40\%$  & $0.32 \pm 0.02 $ & $ 125 \pm 4 $ & $ 1.006 \pm 0.005 $ & $ 1.013 \pm 0.006$ & $118 / 134$ \\
       &    & $40-50\%$  & $0.26 \pm 0.03 $ & $ 116 \pm 4 $ & $ 1.024 \pm 0.006 $ & $  1.023 \pm 0.006$ & $110 / 124$ \\
       &    & $50-60\%$  & $0.17 \pm 0.08 $ & $ 121 \pm 4  $ & $ 1.03\pm  0.01 $ & $ 1.03 \pm 0.01$ & $130 / 121$ \\
        &   & $60-70\%$  & $0 ^{+0.095}_{-0} $ & $ 121 \pm 4 $ & $  1.033 \pm 0.004 $ & $  1.034 \pm 0.003$ & $95 / 116$ \\
       &    & $70-80\%$  & $0 ^{+0.095}_{-0} $ & $ 116 \pm 4  $ & $ 1.032 \pm 0.005  $ & $ 1.025 \pm 0.003$ & $80 / 96$ \\
        &   & $0-80\%$  &   $0.37\pm 0.02$ & $ 105 \pm 5$ & $ 1.026 \pm 0.009 $ & $1.020 \pm 0.008 $ & $46 / 89$ \\ \hline

  $\rm Au+Au$  &$7.7 \;(\rm all)$    & $0-5\%$  & $0.426 \pm 0.005 $ & $105 \pm 2 $ & $1.010 \pm 0.002$ & $ 1.0001 \pm 0.0002 $ & $234 / 171$ \\
        &   & $5-10\%$  & $0.421 \pm 0.005 $ & $106 \pm 2 $ & $1.009 \pm 0.002 $ & $1.0001 \pm 0.0002$ & $ 218 / 172$ \\
        &   & $10-20\%$  & $0.399 \pm 0.005 $ & $112 \pm 2 $ & $1.008 \pm 0.001 $ & $1.0001 \pm 0.0002 $ & $186 / 176$ \\
        &  & $20-30\%$  & $0.377 \pm 0.006 $ & $117 \pm 3 $ & $1.004 \pm 0.001 $ & $1.0001 ^{+0.0003}_{-0.0001}$ & $ 207 / 174$ \\
        &   & $30-40\%$  & $0.362 \pm 0.007 $ & $118 \pm 3 $ & $1.004 \pm 0.001 $ & $1.000 ^{+0.001}_{-0} $ & $180 / 173$ \\
        &    & $40-60\%$  & $0.32 \pm 0.01$ & $ 117 \pm 3$ & $ 1.009 \pm 0.003 $ & $ 1.003 \pm 0.002 $ & $210 / 157$ \\
        &    & $60-80\%$  & $0.24 \pm 0.02 $ & $112 \pm 3 $ & $1.022 \pm 0.004$ & $ 1.011 \pm 0.004 $ & $122 / 128$ \\ \hline

  $\rm Au+Au$  &$11.5 \;(\pi, K, p)$    & $0-5\%$  &  $0.429 \pm 0.006$ & $ 112 \pm 3 $ & $ 1.005\pm 0.002 $ &  $1.000 ^{+0.001}_{-0}$ & $ 100 / 141$ \\
        &    & $5-10\%$  &  $0.417 \pm 0.006 $ & $ 117  \pm 3 $ & $ 1.001 ^{+0.002}_{-0.001} $ & $1.000 ^{+0.002}_{-0}$   &$ 79 / 144$ \\
        &    & $10-20\%$  &   $0.404 \pm 0.006 $ & $ 118 \pm 3  $ & $ 1.004 \pm 0.002  $ & $ 1.000 ^{+0.003}_{-0}$ & $  88 / 144$ \\
        &    & $20-30\%$  & $0.36 \pm 0.02  $ & $ 121 \pm 3 $ & $ 1.012 \pm 0.005 $ &$ 1.009 \pm 0.006$  &  $ 88 / 144$ \\
       &     & $30-40\%$  &   $0.37 \pm 0.01 $ & $  112 \pm 3 $ & $  1.019 \pm 0.005  $ &$ 1.008 \pm 0.005$&  $ 90 / 143$ \\
        &    & $40-50\%$  & $0.24 \pm 0.03  $ & $ 121 \pm 3  $ & $ 1.031 \pm 0.005  $ & $1.025 \pm 0.006$ & $ 104 / 139$ \\
        &    & $50-60\%$  &   $0.24 \pm 0.03  $ & $ 116 \pm 3  $ & $  1.032 \pm 0.005  $ &$1.023 \pm 0.005$& $ 79 / 137$ \\
       &     & $60-70\%$  &  $0^{+0.14}_{-0}  $ & $ 118 \pm 3 $ & $  1.044 \pm 0.003 $ &$1.037 \pm 0.002$ &$  86 / 123$ \\
        &    & $70-80\%$  &  $0^{+0.08}_{-0} $ & $  116 \pm 3  $ & $ 1.041\pm  0.003  $ &$1.034 \pm 0.003$& $ 101 / 119$ \\
        &    & $0-80\%$  &   $0.38 \pm 0.01 $ & $109 \pm 4 $ & $1.022 \pm 0.006 $ & $1.017 \pm 0.006 $ & $30 / 117$ \\ \hline

  $\rm Au+Au$  &$11.5 \;(\rm all)$    & $0-5\%$  & $0.433 \pm 0.004 $ & $107 \pm 2$ & $ 1.010 \pm 0.001$ & $ 1.0001 ^{+0.0002}_{-0.0001}$ & $ 169 / 182$ \\
    &   & $5-10\%$  & $0.427 \pm 0.004$ & $ 108 \pm 2$ & $ 1.011 \pm 0.001$ & $ 1.0001 ^{+0.0003}_{-0.0001} $ & $163 / 185$ \\
    &   & $10-20\%$  & $0.418 \pm 0.004 $ & $110 \pm 2 $ & $1.010 \pm 0.001$ & $ 1.0001 ^{+0.0009}_{-0.0001} $ & $152 / 185$ \\
    &  & $20-30\%$  & $0.394 \pm 0.007 $ & $116 \pm 2$ & $ 1.010 \pm 0.002$ & $ 1.001 ^{+0.002}_{-0.001} $ & $133 / 185$ \\
    &   & $30-40\%$  & $0.376 \pm 0.007 $ & $113 \pm 2$ & $ 1.015 \pm 0.002$ & $ 1.004 \pm 0.002 $ & $124 / 184$ \\
    &    & $40-60\%$  & $0.30 \pm 0.01 $ & $118 \pm 3 $ & $1.022 \pm 0.002 $ & $1.013 \pm 0.002 $ & $154 / 178$ \\
    &    & $60-80\%$  & $0.25 \pm 0.02$ & $ 113 \pm 3$ & $ 1.028 \pm 0.003 $ & $1.014 \pm 0.003 $ & $131 / 157$ \\ \hline

  $\rm Au+Au$  &$14.5 \;(\pi, K, p)$    & $0-5\%$  &  $0.43 \pm 0.02$ & $ 116 \pm 4$ & $ 1.005^{+0.007}_{-0.005}  $ & $1.002 ^{+0.008}_{-0.002} $ & $55 / 148$ \\
       &     & $5-10\%$  &   $0.41 \pm 0.02$ & $ 117 \pm 4$ & $ 1.010 \pm 0.007 $ & $ 1.008 \pm 0.008 $ & $57 / 148$ \\
       &     & $10-20\%$  &   $0.39 \pm 0.02$ & $ 116 \pm 4$ & $ 1.016 \pm 0.006 $ & $ 1.015 \pm 0.007 $ & $53 / 148$ \\
        &    & $20-30\%$  &  $0.39 \pm 0.02 $ & $114 \pm 4$ & $ 1.020 \pm 0.006 $ & $ 1.015 \pm 0.007 $ & $30 / 148$ \\
        &    & $30-40\%$  &   $0.32 \pm 0.03$ & $ 122 \pm 4$ & $ 1.023 \pm 0.006 $ & $ 1.022 \pm 0.007$ & $ 57 / 148$ \\
        &    & $40-50\%$  &   $0.1 \pm 0.1$ & $ 116 \pm 4$ & $ 1.054 \pm 0.009 $ & $ 1.055 \pm 0.011 $ & $83 / 142$ \\

        &    & $50-60\%$  &   $0.27 \pm 0.03$ & $ 111 \pm 4 $ & $1.037 \pm 0.006 $ & $ 1.028 \pm 0.007$ & $ 99 / 138$ \\
        &    & $60-70\%$  &   $0.18 \pm 0.07$ & $ 118 \pm 4$ & $ 1.036 \pm 0.008 $ & $ 1.03 \pm 0.01$ & $ 79 / 130$ \\

       &     & $70-80\%$  &   $0.26 \pm 0.03$ & $ 111 \pm 4$ & $ 1.026 \pm 0.006 $ & $ 1.012 \pm 0.007 $ & $68 / 126$ \\\hline

  $\rm Au+Au$  &$19.6\; (\pi, K, p)$    & $0-5\%$  &  $0.44 \pm 0.01 $ & $  110 \pm 3 $ & $  1.013 \pm 0.005 $ &$1.010 \pm 0.006 $&$  51 / 145$ \\
       &     & $5-10\%$  &   $0.42 \pm 0.01 $ & $  110 \pm 3 $ & $  1.016 \pm 0.005  $ &$1.010 \pm 0.006$& $ 70 / 141$ \\
       &     & $10-20\%$  &   $0.40 \pm 0.01 $ & $  116 \pm 3 $ & $  1.015 \pm 0.005 $ &$1.015 \pm 0.006$ &$  73 / 141$ \\
        &    & $20-30\%$  &  $0.36 \pm 0.02$ & $   114 \pm 3 $ & $  1.028 \pm 0.005 $ &$1.022 \pm 0.005$& $  65 / 141$ \\
        &    & $30-40\%$  &   $0.30 \pm 0.02$ & $   120 \pm 3 $ & $  1.031 \pm 0.005  $ & $1.027 \pm 0.006$&$ 81 / 142$ \\
        &    & $40-50\%$  &   $0.15 \pm 0.09 $ & $  125 \pm 3$ & $  1.045\pm  0.009 $ & $1.05 \pm 0.01$&$  87 / 140$ \\
        &    & $50-60\%$  &   $0^{+0.08}_{-0} $ & $  120 \pm 3$ & $  1.053 \pm 0.002  $ &$1.047 \pm 0.002$& $ 106 / 140$ \\
        &    & $60-70\%$  &   $0^{+0.05}_{-0} $ & $  115 \pm 3 $ & $  1.055 \pm 0.003 $ &$1.044 \pm 0.002$ &$  135 / 134$ \\
       &     & $70-80\%$  &   $0^{+0.05}_{-0} $ & $  113\pm  3 $ & $  1.053 \pm 0.003  $ &$1.039 \pm 0.002$ &$ 125 / 129$ \\
       &     & $0-80\%$  &   $0.36 \pm 0.02 $ &$108 \pm 4 $ &$1.036 \pm 0.006 $ &$1.032 \pm 0.006 $ &$31 / 126$ \\ \hline

  $\rm Au+Au$ &  $19.6 \;(\rm all)$     & $0-5\%$  & $0.453 \pm 0.003$ & $ 105 \pm 2$ & $ 1.013 \pm 0.001 $ & $1.0001^{+0.0002}_{-0.0001} $ & $162 / 186$ \\
          &  & $5-10\%$  & $ 0.446 \pm 0.004 $ & $107 \pm 2$ & $ 1.013 \pm 0.002$ & $ 1.000^{+0.002}_{-0} $ & $158 / 182$ \\
           &  & $10-20\%$  & $ 0.431 \pm 0.004 $ & $110 \pm 2$ & $ 1.015 \pm 0.002$ & $ 1.003 \pm 0.002 $ & $165 / 182$ \\
           &  & $20-30\%$  & $ 0.406 \pm 0.005 $ & $109 \pm 2$ & $ 1.022 \pm 0.002$ & $ 1.009 \pm 0.001 $ & $146 / 182$ \\
           &   & $30-40\%$  & $ 0.370 \pm 0.006 $ & $110 \pm 2 $ & $1.027 \pm 0.002 $ & $1.015 \pm 0.002 $ & $198 / 183$ \\
           &   & $40-60\%$  & $ 0.305 \pm 0.009 $ & $113 \pm 2 $ & $1.035 \pm 0.002 $ & $ 1.023 \pm 0.002 $ & $230 / 181$ \\
           &   & $60-80\%$  & $ 0.17 \pm 0.03 $ & $113 \pm 2 $ & $1.048 \pm 0.003 $ & $ 1.034 \pm 0.003 $ & $197 / 169$\\

\end{tabular}
\end{ruledtabular}
\end{table*}

\begin{table*}[htbp]
\caption{\label{tabTBW4pa2}
Same as \cref{tabTBW4pa1}, but for $\sqrt{s_{\rm{NN}}}=$ 27, 39, 62.4, and 200~GeV.
}
\begin{ruledtabular}
\begin{tabular}{cccccccc}

  $\rm system$  & $\rm \sqrt{s_{NN}} \;(\rm{GeV})$  & centrality  & $\langle\beta\rangle \;$ &  $T \;(\rm{MeV})$  & $q_{M}$ &  $q_{B}$ &  $\chi^{2}/nDoF$ \\
  \hline
    $\rm Au+Au$  &$27 \;(\pi, K, p)$      & $0-5\%$  &   $0.45 \pm 0.01 $ & $  114 \pm 3 $ & $   1.004 ^{+0.006}_{-0.004} $ &$1.004 \pm 0.007$ &$   86 / 138$ \\
       &     & $5-10\%$  &   $0.43 \pm 0.01 $ & $  111 \pm 3  $ & $ 1.016 \pm 0.006  $ &$1.013  \pm 0.006$& $ 65 / 139$ \\
       &     & $10-20\%$  &   $0.41 \pm 0.01 $ & $  114 \pm 3 $ & $  1.019 \pm 0.005 $ &$ 1.017 \pm 0.006$&$  61 / 139$ \\
        &    & $20-30\%$  &  $0.38 \pm 0.02 $ & $  112 \pm 3 $ & $  1.031 \pm 0.005 $ &$1.026 \pm 0.006$ &$  50 / 139$ \\
       &     & $30-40\%$  &  $0.34 \pm 0.02 $ & $  112 \pm 3 $ & $  1.037 \pm 0.005  $ & $1.028 \pm 0.005$&$ 45 / 139$ \\
       &     & $40-50\%$  &  $0.25 \pm 0.03 $ & $  114 \pm 3 $ & $  1.051 \pm 0.005 $ &$1.043 \pm 0.006$ &$  36 / 139$ \\
       &     & $50-60\%$  &   $0.19 \pm 0.05  $ & $ 114\pm 3 $ & $  1.055 \pm 0.006  $ &$ 1.044 \pm 0.007$&$ 47 / 139$ \\
       &     & $60-70\%$  &   $0^{+0.06}_{-0}  $ & $ 114 \pm 3 $ & $  1.063 \pm 0.002 $ &$1.049 \pm 0.002 $&$  57 / 139$ \\
       &     & $70-80\%$  &   $0^{+0.05}_{-0} $ & $  113 \pm 3 $ & $  1.060 \pm 0.002  $ &$1.042 \pm 0.002$& $ 83 / 137$ \\
       &     & $0-80\%$  &   $0.39 \pm 0.01 $ & $113 \pm 3 $ & $1.026 \pm 0.005 $ & $1.024 \pm 0.006 $ & $48 / 136$ \\\hline

  $\rm Au+Au$ &  $27 \;(\rm all)$     & $0-5\%$  & $0.464 \pm 0.003$ &$ 104 \pm 2 $ &$1.014 \pm 0.001$ &$ 1.0001 ^{+0.0003}_{-0.0001}$ &$ 211 / 179$ \\
          &  & $5-10\%$  & $0.458 \pm 0.003$ &$ 105 \pm 2 $ &$1.014 \pm 0.001$ &$ 1.000 ^{+0.002}_{-0} $ &$158 / 180$ \\
          &  & $10-20\%$  & $ 0.443 \pm 0.004$ &$ 108 \pm 2$ &$ 1.017 \pm 0.002$ &$ 1.004 \pm 0.002$ &$ 139 / 180$ \\
          &  & $20-30\%$  & $0.418 \pm 0.004$ &$ 109 \pm 2$ &$ 1.024 \pm 0.002$ &$ 1.010 \pm 0.001 $ &$127 / 180$ \\
         &   & $30-40\%$  & $0.385 \pm 0.006 $ &$110 \pm 2 $ &$1.030 \pm 0.002$ &$ 1.016 \pm 0.001 $ &$101 / 180$ \\
         &   & $40-60\%$  & $   0.329 \pm 0.008 $ &$110 \pm 2$ &$ 1.040 \pm 0.002 $ &$ 1.024 \pm 0.002$ &$ 100 / 180$ \\
         &   & $60-80\%$  & $0.16 \pm 0.03 $ &$115 \pm 2 $ &$1.055 \pm 0.002 $ &$1.040 \pm 0.003 $ &$124 / 178$ \\\hline

  $\rm Au+Au$  &$39\; (\pi, K, p)$      & $0-5\%$  &   $0.47  \pm0.01 $ & $  115 \pm 3 $ & $  1.004 ^{+0.007}_{-0.004}  $ & $1.003 ^{+0.008}_{-0.003}$&$ 58 / 139$ \\
       &     & $5-10\%$  &   $0.45 \pm 0.01 $ & $  117 \pm 4 $ & $  1.009 \pm 0.006  $ &$1.005^{+0.007}_{-0.005}  $&$ 58 / 139$ \\
       &     & $10-20\%$  &   $0.42 \pm 0.01 $ & $  110 \pm 3 $ & $  1.026 \pm 0.006 $ &$ 1.020  \pm0.006$&$  43 / 139$ \\
        &    & $20-30\%$  &   $0.39 \pm 0.01 $ & $  109 \pm 3 $ & $  1.036 \pm 0.005  $ &$ 1.027 \pm 0.006$&$ 52 / 139$ \\
        &    & $30-40\%$  &   $0.36 \pm 0.02 $ & $  110 \pm 3 $ & $  1.043 \pm 0.005  $ &$1.032 \pm 0.006 $&$ 33 / 139$ \\
        &    & $40-50\%$  &   $0.27\pm  0.03 $ & $  111 \pm 3 $ & $  1.056 \pm 0.005 $ &$ 1.048 \pm 0.006$&$  43 / 139$ \\
        &    & $50-60\%$  &  $0.16 \pm 0.06 $ & $  112 \pm 3 $ & $  1.066 \pm 0.006 $ &$1.055 \pm 0.007$ &$  55 / 139$ \\
        &    & $60-70\%$  &   $0.20 \pm 0.04  $ & $ 106 \pm 3 $ & $  1.067 \pm 0.005 $ &$1.045\pm  0.006 $&$  41 / 139$ \\
        &    & $70-80\%$  &   $0^{+0.1}_{-0}  $ & $ 108 \pm 3 $ & $  1.074 \pm 0.003 $ & $1.051 \pm 0.002$&$  59 / 139 $ \\
        &    & $0-80\%$  &   $0.40 \pm 0.01 $ & $111 \pm 3 $ & $1.030 \pm 0.005 $ & $1.023 \pm 0.006 $ & $40 / 139$ \\\hline

  $\rm Au+Au$ &  $39 \; (\rm all)$     & $0-5\%$\footnotemark[1]  & $0.470 \pm 0.004$ & $ 109 \pm 2$ & $ 1.011 \pm0.002$ & $ 1.000 ^{+0.001}_{-0}$ & $ 90 / 180$ \\
           &  & $5-10\%$\footnotemark[1]  & $0.458 \pm 0.006 $ & $111 \pm 3 $ & $1.014 \pm 0.003$ & $ 1.002 ^{+0.003}_{-0.002}$ & $ 86 / 180$ \\
            &  & $10-20\%$  & $0.447 \pm 0.006$ & $ 107 \pm 2 $ & $1.023 \pm 0.002 $ & $1.008 \pm 0.002$ & $ 83 / 190$ \\
            &  & $20-30\%$\footnotemark[1]  & $0.427 \pm 0.006$ & $ 107 \pm 2$ & $ 1.029 \pm 0.003$ & $ 1.012 \pm0.002$ & $ 88 / 180$ \\
             &   & $30-40\%$\footnotemark[1]  & $0.389 \pm 0.007$ & $ 109 \pm 2$ & $ 1.036 \pm 0.002$ & $ 1.020 \pm 0.002$ & $ 58 / 180$ \\
              &   & $40-60\%$  & $ 0.33 \pm 0.01$ & $ 109 \pm 2 $ & $1.046 \pm 0.002 $ & $ 1.029 \pm 0.002 $ & $90 / 190$ \\
              &   & $60-80\%$\footnotemark[1]  & $0.21 \pm 0.02 $ & $110 \pm 2 $ & $1.060 \pm 0.002 $ & $ 1.039 \pm 0.003 $ & $75 / 180$\\ \hline

  $\rm Au+Au$  &$62.4\; (\pi, K, p)$    & $0-10\%$  &   $0.522 \pm 0.005$ & $ 85 \pm 4 $ & $1.035 \pm 0.004$ & $ 1.000 ^{+0.002}_{-0}$ & $ 26 / 64$ \\
          &    & $10-20\%$  &   $ 0.512 \pm 0.009 $ &$89 \pm 5$ &$ 1.035\pm 0.006 $ &$1.002 ^{+0.005}_{-0.002} $ &$34 / 64$ \\
          &    & $20-40\%$  &   $ 0.48 \pm 0.01 $ & $88 \pm 5 $ & $1.047 \pm 0.005 $ & $1.016 \pm 0.005 $ & $19 / 64$ \\
           &    & $40-80\%$  &   $ 0.40 \pm 0.02 $ & $91 \pm 5$ & $ 1.061 \pm 0.004$ & $ 1.033 \pm 0.004$ & $ 31 / 64 $ \\\hline

  $\rm Au+Au$  &$62.4 \;(\rm all)$ & $0-20\%$&$0.487 \pm 0.005$ & $ 104 \pm 3 $ & $1.024 \pm 0.002$&$1.000 ^{+0.001}_{-0}$ & $113 / 132$ \\

       &    & $20-40\%$  &   $0.451 \pm 0.009$ & $ 106 \pm 4$ & $ 1.034 \pm 0.004 $ & $1.011 \pm 0.004 $ & $84 / 132$ \\

   &    & $40-80\%$\footnotemark[1]  &   $0.38 \pm 0.01$ & $ 101 \pm 4 $ & $1.056 \pm 0.004 $ & $1.030 \pm 0.004 $ & $64 / 116$ \\\hline

  $\rm Au+Au$  &  $200\; (\pi, K, p)$     & $0-10\%$\footnotemark[2]  &   $0.544 \pm 0.008 $ & $79 \pm 4 $ & $1.045 \pm 0.004$ & $ 1.006 \pm 0.005 $ & $24 / 78$ \\
            &  & $10-20\%$  &   $ 0.534 \pm 0.009 $ & $80 \pm 4 $ & $1.050 \pm 0.004 $ & $1.012 \pm0.005$ & $ 24 / 78$ \\
         &  & $20-40\%$  &   $ 0.51 \pm 0.01 $ & $78 \pm 4 $ & $1.063 \pm 0.004 $ & $1.025 \pm 0.005 $ & $24 / 80$ \\
        &  & $40-60\%$  &   $ 0.44 \pm 0.02 $ & $83 \pm 4 $ & $1.074 \pm 0.003 $ & $1.043 \pm 0.005$ & $ 33 / 80$ \\
       &  & $60-80\%$  &   $ 0.31 \pm 0.04 $ & $88 \pm 5 $ & $1.088 \pm 0.003$ & $ 1.062\pm0.005 $ & $21 / 80$ \\\hline

  $\rm Au+Au$  &  $200 \;(\rm all)$     & $0-10\%$\footnotemark[3]  &   $0.458 \pm 0.006 $ & $104 \pm 3 $ & $1.044 \pm 0.003$ & $ 1.032 \pm 0.003 $ & $140 / 174$ \\
       &  & $10-20\%$\footnotemark[4]  &   $0.458 \pm 0.006$ & $ 101 \pm 3 $ & $1.048 \pm 0.003$ & $ 1.033 \pm 0.003$ & $ 119 / 172$ \\
      &  & $20-40\%$  &   $0.425 \pm 0.007 $ & $95 \pm 2$ & $ 1.063 \pm 0.003 $ & $1.044 \pm 0.003$ & $ 136 / 176$ \\
       &  & $40-60\%$  &   $0.30 \pm 0.01$ & $ 96 \pm 3$ & $ 1.083 \pm 0.002 $ & $1.070 \pm 0.003 $ & $173 / 176$ \\
      &  & $60-80\%$\footnotemark[5]  &   $0.28 \pm 0.02 $ & $92 \pm 3 $ & $1.088 \pm 0.003 $ & $1.068 \pm 0.004$ & $  56 / 122$ \\
\end{tabular}
\end{ruledtabular}

\footnotetext[1]{Lack of measurements of $\pi^{0}$ at this centrality class \cite{Adare:2012uk}.}
\footnotetext[2]{The measurements of $\pi^{\pm}$, $p$ and $\bar{p}$ for centrality 0-12\% \cite{Abelev:2006jr} are used as 0-10\%.}
\footnotetext[3]{The measurements of $\pi^{\pm}$, $p$ and $\bar{p}$ for centrality 0-12\% \cite{Abelev:2006jr} are used as 0-10\%. $\Lambda$, $\bar{\Lambda}$, $\Xi^{+}$, $\Xi^{-}$ and $\Omega$ for centrality 0-5\% \cite{Adams:2006ke} are used as 0-10\%.}
\footnotetext[4]{Lack of measurements of $\Omega$ at this centrality class \cite{Adams:2006ke}.}
\footnotetext[5]{Lack of measurements of $\Omega$ \cite{Adams:2006ke} and intermediate $p_{T}$ $K^{\pm}$ \cite{Adare:2013esx} at this centrality class.}

\end{table*}

\begin{table*}[htbp]
\caption{\label{tabTBW4pa3}
Same as \cref{tabTBW4pa1}, but for $\sqrt{s_{\rm{NN}}}=$ 2.76 and 5.02~TeV.
}
\begin{ruledtabular}
\begin{tabular}{cccccccc}

  $\rm system$  & $\rm\sqrt{s_{NN}} \;(\mathrm{TeV})$  & centrality  & $\langle\beta\rangle \;$ &  $T \;(\rm{MeV})$  & $q_{M}$ &  $q_{B}$ &  $\chi^{2}/nDoF$ \\
  \hline

  $\rm Pb+Pb$  &$2.76\; (\pi, K, p)$    & $0-5\%$  &   $     0.590 \pm 0.004$ & $ 92 \pm 2$ & $ 1.024 \pm 0.005 $ & $1.026 \pm 0.006$ & $ 246 / 212$ \\
           &     & $5-10\%$  &   $0.588 \pm 0.004 $ & $91 \pm2$ & $ 1.030 \pm 0.005$ & $ 1.028 \pm 0.006$ & $ 247 / 212$ \\
            &     & $10-20\%$  &   $ 0.584 \pm 0.004$ & $ 90 \pm 2 $ & $1.035 \pm 0.005 $ & $1.029 \pm 0.006$ & $ 225 / 212$ \\
             &    & $20-30\%$  &   $ 0.574 \pm 0.005$ & $ 88 \pm 2 $ & $1.046 \pm 0.005 $ & $1.034 \pm 0.006 $ & $191 / 212$ \\
             &    & $30-40\%$  &   $ 0.557 \pm 0.005$ & $ 84 \pm2$ & $ 1.061 \pm 0.004$ & $ 1.044 \pm 0.005$ & $ 162 / 212$ \\
             &    & $40-50\%$  &   $ 0.525 \pm 0.006$ & $ 80 \pm 2 $ & $1.079\pm 0.003 $ & $ 1.060 \pm 0.004$ & $ 127 / 212$ \\
             &    & $50-60\%$  &  $  0.485 \pm0.007$ & $ 78 \pm 2 $ & $1.094 \pm 0.003 $ & $1.073 \pm 0.004$ & $ 117 / 212$ \\
             &    & $60-70\%$  &   $  0.441 \pm 0.008$ & $ 74 \pm 2$ & $ 1.107 \pm 0.002 $ & $1.082 \pm 0.003$ & $ 105 / 212$ \\
             &    & $70-80\%$  &   $  0.40 \pm0.01$ & $ 71 \pm 2 $ & $1.117 \pm 0.002 $ & $1.088 \pm 0.003 $ & $94 / 212$ \\
             &    & $80-90\%$  &   $   0.33\pm 0.02$ & $ 69 \pm 2 $ & $1.124 \pm 0.002$ & $ 1.093 \pm 0.003$ & $ 85 / 212$ \\ \hline






  $\rm Pb+Pb$  &$2.76 \;(\rm all)$    & $0-10\%$  &   $0.577 \pm 0.003$ & $ 100 \pm 2$ & $ 1.025 \pm 0.004 $ & $1.025 \pm 0.005 $ & $513 / 284$ \\
   &  & $10-20\%$  &   $ 0.570 \pm 0.004 $ & $98 \pm 2$ & $ 1.034 \pm 0.004 $ & $1.028 \pm 0.005 $ & $ 462 / 284$ \\
   &  & $20-40\%$  &   $  0.549 \pm 0.004 $ & $94 \pm 2$ & $ 1.051 \pm 0.004 $ & $1.039 \pm 0.004 $ & $439 / 284$ \\
   &  & $40-60\%$  &   $0.498 \pm 0.005 $ & $86 \pm 2$ & $ 1.081 \pm 0.003 $ & $1.062 \pm 0.003 $ & $273 / 284$ \\
   &  & $60-80\%$  &   $0.421 \pm 0.008 $ & $77 \pm 2 $ & $1.108 \pm 0.002 $ & $1.082 \pm 0.003 $ & $162 / 282$ \\\hline

  $\rm Pb+Pb$  &$5.02\; (\pi, K, p)$    & $0-5\%$  &   $ 0.596 \pm 0.003 $ & $99 \pm 2 $ & $1.021 \pm 0.005 $ & $1.041 \pm 0.006 $ & $274 / 89$ \\
      &     & $5-10\%$  &   $ 0.596 \pm 0.003 $ & $95 \pm 2 $ & $1.028 \pm 0.005$ & $ 1.040 \pm 0.005$ & $ 286 / 89$ \\
&     & $10-20\%$  &   $ 0.591 \pm 0.003 $ & $96 \pm 2 $ & $1.031 \pm 0.005 $ & $1.040 \pm 0.005 $ & $306 / 89$ \\
&    & $20-30\%$  &   $ 0.580 \pm 0.004 $ & $95 \pm 2 $ & $1.042 \pm 0.004 $ & $1.044 \pm 0.005$ & $ 267 / 89$ \\
&    & $30-40\%$  &   $ 0.565 \pm 0.004 $ & $91 \pm 2 $ & $1.058 \pm 0.004 $ & $1.050 \pm 0.004 $ & $207 / 89$ \\
&    & $40-50\%$  &   $ 0.535 \pm 0.005 $ & $86 \pm 2 $ & $1.077 \pm 0.003$ & $ 1.064 \pm 0.004 $ & $156 / 89$ \\
&    & $50-60\%$  &  $ 0.492 \pm 0.006 $ & $83 \pm 2 $ & $1.094 \pm 0.003 $ & $1.078 \pm 0.003 $ & $128 / 89$ \\
&    & $60-70\%$  &   $ 0.447 \pm 0.008 $ & $75 \pm 2 $ & $1.112 \pm 0.003 $ & $1.089 \pm 0.003 $ & $69 / 89$ \\
&    & $70-80\%$  &   $ 0.38 \pm 0.01 $ & $73 \pm 2$ & $ 1.124 \pm 0.002 $ & $1.099 \pm 0.003 $ & $54 / 89$ \\
&    & $80-90\%$  &   $  0.32 \pm 0.02 $ & $72 \pm 2$ & $ 1.130 \pm 0.002 $ & $1.104 \pm 0.003 $ & $51 / 89$ \\

\end{tabular}
\end{ruledtabular}
\end{table*}

\begin{table*}[htbp]
\caption{\label{tabBW1}
Extracted kinetic freeze-out parameters and $\chi^{2}/nDoF$ from BGBW fits to identified particle transverse spectra in heavy ion collisions of different centralities at $\sqrt{s_{\rm{NN}}}=$ 7.7, 11.5, 14.5, and 19.6~GeV.  Results for charged pions, kaons, and protons have labels `$(\pi, K, p)$' behind their collision energy. All available hadrons including strange and multi-strange particles are labeled as `(all)'.  
}
\begin{ruledtabular}
\begin{tabular}{cccccc}

   $\rm system$  &$\rm \sqrt{s_{NN}} \;(\rm{GeV})$  & centrality  & $\langle\beta\rangle \;$ &  $T \;(\rm{MeV})$  &  $\chi^{2}/nDoF$ \\
  \hline
  $\rm Au+Au$  &$7.7 \;(\pi, K, p)$    & $0-5\%$  & $0.437 \pm 0.005 $ & $	110  \pm  2$ & $114 / 134$ \\
       &    & $5-10\%$  &   $0.428 \pm 0.006 	$ & $ 110  \pm  2$ & $107 / 135$ \\
       &    & $10-20\%$  &   $0.395\pm  0.007 $ & $	 119  \pm  2$ & $87 / 139$ \\
       &    & $20-30\%$  & $0.378 \pm 0.007 	$ & $ 120 \pm   2$ & $132 / 137$ \\
       &    & $30-40\%$  &   $0.357\pm  0.008 $ & $	 122 \pm   2$ & $127 / 136$ \\
       &    & $40-50\%$  & $0.328\pm  0.009 $ & $	 124  \pm  2$ & $127 / 126$ \\
      &     & $50-60\%$  & $0.30\pm  0.01 	$ & $ 123  \pm  2$ & $149 / 123$ \\
       &    & $60-70\%$  & $0.26\pm  0.01 	$ & $ 126  \pm  2$ & $107 / 118$ \\
       &    & $70-80\%$  & $0.19 \pm 0.02	$ & $ 131  \pm  2  $ & $93 / 98$ \\
       &     & $0-80\%$  &   $0.401 \pm 0.008 $ & $ 116 \pm 2 $ & $53 / 91$ \\ \hline

  $\rm Au+Au$  &$7.7 \;(\rm all)$    & $0-5\%$  & $   0.407 \pm 0.005 $ & $118 \pm 2$ & $ 274 / 173$ \\
      &    & $5-10\%$  &   $ 0.403 \pm 0.005$ & $ 118 \pm 2 $ & $251 / 174$ \\
      &    & $10-20\%$  &   $ 0.378 \pm 0.005$ & $ 124 \pm 2 $ & $218 / 178$ \\
      &    & $20-30\%$  & $0.365 \pm 0.005 $ & $124 \pm 2 $ & $217 / 176$ \\
      &    & $30-40\%$  &   $0.349 \pm 0.006 $ & $125 \pm 2 $ & $190 / 175$ \\
      &    & $40-60\%$  & $0.312 \pm 0.006 $ & $127 \pm 2 $ & $230 / 159$ \\
      &    & $60-80\%$  & $0.259 \pm 0.009 $ & $127 \pm 2 $ & $165 / 130$ \\ \hline

  $\rm Au+Au$  &$11.5\; (\pi, K, p)$    & $0-5\%$  &   $0.423 \pm 0.005 $ & $	 118  \pm 2$ & $105 / 143$ \\
       &     & $5-10\%$  &   $0.416 \pm 0.005 	$ & $ 119 \pm  2$ & $80 / 146$ \\
        &    & $10-20\%$  &   $0.398  \pm0.006	$ & $ 122 \pm  2$ & $92 / 146$ \\
        &    & $20-30\%$  & $0.375  \pm0.007 $ & $	 128  \pm 2$ & $95 / 146$ \\
        &    & $30-40\%$  &   $0.361  \pm0.007 	$ & $ 129  \pm 2$ & $121 / 145$ \\
       &     & $40-50\%$  & $0.302  \pm0.009 	 $ & $138  \pm 2$ & $153 / 141$ \\
        &    & $50-60\%$  &  $0.291  \pm0.009 	$ & $ 136 \pm  2$ & $139 / 139$ \\
        &    & $60-70\%$  & $0.25  \pm0.01 	$ & $ 137  \pm 2$ & $130 / 125$ \\
        &    & $70-80\%$  & $0.23  \pm0.01 $ & $	 136  \pm 2$ & $145 / 121$ \\
        &    & $0-80\%$  &   $0.403 \pm 0.007 $ & $120 \pm 2 $ & $40 / 119$ \\ \hline

  $\rm Au+Au$  &$11.5 \;(\rm all)$    & $0-5\%$  &   $0.410 \pm 0.004 $ & $122 \pm 1 $ & $228 / 184$ \\
          &    & $5-10\%$  &   $0.402 \pm 0.004 $ & $124 \pm 2 $ & $228 / 187$ \\
          &    & $10-20\%$  &   $ 0.392 \pm 0.004 $ & $126 \pm 1 $ & $215 / 187$ \\
          &    & $20-30\%$  & $0.369 \pm 0.005 $ & $131 \pm 2 $ & $186 / 187$ \\
          &    & $30-40\%$  &   $0.350 \pm 0.005 $ & $133 \pm 2 $ & $207 / 186$ \\
         &     & $40-60\%$  & $0.307 \pm 0.006 $ & $138 \pm 2 $ & $276 / 180$ \\
          &    & $60-80\%$  & $0.259 \pm 0.008 $ & $137 \pm 2 $ & $276 / 159$ \\\hline

  $\rm Au+Au$  &$14.5\; (\pi, K, p)$    & $0-5\%$  &   $0.427 \pm 0.006 $ & $120 \pm 2$ & $ 58 / 150$ \\
        &    & $5-10\%$  &   $0.416 \pm 0.007 $ & $123 \pm 2 $ & $61 / 150$ \\
        &    & $10-20\%$  &   $0.415 \pm 0.007 $ & $123 \pm 2 $ & $59 / 150$ \\
        &    & $20-30\%$  & $0.403 \pm 0.007 $ & $125 \pm 2 $ & $44 / 150$ \\
        &    & $30-40\%$  &   $0.373 \pm 0.008 $ & $130 \pm 2 $ & $71 / 150$ \\
         &   & $40-50\%$  & $0.344 \pm 0.009 $ & $133 \pm 3 $ & $139 / 144$ \\
        &    & $50-60\%$  &  $0.32 \pm 0.01 $ & $134 \pm 3 $ & $149 / 140$ \\
        &    & $60-70\%$  & $0.31 \pm 0.01 $ & $130 \pm 3 $ & $108 / 132$ \\
        &    & $70-80\%$  & $0.26\pm 0.01 $ & $133 \pm 3 $ & $107 / 128$ \\\hline

  $\rm Au+Au$  &$19.6 \;(\pi, K, p)$    & $0-5\%$  &   $0.446 \pm 0.005 $ & $	 117 \pm 2$ & $ 57 / 147$ \\
        &    & $5-10\%$  &   $0.433 \pm 0.005 $ & $	 120 \pm 2$ & $82 / 143$ \\
        &    & $10-20\%$  &   $0.421 \pm 0.006 $ & $	 122 \pm  2$ & $82 / 143$ \\
        &    & $20-30\%$  & $0.393  \pm0.006	$ & $ 129 \pm 2$ & $ 99 / 143$ \\
        &    & $30-40\%$  &   $0.357 \pm 0.007 $ & $	 135  \pm 2$ & $120 / 144$ \\
         &   & $40-50\%$  & $0.338  \pm0.008 	$ & $ 136 \pm 2$ & $ 143 / 142$ \\
        &    & $50-60\%$  &  $0.289 \pm 0.008 $ & $	 144 \pm  2$ & $ 214 / 142$ \\
        &    & $60-70\%$  & $0.257\pm  0.009 $ & $	145 \pm  2$ & $279 / 136$ \\
        &    & $70-80\%$  & $0.22  \pm0.01 $ & $	 146  \pm 2$ & $277 / 131$ \\
        &    & $0-80\%$  &   $0.409 \pm 0.006 $ & $124 \pm 2$ & $ 65 / 128$ \\ \hline

  $\rm Au+Au$  &$19.6 \;(\rm all)$    & $0-5\%$  &   $0.421  \pm 0.003$ & $126  \pm 1 $ & $293 / 188$ \\
                 &    & $5-10\%$  &   $0.414  \pm 0.003 $ & $128  \pm 1 $ & $283 / 184$ \\
                 &    & $10-20\%$  &   $0.404  \pm 0.003 $ & $131  \pm 1 $ & $279 / 184$ \\
                  &    & $20-30\%$  & $ 0.382  \pm 0.003 $ & $137  \pm 1 $ & $322 / 184$ \\
                 &    & $30-40\%$  &   $0.363  \pm 0.004 $ & $138  \pm 1 $ & $394 / 185$ \\
                 &     & $40-60\%$  & $0.330 \pm 0.004 $ & $142 \pm 1 $ & $550 / 183$ \\
                 &    & $60-80\%$  & $0.269 \pm 0.006 $ & $146 \pm 2$ & $ 644 / 171$ \\

\end{tabular}
\end{ruledtabular}
\end{table*}

\begin{table*}[htbp]
\caption{\label{tabBW2}
Same as \cref{tabBW1}, but for $\sqrt{s_{\rm{NN}}}=$ 27, 39, 62.4, and 200~GeV.
}
\begin{ruledtabular}
\begin{tabular}{cccccc}

   $\rm system$  &$\rm\sqrt{s_{NN}} \;(\rm{GeV})$  & centrality  & $\langle\beta\rangle \;$ &  $T \;(\rm{MeV})$  &  $\chi^{2}/nDoF$ \\
  \hline

  $\rm Au+Au$  &$27\; (\pi, K, p)$      & $0-5\%$  &   $0.456 \pm 0.005 $ & $	 116\pm 2$ & $87 / 140$ \\
        &    & $5-10\%$  &   $0.448 \pm 0.005 	$ & $ 118\pm 2$ & $73 / 141$ \\
        &    & $10-20\%$  &   $0.434 \pm 0.005 	$ & $ 122 \pm 2$ & $ 73 / 141$ \\
       &     & $20-30\%$  & $0.415 \pm 0.006 $ & $	 127 \pm 2$ & $86 / 141$ \\
       &     & $30-40\%$  &   $0.387  \pm0.007 	$ & $ 133\pm 2$ & $105 / 141$ \\
        &    & $40-50\%$  & $0.354 \pm 0.007 	$ & $ 139 \pm 2$ & $145 / 141$ \\
        &    & $50-60\%$  &  $0.314 \pm 0.008 	$ & $ 146 \pm 2$ & $201 / 141$ \\
        &    & $60-70\%$  & $0.274 \pm 0.009	 $ & $150 \pm 2$ & $283 / 141$ \\
        &    & $70-80\%$  & $0.23  \pm0.01 	$ & $ 153 \pm 2$ & $366 / 139$ \\
         &   & $0-80\%$  &   $0.422 \pm 0.006 $ & $125 \pm 2 $ & $73 / 138$ \\\hline

  $\rm Au+Au$  &$27 \;(\rm all)$      & $0-5\%$  &   $0.434 \pm 0.003 $ & $125 \pm 1$ & $ 351 / 181$ \\
          &    & $5-10\%$  &   $0.426 \pm 0.003 $ & $128 \pm 1 $ & $302 / 182$ \\
         &    & $10-20\%$  &   $0.414 \pm 0.003$ & $ 132 \pm 1 $ & $291 / 182$ \\
         &    & $20-30\%$  & $ 0.394 \pm0.003 $ & $139 \pm 1 $ & $329 / 182$ \\
        &    & $30-40\%$  &   $0.372\pm 0.004$ & $ 143 \pm 1 $ & $365 / 182$ \\
          &     & $40-60\%$  & $0.337 \pm 0.004 $ & $149 \pm 1 $ & $546 / 182$ \\
        &    & $60-80\%$  & $0.283 \pm 0.005 $ & $152 \pm 2 $ & $812 / 180$ \\ \hline

  $\rm Au+Au$  &$39\; (\pi, K, p)$      & $0-5\%$  &   $0.468 \pm 0.005 $ & $	117 \pm  2$ & $58 / 141$ \\
        &    & $5-10\%$  &   $0.449 \pm 0.005 $ & $	123  \pm 2$ & $63 / 141$ \\
       &     & $10-20\%$  &   $0.446 \pm 0.005 $ & $	 124  \pm 2$ & $65 / 141$ \\
        &    & $20-30\%$  & $0.425 \pm 0.006 $ & $	 129 \pm  2$ & $98 / 141$ \\
        &    & $30-40\%$  &   $0.395 \pm 0.007 $ & $	 137 \pm  2$ & $111 / 141$ \\
        &    & $40-50\%$  & $0.372  \pm0.007 $ & $	 140  \pm 2$ & $162 / 141$ \\
        &    & $50-60\%$  &  $0.330 \pm 0.008	$ & $ 147 \pm  2$ & $235 / 141$ \\
        &    & $60-70\%$  & $0.292 \pm 0.009 $ & $	 155   \pm 2$ & $339 / 141$ \\
        &    & $70-80\%$  & $0.254 \pm 0.009 	$ & $ 159   \pm 2$ & $434 / 141$ \\
        &    & $0-80\%$  &   $0.430 \pm 0.006$ & $ 128 \pm 2 $ & $70 / 141$ \\\hline

  $\rm Au+Au$  &$39 \;(\rm all)$      & $0-5\%$\footnotemark[1]  &   $0.454 \pm 0.004 $ & $123\pm 2$ & $ 132 / 182$ \\
       &    & $5-10\%$\footnotemark[1]  &   $0.442 \pm 0.004$ & $ 127 \pm 2$ & $ 134 / 182$ \\
         &    & $10-20\%$  &   $0.431 \pm 0.004$ & $ 132 \pm 2 $ & $211 / 192$ \\
        &    & $20-30\%$\footnotemark[1]  & $ 0.413 \pm 0.004$ & $ 137 \pm 2 $ & $264 / 182$ \\
         &    & $30-40\%$\footnotemark[1]  &   $0.392\pm 0.004$ & $ 143 \pm 2 $ & $312 / 182$ \\
         &     & $40-60\%$  & $0.355 \pm 0.005 $ & $151 \pm 2 $ & $632 / 192$ \\
        &    & $60-80\%$\footnotemark[1]  & $0.296 \pm 0.006$ & $ 159 \pm 2 $ & $879 / 182$ \\ \hline

  $\rm Au+Au$  &$62.4\; (\pi, K, p)$    & $0-10\%$  &   $0.474 \pm 0.006 $ & $125 \pm 3 $ & $105 / 66$ \\
               &    & $10-20\%$  &   $ 0.462 \pm 0.007 $ & $129 \pm 3 $ & $102 / 66$ \\
                &    & $20-40\%$  &   $ 0.444 \pm 0.008 $ & $135 \pm 3$ & $ 109 / 66$ \\
                &    & $40-80\%$  &   $ 0.39 \pm 0.01 $ & $148 \pm 4$ & $ 193 / 66$ \\\hline

  $\rm Au+Au$  &$62.4 \;(\rm all)$    & $0-20\%$  &   $0.445 \pm 0.005 $ & $138 \pm 2$ & $ 216 / 134$ \\

               &    & $20-40\%$  &   $0.422 \pm 0.006 $ & $145 \pm 3$ & $ 205 / 134$ \\

             &    & $40-80\%$\footnotemark[1]  &   $0.376 \pm 0.008$ & $ 155 \pm 3$ & $ 270 / 118$ \\\hline

  $\rm Au+Au$  &  $200 \;(\pi, K, p)$     & $0-10\%$\footnotemark[2]  &   $0.506 \pm 0.005 $  &   $125 \pm 2 $  &   $175 / 80$ \\
           &  & $10-20\%$  &   $ 0.503 \pm 0.006 $  &   $125 \pm 2 $  &   $153 / 80$ \\
          &  & $20-40\%$  &   $ 0.483 \pm 0.006 $  &   $134 \pm 3 $  &   $281 / 82$ \\
         &  & $40-60\%$  &   $ 0.456 \pm 0.008 $  &   $141\pm 3 $  &   $387 / 82$ \\
         &  & $60-80\%$  &   $ 0.43 \pm 0.01$  &   $ 147 \pm 4 $  &   $549 / 82$ \\\hline

  $\rm Au+Au$  &  $200 \;(\rm all)$     & $0-10\%$\footnotemark[3]  &   $0.484 \pm 0.004$ & $ 134 \pm 2 $ & $300 / 176$ \\
       &  & $10-20\%$\footnotemark[4]  &   $0.488 \pm 0.004 $ & $132 \pm 2 $ & $297 / 174$ \\

        &  & $20-40\%$  &   $0.467 \pm 0.004 $ & $140 \pm 2 $ & $519 / 178$ \\
        &  & $40-60\%$  &   $0.439 \pm 0.005$ & $ 144 \pm 2$ & $ 931 / 178$ \\
        &  & $60-80\%$\footnotemark[5]  &   $0.422 \pm  0.007$ & $ 149 \pm  3 $ & $621 / 124$ \\\hline

\end{tabular}
\end{ruledtabular}

\footnotetext[1]{Lack of measurements of $\pi^{0}$ at this centrality class \cite{Adare:2012uk}.}
\footnotetext[2]{The measurements of $\pi^{\pm}$, $p$ and $\bar{p}$ for centrality 0-12\% \cite{Abelev:2006jr} are used as 0-10\%.}
\footnotetext[3]{The measurements of $\pi^{\pm}$, $p$ and $\bar{p}$ for centrality 0-12\% \cite{Abelev:2006jr} are used as 0-10\%. $\Lambda$, $\bar{\Lambda}$, $\Xi^{+}$, $\Xi^{-}$ and $\Omega$ for centrality 0-5\% \cite{Adams:2006ke} are used as 0-10\%.}
\footnotetext[4]{Lack of measurements of $\Omega$ at this centrality class \cite{Adams:2006ke}.}
\footnotetext[5]{Lack of measurements of $\Omega$ \cite{Adams:2006ke} and intermediate $p_{T}$ $K^{\pm}$ \cite{Adare:2013esx} at this centrality class.}

\end{table*}

\begin{table*}[htbp]
\caption{\label{tabBW3}
Same as \cref{tabBW1}, but for $\sqrt{s_{\rm{NN}}}=$ 2.76, and 5.02~TeV.
}
\begin{ruledtabular}
\begin{tabular}{cccccc}

   $\rm system$  &$\rm\sqrt{s_{NN}} \;(\rm{TeV})$  & centrality  & $\langle\beta\rangle \;$ &  $T \;(\rm{MeV})$  &  $\chi^{2}/nDoF$ \\
  \hline

  $\rm Pb+Pb$  &$2.76\; (\pi, K, p)$    & $0-5\%$  &   $     0.602 \pm 0.001 $ & $99 \pm 1 $ & $265 / 214$ \\
              &    & $5-10\%$  &   $ 0.600 \pm 0.001 $ & $101 \pm 1 $ & $274 / 214$ \\
         &     & $10-20\%$  &   $ 0.597 \pm 0.002 $ & $104 \pm 1 $ & $266 / 214$ \\
         &    & $20-30\%$  & $0.590 \pm 0.002 $ & $108 \pm 1 $ & $272 / 214$ \\
        &    & $30-40\%$  &   $0.580 \pm 0.002$ & $ 114 \pm 1 $ & $334 / 214$ \\
         &    & $40-50\%$  & $ 0.566 \pm 0.002$ & $ 120 \pm 1 $ & $472 / 214$ \\
           &    & $50-60\%$  &  $ 0.549 \pm 0.003$ & $ 127 \pm 2 $ & $700 / 214$ \\
          &    & $60-70\%$  & $ 0.526 \pm 0.003 $ & $135 \pm 2 $ & $1039 / 214$ \\
           &    & $70-80\%$  & $ 0.505 \pm 0.004 $ & $142 \pm 2 $ & $1371 / 214$ \\
          &    & $80-90\%$  &   $ 0.484 \pm 0.005$ & $ 143 \pm 2 $ & $1661 / 214$ \\  \hline

  $\rm Pb+Pb$  &$2.76 \;(\rm all)$    & $0-10\%$  &   $0.589 \pm0.001$ & $ 108 \pm 1 $  &   $541 / 286$ \\
       &  & $10-20\%$  &   $ 0.584 \pm 0.001 $  &   $113 \pm 1 $  &   $519 / 286$ \\
      &  & $20-40\%$  &   $0.569 \pm 0.002 $  &   $122 \pm 1 $  &   $601 / 286$ \\
      &  & $40-60\%$  &   $0.542 \pm 0.002$  &   $ 134 \pm 1 $  &   $  816 / 286$ \\
      &  & $60-80\%$  &   $0.507 \pm 0.003 $  &   $146 \pm 2 $  &   $1496 / 284$ \\\hline

    $\rm Pb+Pb$  &$5.02 \; (\pi, K, p)$    & $0-5\%$  &   $  0.613 \pm 0.001$  &   $ 99 \pm 1 $  &   $334 / 91$ \\
&    & $5-10\%$  &   $ 0.613 \pm 0.001 $  &   $100 \pm 1 $  &   $338 / 91$ \\
&     & $10-20\%$  &   $ 0.609 \pm 0.001 $  &   $103 \pm 1$  &   $ 356 / 91$ \\
&    & $20-30\%$  & $0.602 \pm 0.001 $  &   $108 \pm 1 $  &   $336 / 91$ \\
&    & $30-40\%$  &   $ 0.593 \pm 0.001$  &   $ 114 \pm 1$  &   $ 377 / 91$ \\
 &    & $40-50\%$  & $ 0.579 \pm 0.002$  &   $ 121 \pm 1$  &   $ 519 / 91$ \\
&    & $50-60\%$  &  $ 0.559 \pm 0.002 $  &   $131 \pm 1$  &   $ 830 / 91$ \\
&    & $60-70\%$  & $ 0.545 \pm 0.003 $  &   $132 \pm 2$  &   $ 903 / 91$ \\
&    & $70-80\%$  & $ 0.521 \pm 0.004 $  &   $140 \pm 2$  &   $ 1215 / 91$ \\
&    & $80-90\%$  &   $ 0.502 \pm 0.005 $  &   $140 \pm 2$  &   $ 1324 / 91$ \\

\end{tabular}
\end{ruledtabular}
\end{table*}


\begin{thebibliography}{99}

\bibitem{Schnedermann:1993ws}
  E.~Schnedermann, J.~Sollfrank and U.~W.~Heinz,
  Phys.\ Rev.\ C {\bf 48}, 2462 (1993)

\bibitem{Schnedermann:1994gc}
  E.~Schnedermann and U.~W.~Heinz,
  Phys.\ Rev.\ C {\bf 50}, 1675 (1994)

\bibitem{Retiere:2003kf}
F.~Retiere and M.~A.~Lisa,
Phys. Rev. C \textbf{70}, 044907 (2004)

\bibitem{De:2007zza}
  B.~De, S.~Bhattacharyya, G.~Sau and S.~K.~Biswas,
  Int.\ J.\ Mod.\ Phys.\ E {\bf 16}, 1687 (2007).

\bibitem{Wilk:2008ue}
  G.~Wilk and Z.~Wlodarczyk,
  Eur.\ Phys.\ J.\ A {\bf 40}, 299 (2009)

\bibitem{Alberico:1999nh}
  W.~M.~Alberico, A.~Lavagno and P.~Quarati,
  Eur.\ Phys.\ J.\ C {\bf 12}, 499 (2000)

\bibitem{Osada:2008sw}
  T.~Osada and G.~Wilk,
  Phys.\ Rev.\ C {\bf 77}, 044903 (2008)
  Erratum: [Phys.\ Rev.\ C {\bf 78}, 069903(E) (2008)]

\bibitem{Biro:2003vz}
  T.~S.~Biro and B.~Muller,
  Phys.\ Lett.\ B {\bf 578}, 78 (2004)

\bibitem{Bhattacharyya:2015hya}
T.~Bhattacharyya, J.~Cleymans, A.~Khuntia, P.~Pareek and R.~Sahoo,
Eur. Phys. J. A \textbf{52}, 30 (2016)

\bibitem{Jiang:2013gxa}
  K.~Jiang, Y.~Zhu, W.~Liu, H.~Chen, C.~Li, L.~Ruan, Z.~Tang and Z.~Xu,
  Phys.\ Rev.\ C {\bf 91}, 024910 (2015)


\bibitem{Tang:2008ud}
  Z.~Tang, Y.~Xu, L.~Ruan, G.~van Buren, F.~Wang and Z.~Xu,
  Phys.\ Rev.\ C {\bf 79}, 051901(R) (2009)

\bibitem{STAR:2004bgh}
J.~Adams \textit{et al.} [STAR Collaboration],
Phys. Rev. C \textbf{71}, 064902 (2005)

\bibitem{Wilk:1999dr}
  G.~Wilk and Z.~Wlodarczyk,
  Phys.\ Rev.\ Lett.\  {\bf 84}, 2770 (2000)


\bibitem{Wong:2015mba}
C.~Y.~Wong, G.~Wilk, L.~J.~L.~Cirto and C.~Tsallis,
Phys. Rev. D \textbf{91}, 114027 (2015)


\bibitem{ALICE:2012aqc}
B.~Abelev \textit{et al.} [ALICE Collaboration],
Phys. Lett. B \textbf{720}, 52-62 (2013)

\bibitem{Urmossy:2014gpa}
K.~Urmossy, T.~S.~Bir\'o, G.~G.~Barnaf\"oldi and Z.~Xu,
[arXiv:1405.3963 [hep-ph]].

\bibitem{Urmossy:2015kva}
K.~Urmossy, G.~G.~Barnaf\"oldi, S.~Harangoz\'o, T.~S.~Bir\'o and Z.~Xu,
J. Phys. Conf. Ser. \textbf{805}, 012010 (2017)

\bibitem{Rybczynski:2014ura}
M.~Rybczy\'nski, G.~Wilk and Z.~W\l{}odarczyk,
EPJ Web Conf. \textbf{90}, 01002 (2015)



\bibitem{Urmossy:2011xk}
K.~Urmossy, G.~G.~Barnafoldi and T.~S.~Biro,
Phys. Lett. B \textbf{701}, 111-116 (2011)




\bibitem{Broniowski:2001we}
W.~Broniowski and W.~Florkowski,
Phys. Rev. Lett. \textbf{87}, 272302 (2001)



\bibitem{Mazeliauskas:2019ifr}
A.~Mazeliauskas and V.~Vislavicius,
Phys. Rev. C \textbf{101}, 014910 (2020)







\bibitem{AggarwalPRC.84.034909}
M.~M.~Aggarwal  \textit{et al.} [STAR Collaboration],
Phys. Rev. C. \textbf{84}, 034909 (2011)

\bibitem{AbelevPRL.97.132301}
B.~I.~Abelev  \textit{et al.} [STAR Collaboration],
Phys. Rev. Lett.  \textbf{97}, 132301 (2006)


\bibitem{AcharyaPRC.99.024905}
S.~Acharya\textit{et al.} [ALICE Collaboration],
Phys. Rev. C. \textbf{99}, 024905 (2019)


\bibitem{Abelev:2008ab}
  B.~I.~Abelev {\it et al.} [STAR Collaboration],
  Phys.\ Rev.\ C {\bf 79}, 034909 (2009)


\bibitem{Motornenko:2019jha}
A.~Motornenko, V.~Vovchenko, C.~Greiner and H.~Stoecker,
Phys. Rev. C \textbf{102}, 024909 (2020)


\bibitem{Andronic:2014zha}
  A.~Andronic,
  Int.\ J.\ Mod.\ Phys.\ A {\bf 29}, 1430047 (2014)

\bibitem{Zhang:2016tbf}
  S.~Zhang, Y.~G.~Ma, J.~H.~Chen and C.~Zhong,
  Adv.\ High Energy Phys.\  {\bf 2016}, 9414239 (2016)

\bibitem{Adamczyk:2017iwn}
  L.~Adamczyk {\it et al.} [STAR Collaboration],
  Phys.\ Rev.\ C {\bf 96}, 044904 (2017)



\bibitem{Abelev:2012wca}
  B.~Abelev {\it et al.} [ALICE Collaboration],
  Phys.\ Rev.\ Lett.\  {\bf 109}, 252301 (2012)






\bibitem{Lao:2016gxv}
H.~L.~Lao, F.~H.~Liu and R.~A.~Lacey,
Eur. Phys. J. A \textbf{53}, 44 (2017)
Erratum: [Eur. Phys. J. A \textbf{53}, 143 (2017)]


\bibitem{Lao:2017skd}
  H.~L.~Lao, F.~H.~Liu, B.~C.~Li and M.~Y.~Duan,
  Nucl.\ Sci.\ Tech.\  {\bf 29}, 82 (2018)

\bibitem{Zhang:2014jug}
  S.~Zhang, Y.~G.~Ma, J.~H.~Chen and C.~Zhong,
  Adv.\ High Energy Phys.\  {\bf 2015}, 460590 (2015)


\bibitem{Das:2014qca}
  S.~Das [STAR Collaboration],
  EPJ Web Conf.\  {\bf 90}, 08007 (2015)

\bibitem{Luo:2015doi}
  X.~Luo,
  Nucl.\ Phys.\ A {\bf 956}, 75 (2016)

\bibitem{Chatterjee:2015fua}
  S.~Chatterjee, S.~Das, L.~Kumar, D.~Mishra, B.~Mohanty, R.~Sahoo and N.~Sharma,
  Adv.\ High Energy Phys.\  {\bf 2015}, 349013 (2015).






\bibitem{Shao:2009mu}
  M.~Shao, L.~Yi, Z.~Tang, H.~Chen, C.~Li and Z.~Xu,
  J.\ Phys.\ G {\bf 37}, 085104 (2010)

\bibitem{Tang:2011xq}
  Z.~Tang {\it et al.},
  Chin.\ Phys.\ Lett.\  {\bf 30}, 031201 (2013)




\bibitem{Abelev:2009bw}
  B.~I.~Abelev {\it et al.} [STAR Collaboration],
  Phys.\ Rev.\ C {\bf 81}, 024911 (2010)



\bibitem{Adam:2019dkq}
  J.~Adam {\it et al.} [STAR Collaboration],
  Phys.\ Rev.\ C\  {\bf 101}, 024905 (2020)



\bibitem{Abelev:2007ra}
  B.~I.~Abelev {\it et al.} [STAR Collaboration],
  Phys.\ Lett.\ B {\bf 655}, 104 (2007)

\bibitem{Adams:2003xp}
  J.~Adams {\it et al.} [STAR Collaboration],
  Phys.\ Rev.\ Lett.\  {\bf 92}, 112301 (2004)

\bibitem{Abelev:2006jr}
  B.~I.~Abelev {\it et al.} [STAR Collaboration],
  Phys.\ Rev.\ Lett.\  {\bf 97}, 152301 (2006)


\bibitem{Adam:2019koz}
  J.~Adam {\it et al.} [STAR Collaboration],
  Phys.\ Rev.\ C {\bf 102}, 034909 (2020)




\bibitem{Aggarwal:2010ig}
  M.~M.~Aggarwal {\it et al.} [STAR Collaboration],
  Phys.\ Rev.\ C {\bf 83}, 024901 (2011)

\bibitem{Abelev:2008aa}
  B.~I.~Abelev {\it et al.} [STAR Collaboration],
  Phys.\ Rev.\ C {\bf 79}, 064903 (2009)


\bibitem{Adare:2012uk}
  A.~Adare {\it et al.} [PHENIX Collaboration],
  Phys.\ Rev.\ Lett.\  {\bf 109}, 152301 (2012)


\bibitem{Adare:2013esx}
A.~Adare \textit{et al.} [PHENIX Collaboration],
Phys. Rev. C \textbf{88}, 024906 (2013)


\bibitem{Adams:2006ke}
  J.~Adams {\it et al.} [STAR Collaboration],
  Phys.\ Rev.\ Lett.\  {\bf 98}, 062301 (2007)



\bibitem{Abelev:2007rw}
B.~I.~Abelev \textit{et al.} [STAR Collaboration],
Phys. Rev. Lett. \textbf{99}, 112301 (2007)





\bibitem{Abelev:2013vea}
  B.~Abelev {\it et al.} [ALICE Collaboration],
  Phys.\ Rev.\ C {\bf 88}, 044910 (2013)


\bibitem{Abelev:2013xaa}
  B.~B.~Abelev {\it et al.} [ALICE Collaboration],
  Phys.\ Rev.\ Lett.\  {\bf 111}, 222301 (2013)

\bibitem{ABELEV:2013zaa}
  B.~B.~Abelev {\it et al.} [ALICE Collaboration],
  Phys.\ Lett.\ B {\bf 728}, 216 (2014)
  Erratum: [Phys.\ Lett.\ B {\bf 734}, 409 (2014)]


\bibitem{Acharya:2019yoi}
S.~Acharya \textit{et al.} [ALICE Collaboration],
Phys. Rev. C \textbf{101}, 044907 (2020)



\bibitem{Adams:2005dq}
  J.~Adams {\it et al.} [STAR Collaboration],
  Nucl.\ Phys.\ A {\bf 757}, 102 (2005)

\bibitem{vanHecke:1998yu}
  H.~van Hecke, H.~Sorge and N.~Xu,
  Phys.\ Rev.\ Lett.\  {\bf 81}, 5764 (1998)

\bibitem{Adams:2003fy}
  J.~Adams {\it et al.} [STAR Collaboration],
  Phys.\ Rev.\ Lett.\  {\bf 92}, 182301 (2004)

\bibitem{Petrovici:2009pd}
  M.~Petrovici and A.~Pop,
  Rom.\ J.\ Phys.\  {\bf 57}, 419 (2012)

\bibitem{Ristea:2013ara}
  O.~Ristea, A.~Jipa, C.~Ristea, T.~Esanu, M.~Calin, A.~Barzu, A.~Scurtu and I.~Abu-Quoad,
  J.\ Phys.\ Conf.\ Ser.\  {\bf 420}, 012041 (2013).

\bibitem{Wilk:2009nn}
  G.~Wilk and Z.~Wlodarczyk,
  Phys.\ Rev.\ C {\bf 79}, 054903 (2009)


\bibitem{Kharzeev:2007wb}
D.~Kharzeev and K.~Tuchin,
JHEP \textbf{09}, 093 (2008)

\bibitem{Karsch:2007jc}
F.~Karsch, D.~Kharzeev and K.~Tuchin,
Phys. Lett. B \textbf{663}, 217-221 (2008)
















\end{thebibliography}
\end{document}